\documentclass[journal=jacsat,manuscript=article]{achemso}

\usepackage[version=3]{mhchem} 
\usepackage{tikz}
\usetikzlibrary{decorations.pathmorphing, shapes}
\usepackage{subcaption}
\usepackage{braket}
\usepackage{graphicx}
\usepackage{xcolor}
\usepackage{hyperref}
\usepackage{url}
\usepackage{siunitx}
\usepackage{physics}
\usepackage[percent]{overpic}

\AtBeginDocument{\RenewCommandCopy\qty\SI}
\ExplSyntaxOn
\msg_redirect_name:nnn { siunitx } { physics-pkg } { none }
\ExplSyntaxOff

\usepackage{etoolbox}
\usepackage{colortbl}

\newbool{finalversion}
\booltrue{finalversion} 
\newcommand{\oldrow}{\ifbool{finalversion}{\commentoutrow}{\rowcolor{gray!20}}}

\newcommand{\newrow}{\ifbool{finalversion}{}{\rowcolor{red!10}}}

\makeatletter
\newcommand{\commentoutrow}[1][]{%
  \aftergroup\remove@to@nn
}
\newcommand{\remove@to@nn}[1]{\ignorespaces}
\makeatother

\ifbool{finalversion}{
    \newcommand{\remove}[1]{}
    \newcommand{\added}[1]{#1}
    \newcommand{\comment}[1]{}
    \newcommand{\removeTableShort}[2]{#2}
}{
    \newcommand{\remove}[1]{{\color{gray}(REMOVED: #1)}}
    \newcommand{\added}[1]{{\color{red} #1}}
    \newcommand{\comment}[1]{{\color{blue} (Intern) #1}}
    \newcommand{\removeTableShort}[2]{{{\color{red} #2}}}
}

\newcommand{\replaced}[2]{\remove{#1} \added{#2}}

\DeclareSIUnit\angstrom{\text {Å}}

\newcommand{\precision}{2}

\newcommand{\reffigure}[1]{Figure~\ref{#1}}
\newcommand{\reftable}[1]{Table~\ref{#1}}

\newcommand{\symci}{\textit{SymbolicCI}}
\newcommand{\noci}{\textit{NOCI-F}}
\newcommand{\symciShort}{\textit{SymCI}}
\newcommand{\nociShort}{\textit{NOCI}}

\newcommand{\ffNames}[1]{#1}
\newcommand{\ffspin}[2]{\ffNames{\textsuperscript{#1}(#2)}}
\newcommand{\leleR}{\ffNames{LELE}}
\newcommand{\lectR}{\ffNames{LECT}}
\newcommand{\ctctR}{\ffNames{CTCT}}

\newcommand{\ctxR}{\ffNames{CTX}}
\newcommand{\sepleleR}{\ffNames{LE$\ldots$LE}}
\newcommand{\sepctctR}{\ffNames{CT$\ldots$CT}}
\newcommand{\ttR}{\ffNames{TT}}
\newcommand{\tcttR}{\ffNames{TCT$_T$}}
\newcommand{\cttcttR}{\ffNames{CT$_T$CT$_T$}}
\newcommand{\onettR}{\ffspin{1}{TT}}
\newcommand{\seponettR}{\ffspin{1}{T$\ldots$T}}
\newcommand{\threettR}{\ffspin{3}{TT}}
\newcommand{\fivettR}{\ffspin{5}{TT}}
\newcommand{\leR}{\ffNames{LE}}
\newcommand{\ctR}{\ffNames{CT}}
\newcommand{\gsR}{\ffNames{GS}}

\newcommand{\tR}{\ffNames{T}}
\newcommand{\dpR}{\ffNames{D}$^+$}
\newcommand{\dmR}{\ffNames{D}$^-$}
\newcommand{\soneR}{\ffNames{S$_1$}}
\newcommand{\szeroR}{\ffNames{S$_0$}}
\newcommand{\toneR}{\ffNames{T$_1$}}
\newcommand{\dpdmR}{\ffNames{\dpR{}\dmR{}}}
\newcommand{\sepdpdmR}{\ffNames{\dpR{}$\ldots$\dmR{}}}

\newcommand{\sone}{\ensuremath{\mathrm{S}_1}}
\newcommand{\szero}{\ensuremath{\mathrm{S}_0}}

\newcommand{\dpm}{\ensuremath{\mathrm{D}^+}}
\newcommand{\dmm}{\ensuremath{\mathrm{D}^-}}
\newcommand{\oneE}[2]{\ensuremath{\langle #1|\hat{h}|#2\rangle}}
\newcommand{\twoE}[4]{\ensuremath{\langle #1,#2|#3,#4\rangle}}

\definecolor{colorH_agg}{HTML}{a93533}
\definecolor{colorDavydov0_agg}{HTML}{2e60a0}
\definecolor{colorZeroCoupling_agg}{HTML}{a84a89}
\definecolor{colorJ_agg}{HTML}{57a561}
\author{Johannes E. Adelsperger}
\affiliation{Center for Nanosystems Chemistry, Julius-Maximilians University Würzburg, Theodor Boveri-Weg 1, 97074 Würzburg, Germany and Institute for Physical and Theoretical Chemistry, Julius-Maximilians University Würzburg, Emil-Fischer Straße 42, 97074 Würzburg, Germany}

\author{Coen de Graaf}
\email{coen.degraaf@urv.cat}
\affiliation{Departament de Química Física i Inorgànica, Universitat Rovira i Virgili, 43007 Tarragona, Spain}
\alsoaffiliation{ICREA, Pg. Lluís Companys 23, 08010 Barcelona, Spain}

\author{Merle I. S. Röhr}
\email{merle.roehr@uni-wuerzburg.de}
\affiliation{Center for Nanosystems Chemistry, Julius-Maximilians University Würzburg, Theodor Boveri-Weg 1, 97074 Würzburg, Germany and Institute for Physical and Theoretical Chemistry, Julius-Maximilians University Würzburg, Emil-Fischer Straße 42, 97074 Würzburg, Germany}

\title[SYMBOL-CI]
  {Fragment-Based Configuration Interaction: Towards a Unifying Description of Biexcitonic Processes in Molecular Aggregates}

\abbreviations{CI,CASSCF}
\keywords{American Chemical Society, \LaTeX}

\usepackage{xr}
\makeatletter
\newcommand*{\addFileDependency}[1]{
\typeout{(#1)}
\@addtofilelist{#1}
\IfFileExists{#1}{}{\typeout{No file #1.}}
}\makeatother

\addFileDependency{SI.tex}
\addFileDependency{SI.aux}
\begin{document}
\onecolumn

\begin{tocentry}
\includegraphics[height=4.5cm]{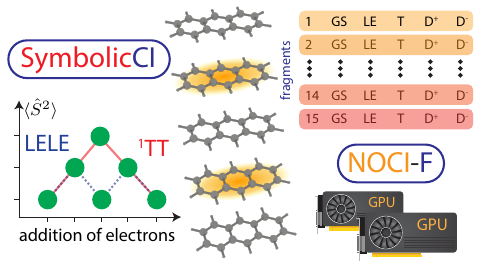}




\end{tocentry}

\begin{abstract}
Biexcitonic states govern singlet fission, triplet–triplet and exciton–exciton annihilation, yet a unified understanding of how these processes compete within a shared electronic manifold remains elusive. We outline a conceptual framework based on fragment-based configuration-interaction that systematically constructs diabatic Hamiltonians spanning the full one-particle (\leR{}, \ctR{}) and two-particle (\leleR{}, \ctctR{}, \ttR{}, \ctxR{} with X = LE, CT, or T) manifolds from monomer-local building blocks, preserving physical interpretability throughout. \symci{} provides analytic Hamiltonian matrix elements for efficient large-scale calculations; \noci{} delivers benchmark-quality couplings. The resulting diabatic Hamiltonians can be coupled to quantum dynamics simulations. Applications to ethylene aggregates and the anthracene crystal reveal \ctxR{} configurations as electronic gateways bridging excitonic manifolds, with CT-mediated relaxation pathways competing with conventional annihilation. In H-type aggregates, \lectR{} admixture stabilizes a ``bi-excimer'' analogous to one-particle excimers. By providing first-principles access to biexciton formation, separation, and transport, we hope to stimulate further exchange between electronic structure and quantum dynamics communities toward a predictive understanding of multiexcitonic photophysics.

\end{abstract}

\section{Introduction}
Biexcitonic states play a central role in light-driven processes such as singlet fission\cite{Smith2010}, exciton--exciton annihilation\cite{Tempelaar2017}, triplet--triplet annihilation\cite{Bossanyi_Nat_chem}, and high-energy charge generation\cite{Schlesinger2020} in electronically coupled molecular aggregates. They arise either from the correlation of two independently formed excitons\cite{Dostal2018} or through direct coupling from the optically accessible single-exciton manifold\cite{Casanova2018}. Once formed, biexcitons exhibit rich and highly system-dependent electronic structure: depending on molecular packing and intermolecular coupling, they may display long-range configuration mixing, pronounced charge-transfer character, or strong spin correlations. As a result, the biexcitonic manifold comprises a diverse set of correlated two-particle states, including Frenkel-type biexcitons (\leleR{})\cite{GuttierrezMeza2021}, charge-transfer biexcitons (\ctctR{})\cite{Schlesinger2020}, singlet-coupled triplet pairs (\onettR{} or here \ttR{})\cite{Miyata2019}, and mixed \ctxR{} configurations\cite{Scholes2015}. Coexisting with single excitons in a shared electronic manifold, these states compete with them in shaping photophysical function and loss pathways.

Despite their importance, biexcitons remain challenging to describe theoretically. Their intrinsic double-excitation character places them beyond the scope of linear-response methods, such as time-dependent density functional theory, which are restricted to the single-exciton manifold.\cite{Dreuw2005} Phenomenological exciton models have been extended to include two-particle terms or annihilation operators,\cite{Spano1991,Tempelaar2017} but these descriptions are typically system specific and often neglect entire classes of configurations, in particular those involving charge transfer. Even within \textit{ab initio} electronic-structure theory, most studies have focused on selected biexcitonic species---most prominently the \ttR{} state relevant for singlet fission\cite{Smith2010,Berkelbach2013a,Berkelbach2013b,Miyata2019}---while a general, chemically interpretable framework that treats \leleR{}, \ctctR{}, and mixed \ctxR{} configurations on equal footing is still lacking. As a consequence, key questions regarding the structure, energetics, and coupling mechanisms of biexcitons in extended systems remain unresolved, particularly beyond the dimer \replaced{limit}{model}.

In this Perspective, we outline a configuration-interaction framework that provides a unified \textit{ab initio} description of the full biexcitonic manifold. Rather than focusing on individual biexciton species, the approach systematically constructs all two-particle configurations arising from monomer-local \leR{}, cation \dpR{}, anion \dmR{} and triplet \tR{} building blocks, including adjacent and spatially separated Frenkel biexcitons, charge-separated double excitons, triplet-pair states, and mixed \ctxR{} configurations. We discuss two complementary fragment-based realizations of this framework. The \symci{} approach enables an efficient construction of diabatic Hamiltonians in fragment-local active spaces, allowing large aggregates to be treated with modest computational cost, while the \noci{} methodology provides benchmark-quality biexciton couplings by combining fully relaxed multiconfigurational fragment states within a nonorthogonal CI formalism. Together, these methods establish a coherent first-principles foundation for biexciton theory.
Beyond the question of how correlated two-particle states are accessed from the single-exciton manifold, a complete understanding of biexcitonic photophysics requires knowledge of how these correlated states propagate through the aggregate. In analogy to exciton transport in the one-particle picture, biexciton 'transport' is governed by the electronic couplings between spatially distinct biexcitonic configurations. The fragment-based CI framework provides direct access to these inter-biexciton couplings, enabling a systematic analysis of both formation and migration pathways.
After briefly summarizing experimental observations that motivate the explicit inclusion of the full biexcitonic manifold in photophysical models,\cite{Schlesinger2020,Dostal2018,Zeng2014} we introduce the conceptual principles underlying the fragment-based CI approach and illustrate its implications using representative ethylene and anthracene aggregates. These examples demonstrate how a unified electronic-structure framework clarifies multiexciton connectivity in extended molecular systems and opens new avenues for the predictive modeling of correlated excited-state phenomena in functional molecular materials.

\begin{figure*}[ht]
    \centering
    \includegraphics[width=\linewidth]{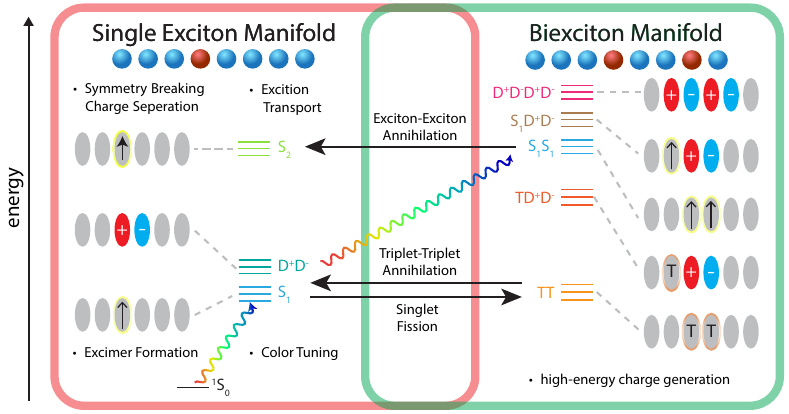}
    \caption{Scheme illustrating the energetic ordering of the one-particle exciton and the biexciton manifold, indicating the photophysical processes that proceed via intermanifold pathways.}
    \label{fig:overview}
\end{figure*}

\subsection{Frenkel--Frenkel Biexcitons (\leleR{})}

Frenkel--Frenkel biexcitons arise from two local electronic excitations residing on different chromophores within a molecular aggregate.\cite{Spano1991} The two excitons interact primarily through Coulomb coupling between transition densities and may become partially delocalized, depending on intermolecular distance, relative orientation, and aggregate packing. As a consequence, the \leleR{} manifold comprises both nearby and spatially separated exciton pairs, with binding energies and splittings that are highly sensitive to aggregate geometry and exciton--exciton interactions.

Despite their conceptual simplicity, \leleR{} biexcitons are rarely treated explicitly within \textit{ab initio} electronic-structure theory. \added{One reason for this is that many standard \textit{ab initio} methods, such as density functional theory in its adiabatic linear-response formulation or single-reference coupled-cluster theory, lack an explicit description of double excitations; alternatively, where such states are accessible---as in multireference configuration interaction---the computational scaling typically restricts their application to small molecular systems.}  \replaced{Instead}{Therefore}, most theoretical insight has been derived from phenomenological Frenkel-exciton models augmented by effective two-particle interaction terms or annihilation operators.\cite{Tempelaar2017} Early analytical and model-Hamiltonian treatments established how dimensionality, Coulomb repulsion, and intermolecular coupling govern biexciton formation and binding,\cite{Vektaris1994,Gallagher1996,Agranovich2003} while later exact-diagonalization studies of extended Hubbard-type models highlighted the strong dependence of biexciton stability on geometry and electronic correlation.\cite{Dixit1991,Chandross1997PRB,Shukla2003,Psiachos2009} Although these approaches capture key qualitative trends, they typically neglect charge-transfer contributions and lack the configurational resolution required to describe multiexciton connectivity in realistic molecular aggregates.

Experimentally, \leleR{} biexcitons are accessed under conditions of high excitation density, such as femtosecond laser experiments and multidimensional spectroscopies. In two-dimensional electronic spectroscopy, ground-to-biexciton coherences manifest as signals in the double-quantum-coherence region.\cite{BittnerSilva2022,maly_wavelike_2020} Two-dimensional photoluminescence excitation studies in polymeric semiconductors further demonstrate the coexistence of bound and unbound \leleR{} biexcitons, with the sign and magnitude of the binding energy governed by aggregate packing motifs.\cite{GuttierrezMeza2021}

While effective Frenkel Hamiltonians parameterized from quantum-chemical input can rationalize these observations in specific cases, they remain inherently system dependent. In particular, couplings between \leleR{} configurations and single or double charge-transfer states are usually neglected, leaving open whether such interactions merely renormalize Frenkel biexciton energies or actively mediate transitions between excitation sectors.

\subsection{Charge-Transfer--Charge-Transfer (\ctctR) Biexcitons}

Charge-transfer (\ctR{}) excitons play a central role in the photophysics of donor--acceptor assemblies. The corresponding \ctctR{} biexcitons, formed by two interacting \ctR{} excitons on distinct \sepdpdmR{} units, constitute a distinct class of two-particle states within the biexciton manifold. Their existence was anticipated in early model analyses of mixed-stack molecular crystals,\cite{Mavroyannis1977} and related many-exciton bound states, often termed “exciton strings”, have been observed experimentally in \ctR{} crystals.\cite{KuwataGonokami1994}

Direct evidence for \ctctR{} biexciton dynamics has recently emerged from studies of the PXX-Ph$_4$PDI co-crystal.\cite{Schlesinger2020} Following high-density photoexcitation, initially formed \leR{} singlets rapidly undergo charge separation, generating \ctR{} excitons that diffuse along one-dimensional \dpdmR{} stacks. When two \ctR{} excitons encounter one another on adjacent \dpdmR{} pairs, an annihilation channel opens in which a correlated \ctctR{} configuration gives rise to a pair of spatially separated, high-energy charges (\sepdpdmR{}) with long lifetimes. These species lie energetically above the original \ctR{} excitons but are stabilized against recombination by their large spatial separation and by their placement in the Marcus inverted regime. The observed early-time $t^{-1/2}$ kinetics are consistent with one-dimensional diffusion-limited annihilation, while the extracted hopping rates and energy profiles point to a superexchange-mediated mechanism.\cite{Schlesinger2020}

Complementary single-crystal measurements on mixed-stack systems reveal fast and anisotropic \ctR{}-exciton formation and transport, reinforcing the view that donor--acceptor aggregates support long-range charge motion even when correlated biexcitonic signatures are not explicitly isolated.\cite{Myong2021} From a theoretical perspective, however, several questions remain unresolved. It is unclear whether \ctctR{} biexcitons are accessed directly from the \leleR{} manifold or predominantly via intermediate mixed \lectR{} configurations, how strongly they mix with other two-particle states, or to what extent they couple back into the single-exciton manifold. Moreover, the dependence of \ctctR{} binding, diffusion, and annihilation pathways on packing geometry and donor--acceptor energetics has yet to be clarified, calling for electronic-structure approaches that can explicitly resolve multiexciton wavefunctions with \ctR{} character and connect them to experimentally accessible nonlinear spectroscopic observables.

\subsection{The \ttR{} State}

The singlet-coupled triplet-pair \ttR{} is the central intermediate of singlet fission. Early theoretical work established the concept of correlated multiexcitonic triplet pairs,\cite{Smith2010} and subsequent multireference electronic-structure studies on acene crystals provided compelling evidence for a low-lying, optically dark \ttR{} state that mediates the singlet fission process.\cite{Zimmerman2010} While the formation of \ttR{} is now well established and comprehensively reviewed,\cite{Casanova2018} its subsequent evolution—most notably the separation into two independent triplets or conversion into other spin configurations—remains an open problem. In particular, the energetic landscape and coupling mechanisms governing the interconversion between adjacent and spatially separated triplet-pair states are still incompletely understood.\cite{Miyata2019,Sanders2019,hudson_what_2022}

Experiments increasingly indicate that triplet pairs can retain spin correlation over nanosecond timescales,\cite{folie_long-lived_2018,Kundu2021} enabling mechanistic pathways beyond simple decoherence into free triplets. Two processes are discussed most prominently. First, spin–spin interactions can admix the singlet-coupled \onettR{} or here \ttR{} state with the quintet-coupled \fivettR{} manifold,\cite{Tayebjee2016,basel_unified_2017,lubert-perquel_identifying_2018,weiss_strongly_2017} thereby opening alternative spin-evolution channels. Second, triplet–triplet energy transfer can produce spatially separated yet still correlated \seponettR{} states: Early experimental evidence for such a separated intermediate was provided by Pensack and co-workers in pentacene,\cite{Pensack2016} motivating the now widely used kinetic scheme
\[
S_0 + S_1 \;\longrightarrow\; \big[\,^{1}(\mathrm{TT}) \;\rightleftharpoons\; ^{1}(\mathrm{T}\ldots\mathrm{T})\,\big] \;\rightleftharpoons\; \mathrm{T}+\mathrm{T}.
\]
In this picture, the \seponettR{} species corresponds to an entangled triplet pair residing on non-nearest-neighbour chromophores, formed from adjacent \onettR{} states via Dexter-type triplet–triplet energy transfer.\cite{Scholes2015,Lee2018} Closely related multiexciton kinetics were subsequently identified in pentacene/fullerene bilayers by femtosecond nonlinear spectroscopy,\cite{Chan2011} and analogous \seponettR{} intermediates have since been reported in rubrene,\cite{Baronas} hexacene,\cite{Qian} and perylene-based molecular systems.\cite{Korovina2020,korovina_singlet_2016}

On the theoretical side, Ambrosio et al.\ employed a phenomenologically parametrized configuration-interaction Hamiltonian in a truncated HOMO/LUMO basis to analyze adjacent and spatially separated multiexciton configurations in pentacene- and tetracene-like trimers.\cite{Ambrosio2014} Within this restricted model, a more localized multiexciton state ($\mathrm{ME}_b$) is stabilized by strong orbital overlap, whereas a more spatially separated configuration ($\mathrm{ME}_u$) lies higher in energy and resembles two weakly interacting triplets. The resulting $\mathrm{ME}_u$–$\mathrm{ME}_b$ energy splitting can be interpreted as a model binding energy, suggesting an effective barrier for triplet-pair separation in that simplified picture. Although $\mathrm{ME}_b$ and $\mathrm{ME}_u$ are not explicitly spin-adapted, they map naturally onto adjacent and non-adjacent analogues of the \onettR{} and \seponettR{} states. The magnitude, and even the sign, of the inferred binding energy is, however, expected to be sensitive to the approximations inherent in the truncated active space.

Complementary theoretical work has focused on the spin structure and exchange interactions of coupled triplet pairs. Tao and Tan developed a modular tensor diagram framework to construct analytic spin eigenfunctions for exciton-pair and exciton-trimer manifolds, yielding compact, symmetry-adapted representations of the \onettR{}, \threettR{}, and \fivettR{} subspaces in generic spin-coupling models.\cite{Tao2020,Tan2021} Taffet et al.\ quantified how orbital overlap controls the exchange splitting between coupled triplets in tetracene dimers,\cite{taffet_overlap-driven_2020} while Abraham et al.\ derived a formal Dexter-type triplet-transfer integral from a Heisenberg spin-Hamiltonian perspective and demonstrated its sensitivity to molecular packing using \textit{ab initio} inputs.\cite{abraham_revealing_2021} More recently, Wang and co-workers employed a fragment particle--hole diabatization scheme to show how high-lying, multi-excited charge-transfer configurations mediate the emergence and separation of \seponettR{} states in tetracene.\cite{Wang2023}

Recent work from the Röhr group introduced a symbolic configuration-interaction framework that enables the systematic construction of diabatic Hamiltonians for multichromophoric systems, exemplified by PDI trimers.\cite{singh2024} The method formulates spin-adapted configurations in second quantization and evaluates Hamiltonian matrix elements symbolically, allowing rapid and chemically transparent generation of large diabatic Hamiltonians. By restricting the basis to local excitations together with low-lying charge-transfer and multiexciton configurations, the approach provides an efficient route to screening correlated-state pathways in extended aggregates. Applied to PDI trimers, it revealed that Dexter-type superexchange mediated by virtual triplet charge-transfer configurations (e.g. \ffspin{1}{T\dmR\dpR} and \ffspin{1}{T\dpR\dmR} ) drives the conversion of adjacent \onettR{} states into spatially separated \seponettR{} configurations, consistent with the mechanism proposed by Scholes and co-workers.\added{\cite{Scholes2015, Lee2018}} In addition, geometry scans uncovered packing regimes in which nearest-neighbour \leR{} and \ctR{} couplings cancel to form null aggregates, opening a distinct one-step pathway in which \leR{} couples virtually to a spatially separated \ctR{} state \ffspin{1}{\dpR$\ldots$\dmR} that subsequently relaxes into \seponettR{}. Over extended regions of configuration space, the \onettR{} and \seponettR{} states were found to be nearly degenerate, enabling interconversion with minimal energetic cost.

Complementary insight is provided by the nonorthogonal configuration-interaction approach developed by Sousa, de Graaf, and co-workers,\cite{sousa2025} which combines fragment multiconfigurational wave functions with a rigorous treatment of nonorthogonality. This framework avoids artificial delocalization while accurately capturing biexcitonic and charge-transfer correlations across multiple chromophores. Applications to crystalline and substituted pentacenes, PDI derivatives, and stacked indolonaphthyridines demonstrated how molecular packing and local distortions govern the energetic ordering and coupling of \onettR{}, \seponettR{}, and \ctR{} states. In several trimer models, the adjacent and spatially separated singlet triplet-pair states were found to be nearly degenerate, particularly when triplets occupy terminal sites, resulting in strong mixing and effectively barrierless dissociation pathways.

Within the same biexcitonic configuration space, triplet--triplet annihilation (TTA) corresponds to the fusion dynamics of correlated triplet-pair states and is commonly viewed as the energetic reverse of singlet fission, in which two \toneR{} excitons encounter one another and, via short-range Dexter-type exchange, regenerate an emissive \soneR{} state.\cite{Monguzzi2008,Gray2014,Zeng2022} This process provides the mechanistic basis for photon upconversion and has attracted considerable interest for applications ranging from solar energy harvesting to bio-imaging and photoredox catalysis under low-intensity illumination.\cite{Zeng2022} Because both TTA and singlet fission proceed through correlated triplet-pair manifolds, insights into the energetics and spin structure of the \onettR{} state developed in the context of singlet fission are directly relevant.\cite{Yong2017} While weakly exchange-coupled triplet pairs are known to influence TTA kinetics,\cite{Bossanyi_Nat_chem} the existence of a long-lived, well-defined \seponettR{} intermediate analogous to that proposed for singlet fission has not yet been unambiguously established.

Magnetic-field-dependent photoluminescence experiments demonstrate that radiative emission in TTA originates exclusively from the singlet component of the triplet-pair manifold, while triplet and quintet configurations act as loss channels, consistent with spin-statistical constraints.\cite{Bossanyi_JACS_AU,Clark2021} Recent electronic-structure studies further show that the interplay of Dexter exchange and charge-transfer admixture controls the binding and separation of triplet pairs,\cite{Gilligan2024} reinforcing the close conceptual link between triplet-pair fusion in TTA and triplet-pair dissociation in singlet fission.

\subsection{Electronic-Structure Strategies for Biexciton States}

The theoretical description of biexcitons in molecular aggregates poses a fundamental challenge for electronic-structure theory. Because biexcitonic states involve correlated two-particle excitations, they lie beyond the scope of linear-response approaches restricted to the single-exciton manifold
\replaced{, while effective exciton models rely on system-specific parametrizations and do not provide a general framework in which all relevant single- and two-particle configurations are generated on equal footing.}{. While effective exciton models---defined here as low-dimensional, parameterized representations of excitonic quasiparticles---offer a simplified view of these processes, they typically rely on system-specific parameters and do not provide a general framework in which all relevant single- and two-particle configurations are generated on equal footing from first principles.}
Once spatially separated and mixed biexcitons are included, the number of relevant configurations grows rapidly and analytic expressions\added{, as algebraic, symbolically generated coupling terms derived from second-quantized operator algebra and Slater–Condon rules,}The results---though preliminary---are promising. for their mutual couplings are no longer available in closed form.
The central difficulty is therefore not the identification of individual biexciton species, but the development of \textit{ab initio} electronic-structure frameworks that generate the full multiexciton configuration space from chemically meaningful building blocks and expose its internal connectivity directly through the Hamiltonian, without recourse to ad hoc assumptions.

In this context, a quasi-diabatic representation of the electronic states is particularly advantageous. Expressing excited states in terms of localized singlet and triplet excitations, charge-separated configurations, and correlated multiparticle states enables direct physical interpretation and provides a natural language for analyzing interconversion pathways. 

Two broad classes of strategies have emerged to address this challenge. The first is the supermolecule approach, in which the entire aggregate is treated as a single electronic system. This strategy has been widely applied in studies of singlet fission, where the biexcitonic \ttR{} state is described together with the single-exciton manifold to which it couples.
In practice, adiabatic excited states of the full aggregate are computed and subsequently transformed into a diabatic basis using projection-based methods\cite{Zeng2014,zirzlmeier_solution-based_2016,tamura_first-principles_2015,accomasso_diabatization_2019,tamura_triplet_2020} or density-based diabatization schemes.\cite{yang_first-principle_2015,manjanath_enhancing_2022,Wang2023}
\added{Because standard single-reference methods like TDDFT cannot access the doubly excited configurations essential for characterizing the multiexcitonic $^{1}(TT)$ state, these studies typically employ multireference strategies. Specifically, the Restricted Active Space Configuration Interaction (RASCI) family of methods is widely used to capture these states at a manageable cost by truncating excitations within a restricted active space \cite{zirzlmeier_solution-based_2016,tamura_first-principles_2015,accomasso_diabatization_2019,manjanath_enhancing_2022,Parker2014}. While RASCI variants like RAS-EE-CI(h,p) provide a robust description of multiexcitonic character \cite{Wang2023}, other approaches incorporate dynamic correlation through multi-reference Møller-Plesset perturbation theory (MRMP2) \cite{tamura_triplet_2020} or use state-averaged CASSCF \cite{zirzlmeier_solution-based_2016} to ensure a balanced treatment of the supersystem.}
While formally complete, this approach becomes prohibitively expensive as system size increases. In particular, an explicit treatment of charge-transfer--charge-transfer biexcitons requires aggregates comprising multiple chromophores, and any analysis of exciton diffusion or spatial separation rapidly exceeds feasible system sizes.

An alternative class of strategies constructs quasi-diabatic states without an explicit supermolecular treatment by exploiting molecular fragments as elementary building blocks. Within this class, two conceptually distinct realizations can be identified. In the first, quasi-diabatic states are constructed as antisymmetrized products of fragment-local many-electron wave functions, typically obtained from multireference calculations on the individual molecules. Representative examples include active space decomposition,\cite{Parker2014} multistate density functional theory,\cite{chan_quantum_2013} and nonorthogonal configuration interaction (\nociShort), as well as its explicitly fragment-based variant \noci{}.\cite{wibowo_rigorous_2017,Straatsma:2022} These approaches retain the internal electronic structure of each fragment, but their computational cost and the need to combine fragment wave functions at the many-electron level limit scalability.

A second realization adopts a more approximate but computationally efficient strategy in which quasi-diabatic configurations are constructed at the level of fragment-local orbitals rather than fragment wave functions. Here, a minimal active space formed from the frontier orbitals on each molecule enables analytic construction of local excitations, charge-separated configurations, and correlated triplet pairs, as exemplified by Michl’s Simple Model.\replaced{\cite{Smith2010}}{\cite{Smith2010, zaykov_singlet_2019}} This approach has proven valuable for rationalizing trends in singlet-fission efficiency and excitonic couplings.\cite{Berkelbach2013b,berkelbach_microscopic_2014,nakano_quantum_2019,ryerson_structure_2019} However, the minimal orbital space restricts the local electronic flexibility, while dimer-based parametrizations preclude spatially separated biexcitonic configurations as well as states extending over more than two monomers, such as \ctxR{} and \ctctR{} states.

Taken together, these strategies expose a central limitation of current electronic-structure treatments of biexcitons: while individual approaches may be accurate, scalable, or chemically transparent, none provides a representation in which all relevant single- and two-particle configurations are generated on equal footing and connected systematically by the Hamiltonian itself. As a result, biexcitonic processes are commonly described in terms of preselected states and assumed coupling pathways, rather than emerging from the electronic structure. A representation that treats the one- and two-particle manifolds within a common Hamiltonian instead makes the intermanifold connectivity explicit, allowing the electronic accessibility of correlated two-particle states to be analyzed directly as a function of molecular packing, charge-transfer admixture, and electronic correlation.

\begin{figure*}[ht]
    \centering
    \includegraphics[width=\linewidth]{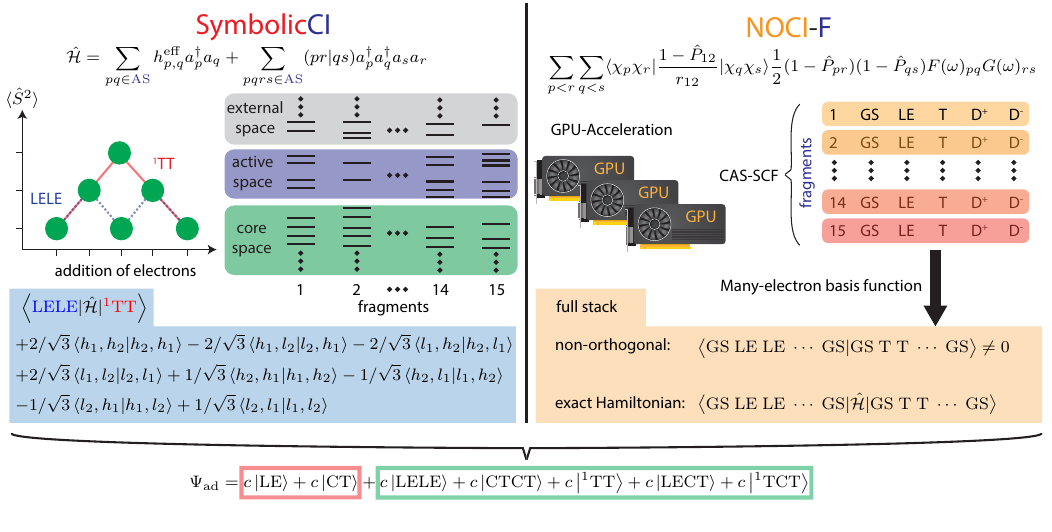}
    \caption{Overview illustrating the key features of the \symci{} and the \noci{} methodology.}
    \label{fig:SymCI-NOCI}
\end{figure*}

\section{Unified CI Description of Biexcitonic Manifolds}
In the following, we introduce two complementary realizations of this unified CI framework. The first is an orbital-based construction that enables the analytic and scalable generation of large Hamiltonians spanning single- and biexcitonic configurations. The second is a wave-function-based formulation that provides benchmark accuracy by combining fully relaxed multireference fragment states within a nonorthogonal CI formalism.

\subsection{\symci{}}
 The conceptual origin of the \symci{} approach can be traced to Michl’s Simple Model,\replaced{\cite{Smith2010}}{\cite{Smith2010, zaykov_singlet_2019}} in which local excitations, charge-separated configurations, and correlated triplet pairs are constructed analytically in a minimal frontier-orbital space for molecular dimers. \symci{} adopts this organizing principle while lifting its most restrictive assumptions by allowing multiple active orbitals per fragment and extending the construction to aggregates of arbitrary size and topology. In this way, local excitons, charge-transfer states, triplet excitations, and their multiexcitonic combinations are generated on equal footing within a single configuration-interaction space, independent of spatial adjacency or predefined interaction range. Depending on the chosen active space and excitation rank, the resulting CI spans frontier-orbital models, fragment-based CISD, or full CAS-type descriptions, while preserving the same fragment-local and spin-adapted structure across system sizes.
 \added{The \symci{} method utilizes naturally localized monomer orbitals which, while internally orthogonal, are non-orthogonal between distinct fragments. Monomer-centered orbitals are obtained either internally at the Hartree–Fock level or imported from external state-averaged CASSCF calculations. To satisfy the CI procedure's requirement for an orthogonal basis while preserving the local interpretability of the fragments, we employ a symmetric Löwdin orthogonalization of the fragment-local core and active orbitals. A critical step in this process is the exclusion of unoccupied external orbitals from the orthogonalization, which is essential for minimizing the perturbation to the spatial character of the original orbitals. As a quality control measure, the program monitors the coefficients between the original and orthogonalized sets; a warning is issued if the deviation exceeds \SI{1}{\percent}, though observed changes are typically orders of magnitude below this threshold.}
 \replaced{In \symci{}, active orbitals are assigned to predefined molecular fragments, and the many-electron basis is constructed from spin-adapted configuration state functions (CSFs). Unlike conventional CI implementations that are formulated in a Slater-determinant basis and obtain spin-pure states only after Hamiltonian diagonalization, the \symci{} method}{Further, \symci{}}constructs spin-adapted configuration state functions \added{(CSFs)} explicitly and symbolically, enabling analytic control over the spin structure and excitation character of each configuration. constructs spin-adapted configuration state functions  explicitly and symbolically, enabling analytic control over the spin structure and excitation character of each configuration. \cite{falesFastTransformationsConfiguration2020a,ugandi2023configuration,chilkuriComparisonManyParticleRepresentations2021} 
Each CSF corresponds to a prescribed occupation pattern of fragment-local orbitals and is uniquely labeled by total spin $S$\remove{, spin projection $S_z$,} and a specific spin-coupling pathway $P$,
\begin{equation}
\ket{\Psi^{S\remove{,S_z},P}} = \sum_I c_I \ket{\Phi_I^{S\remove{,S_z},P}} .
\end{equation}
Spin adaptation is implemented explicitly using Yamanouchi--Kotani branching diagrams,\cite{yamanouchiCalculationAtomicEnergy1936,yamanouchiConstructionUnitaryIrreducible1937,yamanouchiAtomicEnergyLevels1938,kotaniTablesIntegralsUseful1953,grabenstetterGenerationGenealogicalSpin1976,paunczBranchingDiagramSerbertype1977} which generate spin eigenfunctions recursively for a given number of active electrons. As a consequence, configurations with identical orbital occupations but different spin structure, such as singlet-coupled triplet pairs and other double excitations, are distinguished already at the level of the basis. In this sense, the \symci{} representation is quasi-diabatic: fragment occupation and spin character are fixed prior to Hamiltonian evaluation, and intermanifold connectivity is encoded directly in the Hamiltonian matrix elements.
The electronic Hamiltonian is expressed in a standard second-quantized form in the active-space molecular-orbital basis,
\begin{equation}
\hat H =
\sum_{pq} h^{\mathrm{MO}}_{pq} a_p^\dagger a_q
+ \tfrac12 \sum_{pqrs} \replaced{g^{\mathrm{MO}}_{prqs}}{\langle pq|rs\rangle} a_p^\dagger a_q^\dagger a_s a_r ,
\end{equation}
\added{where $p, q, r, s$ denote the indices of the active-space spin-orbitals. The terms $h^{\mathrm{MO}}_{pq}$ are the effective one-electron integrals, accounting for the kinetic energy operator, the nuclear potential, and the interaction with the frozen core electrons. The $\langle pq|rs\rangle$ terms represent the electron-electron repulsion integrals in physicist's notation, while $a_p^\dagger$ and $a_p$ are the creation and annihilation operators for spin-orbital $p$.}
\remove{with one- and two-electron integrals defined after symmetric Löwdin orthogonalization of the fragment-local core and active orbitals. Core--active interactions are incorporated through the effective one-electron contribution
\begin{equation}
G^{\mathrm{core}}_{\mu\nu} =
\sum_{\lambda\sigma} \rho^{\mathrm{core}}_{\lambda\sigma}
\!\left[(\mu\nu|\lambda\sigma) - \tfrac12(\mu\lambda|\sigma\nu)\right],
\end{equation}
leading to the corresponding transformed matrix elements
\begin{align}
h^{\mathrm{MO}}_{ij} &=
\sum_{\mu\nu} C_{\mu i} C_{\nu j}
\big[T_{\mu\nu} + V_{\mu\nu} + G^{\mathrm{core}}_{\mu\nu}\big],\\
g^{\mathrm{MO}}_{ijkl} &=
\sum_{\mu\nu\lambda\sigma}
C_{\mu i} C_{\nu j} C_{\lambda k} C_{\sigma l}
(\mu\nu|\lambda\sigma).
\end{align}}

\added{The} Hamiltonian matrix elements between CSFs are evaluated using the Slater--Condon rules in their determinant expansion\remove{,
\begin{align}
\begin{split}
&\mel{\Psi^{S,S_z,P}_{\mathrm{CSF}}}{\hat H}{\Psi^{S,S_z,P}_{\mathrm{CSF}}}
=\\ \sum_{ab} c_a c_b \Bigg[
&\sum_{pq \in \mathrm{AS}} \mel{a}{a_p^\dagger a_q}{b} h^{\mathrm{MO}}_{pq} \\
&+ \sum_{pqrs \in \mathrm{AS}} \mel{a}{a_p^\dagger a_q^\dagger a_s a_r}{b} g^{\mathrm{MO}}_{prqs}
\Bigg],
\end{split}
\end{align}
where the coefficients $c_a$ and $c_b$ arise from the spin-coupling pathways defined by the branching diagrams}
.\cite{weinbergSPINCETERA2015,helgakerSpinSecondQuantization2000}

Because the CI basis is constructed symbolically from fragment-local CSFs, individual Hamiltonian matrix elements can be expressed analytically in terms of one- and two-electron integrals, allowing direct interpretation of biexcitonic couplings beyond purely numerical evaluation. This provides general insight into how different biexcitonic processes depend on orbital locality and spatial topology. \replaced{For example, the coupling mediating parallel diffusion of two Frenkel excitons,}{To illustrate a symbolic expression for the inter-monomer coupling, we consider the matrix element $\mel{\mathrm{S}_1\mathrm{S}_1\mathrm{S}_0\mathrm{S}_0}{\hat{\mathcal H}}{\mathrm{S}_0\mathrm{S}_1\mathrm{S}_1\mathrm{S}_0}$:}
\begin{equation}
\mel{\mathrm{S}_1\mathrm{S}_1\mathrm{S}_0\mathrm{S}_0}{\hat{\mathcal H}}{\mathrm{S}_0\mathrm{S}_1\mathrm{S}_1\mathrm{S}_0}
= 2\replaced{(l_1,h_1|h_3,l_3) - (l_1,l_3|h_3,h_1)}{\langle l_1, h_3 | h_1, l_3\rangle - \langle l_1, h_3 | l_3, h_1\rangle},
\end{equation}
\added{where $h_i$ and $l_i$ denote the HOMO and LUMO orbitals on site $i$, respectively. The spin indices have been integrated out, and the expression is shown in terms of spatial orbitals, consistent with the use of a restricted orbital basis.}
\remove{exhibits the familiar restricted-Hartree–Fock Coulomb–exchange structure, in which the exchange contribution becomes negligible for spatially separated fragments and the interaction is effectively Coulombic. In contrast, for asymmetric or non-parallel diffusion pathways,
\begin{equation}
\mel{\mathrm{S}_1\mathrm{S}_1\mathrm{S}_0\mathrm{S}_0}{\hat{\mathcal H}}{\mathrm{S}_1\mathrm{S}_0\mathrm{S}_1\mathrm{S}_0}
= 2(l_2,h_2|h_3,l_3) - (l_2,l_3|h_3,h_2),
\end{equation}
the same formal structure applies, but the exchange contribution can become significant due to increased orbital overlap. A direct physical consequence is that different biexcitonic diffusion pathways exhibit qualitatively different sensitivity to molecular packing: Coulomb-dominated parallel diffusion is comparatively robust with respect to intermolecular separation, whereas exchange-assisted asymmetric pathways are short-ranged and strongly geometry-dependent. Within the \symci{} framework, this distinction emerges directly from the analytic structure of the Hamiltonian and provides a microscopic rationale for packing-dependent multiexciton behavior in molecular aggregates.}

\subsection{\noci{}}
Non-orthogonal configuration interaction for fragments provides a complementary \textit{ab initio} realization of the unified CI framework, in which diabatic states are represented by fully optimized \added {all-electron} multiconfigurational fragment wave functions. In contrast to \symci{}, where a common orthogonal orbital basis is employed, \noci{} rigorously accounts for orbital relaxation by allowing each diabatic configuration to be expressed in its own optimal set of molecular orbitals. As a consequence, the resulting many-electron basis functions are non-orthogonal, but retain a transparent interpretation in terms of chemically intuitive fragment-local electronic states.

\replaced{In practice, a set of multiconfigurational wave functions—typically of CASSCF type—is constructed independently for each fragment and electronic state of interest. These fragment states are then combined into antisymmetrized, spin-adapted many-electron basis functions of the aggregate. The separate optimization of the fragment wave functions ensures full orbital relaxation for all diabatic states, but introduces non-orthogonality among the resulting basis functions. As a result, Hamiltonian matrix elements cannot be evaluated using Slater–Condon rules, and all determinant pairs contribute explicitly to
\[
H_{ij} = \langle \mathrm{MEBF}_i | \hat H | \mathrm{MEBF}_j \rangle .
\]
Dynamic electron correlation is incorporated through state-specific energy corrections, applied as shifts to the diagonal matrix elements of the \noci{} Hamiltonian.
The evaluation of non-orthogonal matrix elements is carried out using the General Non-Orthogonal Matrix Elements (GNOME) formalism, which is based on a singular value decomposition of the overlap matrix
\begin{equation}
\mathbf{S} = \langle \phi | \psi \rangle = \mathbf{U}\boldsymbol{\lambda}\mathbf{V},
\end{equation}
allowing the transformation to corresponding orbital sets and the factorization of two-electron contributions into tractable terms.\cite{Broer:1988}
While computationally more demanding than orthogonal CI, this approach yields benchmark-quality diabatic Hamiltonians that explicitly incorporate state-specific orbital relaxation and dynamic correlation. Efficient massively parallel implementations, including GPU acceleration, enable applications to molecular aggregates of moderate size.\cite{Sousa:2023,Lopez:2023a,sousa2025}
Further details on the GNOME formalism and its numerical implementation can be found in Refs.~\citenum{Straatsma:2022,deGraaf:2023}.}{The \noci{} approach consists of four steps. First, state-specific multiconfigurational wave functions are optimized for all the electronic states of interest for the problem under study. This is done for each fragment (molecule) in the aggregate under consideration. The state-specific approach implies that each electronic state is expressed in a different set of orbitals---namely the optimal ones for that particular state---and introduces non-orthogonality among all fragment states, not only the ones localized on different molecules, but also within one fragment. In the second step, the fragment states are combined to form antisymmetric spin-adapted many-electron basis functions (MEBFs) that span the non-orthogonal configuration interaction space. The use of Clebsch-Gordan coefficients during the construction of the MEBFs ensures that the expansion in determinants is an eigenfunction of the total spin operator. Note that the number of MEBFs is  relatively small and completely determined by the physics of the problem, typically including the product of the ground state on each fragments; products of ground state wave functions combined with a local excited state on one of the fragments; cationic and anionic states on two fragments combined with ground state functions on all other molecules in the ensemble to describe a charge transfer state, etc. The third step consists of the calculation of the Hamiltonian and overlap matrix elements of the MEBFs
\begin{equation}
H_{ij} = \langle \mathrm{MEBF}_i | \hat H | \mathrm{MEBF}_j \rangle \qquad S_{ij} = \langle \mathrm{MEBF}_i |\mathrm{MEBF}_j \rangle .    
\end{equation}
Due to the non-orthogonality, the Slater-Condon rules for the evaluation of $H_{ij}$ do not apply and all the \textit{bra}-\textit{ket} determinant pairs have to evaluated. The calculation of these non-orthogonal matrix elements is carried out using the General Non-Orthogonal Matrix Elements (GNOME) algorithm.\cite{Broer:1988} To avoid the calculation of the first- and second-order co-factors of the overlap matrix,\cite{Lowdin:1950} GNOME transforms the orbitals of MEBF$_i$ and MEBF$_j$ to the corresponding orbital basis\cite{Amos:1961,King:1967} by a singular value decomposition of the molecular orbital overlap matrix 
\begin{equation}
\mathbf{S} = \langle \phi | \psi \rangle = \mathbf{U}\boldsymbol{\lambda}\mathbf{V},
\end{equation}
from which the overlap of the two determinants is obtained by the multiplying the singular values $\lambda_i$.  A second aspect of the efficiency of the GNOME algorithm is the factorization of the two-electron contributions, allowing to write the $N^4$-dimensional matrices as the product of two $N^2$-dimensional matrices.\cite{Montfort:1980} More detailed descriptions of the GNOME algorithm can be found in the original papers\cite{Broer:1981,Broer:1988} and recent publications on the \noci{} approach.\cite{Straatsma:2022,deGraaf:2023} Note that contrary to \symci{}, no approximations are made in the calculation of the these matrix elements and that all calculations are based on an all-electron description. In the fourth and final step of the \noci{} approach, the general eigenvalue problem is solved to obtain the NOCI wave functions and energies. Furthermore, the electronic couplings $\gamma_{ij}$ among the MEBFs are calculated through
\begin{equation}
    \gamma_{ij} = \frac{H_{ij}-\tfrac{1}{2}(H_{ii}+H_{jj})S_{ij}}{1-S_{ij}^2}.
\end{equation}
\noci{} has been implemented in the massively parallel and GPU-accelerated open-source code GronOR---available from GitLab--- and enables applications to molecular aggregates of moderate size \cite{Sousa:2023,Lopez:2023a,sousa2025,Stan:2026}. GronOR is interfaced to OpenMolcas\cite{Limanni:2023} to obtain the multiconfigurational fragment wave functions and the one- and two-electron integrals. While computationally more demanding than orthogonal CI, this approach yields benchmark-quality diabatic Hamiltonians that explicitly incorporate state-specific orbital relaxation and dynamic correlation.}

In the present contribution, \noci{} is used for the first time to describe a comprehensive  multiexcitonic manifold, and thereby provides a systematic point of comparison for validating the diabatic couplings and relative energetic trends described by \symci{}.
\section{Results}
\subsection{Comparison of Methods}
We first benchmark \symci{} against \noci{} under conditions where both approaches can be applied on equal footing. \symci{}  systematically generates the complete set of single and double excitations within a monomer-local orbital space, whereas \noci{} constructs the electronic structure from a selected set of optimized fragment states. To enable a direct comparison, we therefore define an identical reduced diabatic subspace that can be constructed consistently in both approaches. This shared subspace is chosen solely to probe representational consistency, ensuring that differences between the resulting Hamiltonians and adiabatic states arise from the electronic-structure formulation rather than from differences in state selection.

The benchmark is carried out for two systems. Ethylene aggregate stacks are used as a minimal $\pi$-conjugated model that \remove{nevertheless} supports Frenkel, charge-transfer, and (mixed) biexcitonic configurations, and is small enough to permit \noci{} calculations for extended stacks of up to 15 monomers. As a second test case, anthracene pentamers are examined, representing a chemically more realistic chromophore class closer to materials of practical interest.

For the linear stacks, we consider four representative packing geometries: (i) an H-aggregate with parallel stacking of monomer transition dipoles (\szeroR{} $\rightarrow$ \soneR{}), (ii) a J-aggregate in head-to-tail configuration, (iii) a null aggregate geometry, in which Frenkel-type coupling and \ctR{} coupling effectively cancel each other, and (iv) an intermediate Zero-Frenkel geometry exhibiting vanishing Frenkel-type coupling. These geometries serve as “neuralgic” points in the packing landscape and allow a systematic comparison of biexciton character and coupling mechanisms in different excitonic regimes. 

\begin{figure*}[ht]
    \centering
    \includegraphics[width=0.9\textwidth]{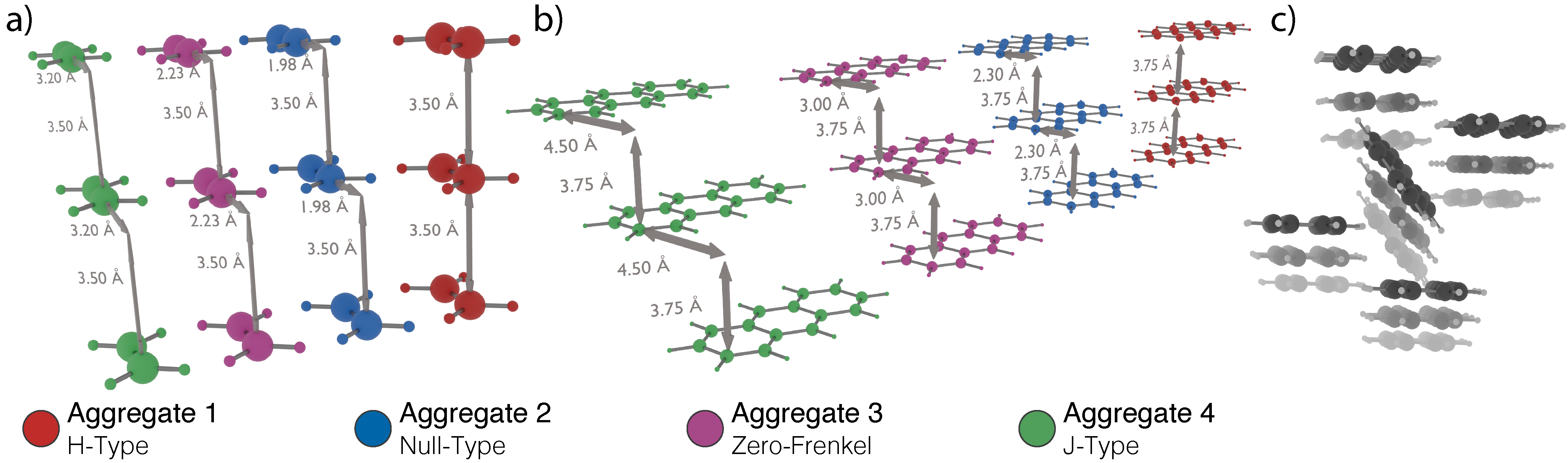}
    \caption{a) Cutout of the ethylene 15-mer aggregates. Aggregate 1, H-type (red): z-translation \SI{3.50}{\angstrom}. Aggregate 2, Null-Type (blue): x-translation \SI{1.98}{\angstrom}, z-translation \SI{3.50}{\angstrom}. Aggregate 3, Zero-Frenkel (violet): x-translation \SI{2.23}{\angstrom}, z-translation \SI{3.50}{\angstrom}. Aggregate 4, J-type (green): x-translation \SI{3.20}{\angstrom}, z-translation \SI{3.50}{\angstrom}. b) Cutout of the anthracene 15-mer aggregates. Aggregate 1, H-type (red): z-translation \SI{3.75}{\angstrom}. Aggregate 2, Null-Type (blue): y-translation \SI{2.30}{\angstrom}, z-translation \SI{3.75}{\angstrom}. Aggregate 3, Zero-Frenkel (violet): y-translation \SI{3.00}{\angstrom}, z-translation \SI{3.75}{\angstrom}. Aggregate 4, J-type (green): y-translation \SI{4.50}{\angstrom}, z-translation \SI{3.75}{\angstrom}. c) Side view of the anthracene crystal cutout containing 15 monomers, as taken from the experimental crystal structure \cite{marciniak_crystal_2002}.}
    \label{fig:geometries}
\end{figure*}

All monomer geometries were optimized at the MP2 level using the \textit{cc-pVTZ}\cite{dunning1989gaussian} basis set for ethylene and anthracene. \symci{} calculations employed SA-CASSCF(2,2)\cite{roos_complete_1980,RN155,RN230} orbitals obtained from \textit{ORCA 6.0}\cite{RN204, RN205, RN84, RN139, RN21, RN171, RN200, RN96, RN178} and included only the frontier orbitals of each monomer in the active space. \replaced{\noci{} calculations for anthracene used the smaller \textit{cc-pVDZ}\cite{dunning1989gaussian} basis.}{For ethene the SA-CASSCF(2,2) as well as the \symci{} calculation used the \textit{cc-pVTZ}\cite{dunning1989gaussian} basis,  for anthracene the smaller \textit{cc-pVDZ}\cite{dunning1989gaussian} basis is used. \noci{} used the same basis as used in the \symci{} for all the comparing calculations.} 

\subsubsection{Ethylene}
We consider four 15-monomer ethylene aggregates, each with a fixed interplanar distance of \SI{3.50}{\angstrom}. The H-type aggregate has no in-plane translation. The Null-Type aggregate introduces an in-plane shift of \SI{1.98}{\angstrom} per monomer along the C–C bond (x-direction), the Zero-Frenkel aggregate applies a slightly larger x-translation of \SI{2.23}{\angstrom}. Finally, the J-type aggregate, with an x-translation of \SI{3.20}{\angstrom}. All geometries were derived from a continuous dimer scan, the resulting packing motifs are illustrated in \reffigure{fig:geometries}.

Without restriction, \symci{} generates all single and double excitations within the active space\added{, which includes the HOMO and LUMO orbitals of each ethylene molecule and hence has 30 electrons in 30 spatial orbitals,} relative to the ground state, yielding a Hamiltonian of dimension $\num{25651} \cross \num{25651}$. Performing a \noci{} calculation of this size is computationally unfeasible. Thus, both approaches were benchmarked on a shared, manually selected 72-state subspace. This includes the ground state, nearest- and next-nearest-neighbor double excitations of Frenkel type (\leleR{}), charge-transfer–Frenkel combinations (\lectR{}), where \ctR{} units flank the \leR{} excitation (e.g., S$_1$D$^+$D$^-$), and nearest-neighbor \ctctR{} biexcitons (e.g., D$^+$D$^-$D$^+$D$^-$). More delocalized configurations such as D$^+$D$^-$S$_0$D$^+$S$_0$D$^-$ or \lectR{} states with central \leR{} character (e.g., D$^+$S$_1$D$^-$) are excluded in this first comparison.
To suppress vacuum artifacts, the two outermost monomers remain in the ground state in all configurations. This is particularly relevant for \ctR{} states, where boundary conditions strongly affect electrostatic interactions and excitation energies.

\begin{figure*}[ht]
    \centering
    \includegraphics[width=\linewidth]{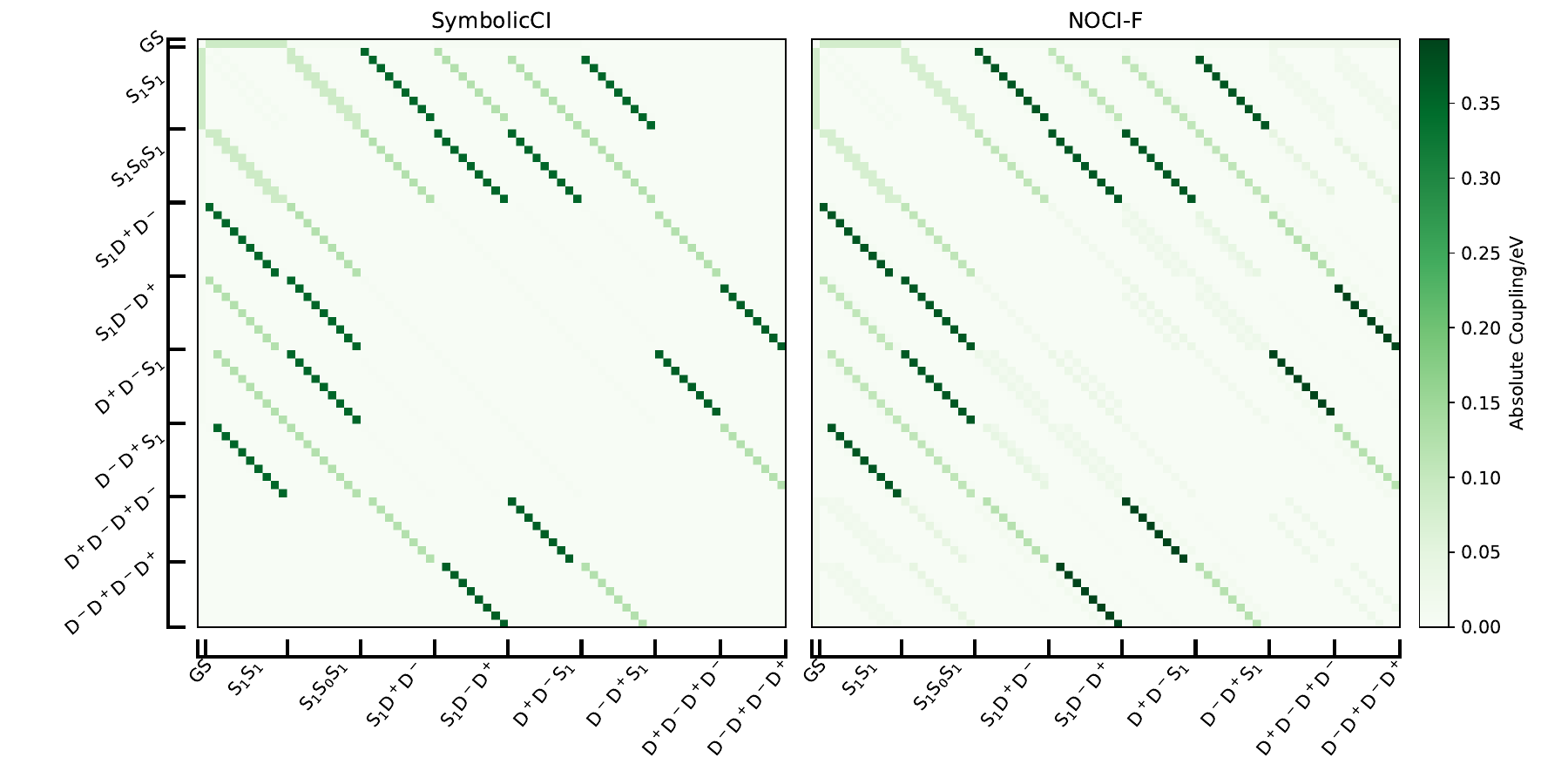}
    \caption{Comparison of the \symci{} Hamiltonian (left) with the \noci{} Hamiltonian (right) of the Aggregate 4, J-type (green): x-translation \SI{3.20}{\angstrom}, z-translation \SI{3.50}{\angstrom}. Diabatic energies (on the diagonal) are set to zero.}
    \label{fig:ethene_comparison_J_agg}
\end{figure*}

For a first visual comparison, the Hamiltonians obtained from \symci{} and \noci{} for the J-type aggregate are presented in \reffigure{fig:ethene_comparison_J_agg}. It can be seen that the most significant couplings are very similarly represented by both programs. However, smaller couplings involving \ctR{} states, as calculated by \noci{}, tend to be underestimated by \symci{}. Nonetheless, the overall structure of the Hamiltonian is highly comparable.

A detailed comparison between the two electronic structure methods is provided in \reftable{tab:ethene_couplings}, where the most significant couplings are listed for all four aggregates. Overall, the agreement between \symci{} and \noci{} is good, with only a few notable deviations. Specifically, discrepancies appear for the coupling associated with parallel biexciton \replaced{diffusion}{transfer}, $\mel{\mathrm{S}_1\mathrm{S}_1\mathrm{S}_0}{\hat{\mathcal{H}}}{\mathrm{S}_0\mathrm{S}_1\mathrm{S}_1}$, in the Zero-Frenkel aggregate, and for the coupling between a Frenkel-type biexciton and an \lectR{} exciton, $\mel{\mathrm{S}_1\mathrm{S}_1\mathrm{S}_0}{\hat{\mathcal{H}}}{\mathrm{D}^+\mathrm{D}^-\mathrm{S}_1}$, in the H-aggregate. The most pronounced underestimation is found for the coupling between Frenkel-type and \ctR{}-type biexcitons, $\mel{\mathrm{S}_0\mathrm{S}_1\mathrm{S}_1\mathrm{S}_0}{\hat{\mathcal{H}}}{\mathrm{D}^+\mathrm{D}^-\mathrm{D}^+\mathrm{D}^-}$ and $\mel{\mathrm{S}_1\mathrm{S}_0\mathrm{S}_1\mathrm{S}_0}{\hat{\mathcal{H}}}{\mathrm{D}^+\mathrm{D}^-\mathrm{D}^+\mathrm{D}^-}$, across all aggregates. For the remaining couplings, both methods yield values of similar magnitude and capture the same trends across different aggregate types.
\begin{table*}[ht]
    \centering  
    \begin{tabular}{cc|rr|rr|rr|rr|}
 && \multicolumn{2}{c|}{ \removeTableShort{H Aggregate}{H-type}} & \multicolumn{2}{c|}{ \removeTableShort{Zero Davydov}{Null-Type}} & \multicolumn{2}{c|}{  \removeTableShort{Zero Coupling}{Zero-Frenkel}} & \multicolumn{2}{c|}{  \removeTableShort{J Aggregate}{J-type}} \\
 $\langle$ bra $|$ &  $|$ ket $\rangle$ & \nociShort{} & \symciShort{} & \nociShort{} & \symciShort{} & \nociShort{} & \symciShort{} & \nociShort{} & \symciShort{} \\ \hline
S$_1$S$_1$S$_0$ & S$_0$S$_1$S$_1$ & \num[round-mode=places,round-precision=\precision]{21.7630909476} & \num[round-mode=places,round-precision=\precision]{49.3486866208} & \num[round-mode=places,round-precision=\precision]{3.1848194571} & \num[round-mode=places,round-precision=\precision]{7.2279150101} & \num[round-mode=places,round-precision=\precision]{0.4145078285} & \num[round-mode=places,round-precision=\precision]{2.6235130527} & \num[round-mode=places,round-precision=\precision]{6.1441899426} & \num[round-mode=places,round-precision=\precision]{7.5550192520}  \\
S$_1$S$_1$S$_0$ & S$_1$S$_0$S$_1$ & \num[round-mode=places,round-precision=\precision]{460.1316573319} & \num[round-mode=places,round-precision=\precision]{452.6576936603} & \num[round-mode=places,round-precision=\precision]{21.4713337064} & \num[round-mode=places,round-precision=\precision]{44.2691377106} & \num[round-mode=places,round-precision=\precision]{17.1769806964} & \num[round-mode=places,round-precision=\precision]{0.4435259087} & \num[round-mode=places,round-precision=\precision]{72.8497356953} & \num[round-mode=places,round-precision=\precision]{87.7334616535}  \\
S$_1$S$_1$S$_0$ & S$_1$D$^+$D$^-$ & \num[round-mode=places,round-precision=\precision]{882.2938059989} & \num[round-mode=places,round-precision=\precision]{852.4966275689} & \num[round-mode=places,round-precision=\precision]{103.4121151744} & \num[round-mode=places,round-precision=\precision]{140.8136050597} & \num[round-mode=places,round-precision=\precision]{215.9268011754} & \num[round-mode=places,round-precision=\precision]{241.5770462509} & \num[round-mode=places,round-precision=\precision]{370.1554128037} & \num[round-mode=places,round-precision=\precision]{350.4621103230}  \\
S$_1$S$_1$S$_0$ & S$_1$D$^-$D$^+$ & \num[round-mode=places,round-precision=\precision]{548.5132289892} & \num[round-mode=places,round-precision=\precision]{589.5600952559} & \num[round-mode=places,round-precision=\precision]{291.2920744328} & \num[round-mode=places,round-precision=\precision]{334.6313578040} & \num[round-mode=places,round-precision=\precision]{244.8058200069} & \num[round-mode=places,round-precision=\precision]{283.5593261256} & \num[round-mode=places,round-precision=\precision]{106.8619188165} & \num[round-mode=places,round-precision=\precision]{123.4767070337}  \\
S$_1$S$_1$S$_0$ & D$^+$D$^-$S$_1$ & \num[round-mode=places,round-precision=\precision]{91.3334639775} & \num[round-mode=places,round-precision=\precision]{8.2845957169} & \num[round-mode=places,round-precision=\precision]{2.9898549310} & \num[round-mode=places,round-precision=\precision]{4.3805461024} & \num[round-mode=places,round-precision=\precision]{7.1108087611} & \num[round-mode=places,round-precision=\precision]{3.5312104719} & \num[round-mode=places,round-precision=\precision]{9.6468054021} & \num[round-mode=places,round-precision=\precision]{0.5671963897}  \\
S$_0$S$_1$S$_1$S$_0$ & D$^+$D$^-$D$^+$D$^-$ & \num[round-mode=places,round-precision=\precision]{157.4109309815} & \num[round-mode=places,round-precision=\precision]{0.8725800614} & \num[round-mode=places,round-precision=\precision]{17.7863098711} & \num[round-mode=places,round-precision=\precision]{0.0177291385} & \num[round-mode=places,round-precision=\precision]{23.4067094319} & \num[round-mode=places,round-precision=\precision]{0.0027340282} & \num[round-mode=places,round-precision=\precision]{12.5421182327} & \num[round-mode=places,round-precision=\precision]{0.0026599368}  \\
S$_1$S$_0$S$_1$S$_0$ & D$^+$D$^-$D$^+$D$^-$ & \num[round-mode=places,round-precision=\precision]{261.3463942369} & \num[round-mode=places,round-precision=\precision]{1.2493962134} & \num[round-mode=places,round-precision=\precision]{9.2874171142} & \num[round-mode=places,round-precision=\precision]{0.0016584348} & \num[round-mode=places,round-precision=\precision]{25.4163636835} & \num[round-mode=places,round-precision=\precision]{0.0388581813} & \num[round-mode=places,round-precision=\precision]{44.3888557591} & \num[round-mode=places,round-precision=\precision]{0.0861474450}  \\
\newrow S$_0$S$_1$D$^+$D$^-$ & D$^+$D$^-$D$^+$D$^-$ & \num[round-mode=places,round-precision=\precision]{531.9280269659} & \num[round-mode=places,round-precision=\precision]{563.9691725127} & \num[round-mode=places,round-precision=\precision]{311.5831496389} & \num[round-mode=places,round-precision=\precision]{328.8614480701} & \num[round-mode=places,round-precision=\precision]{267.5611130628} & \num[round-mode=places,round-precision=\precision]{279.9534986690} & \num[round-mode=places,round-precision=\precision]{120.4456391617} & \num[round-mode=places,round-precision=\precision]{123.4277193018}  \\
\end{tabular}
    \caption{Absolute electronic couplings in $\si{\milli\eV}$ from important selected transitions. The couplings are compared for the four different ethene aggregates and between the two methods.}
    \label{tab:ethene_couplings}
\end{table*}

For the parallel biexciton \replaced{diffusion}{transfer} coupling, $\mel{\mathrm{S}_1\mathrm{S}_1\mathrm{S}_0}{\hat{\mathcal{H}}}{\mathrm{S}_0\mathrm{S}_1\mathrm{S}_1}$, \symci{} predicts values roughly two to three times larger than \noci{} in all aggregates except the J-aggregate, where both methods yield comparable results. The non-parallel \replaced{diffusion}{transfer} coupling, $\mel{\mathrm{S}_1\mathrm{S}_1\mathrm{S}_0}{\hat{\mathcal{H}}}{\mathrm{S}_1\mathrm{S}_0\mathrm{S}_1}$, agrees well between the two methods in the H- and J-aggregates and is only overestimated by a factor of two with \symci{} in the Null-Type aggregate. In the Zero-Frenkel aggregate, however, this interaction differs markedly between the two approaches.
The couplings between Frenkel-type biexcitons and \lectR{} states, $\mel{\mathrm{S}_1\mathrm{S}_1\mathrm{S}_0}{\hat{\mathcal{H}}}{\mathrm{S}_1\mathrm{D}^+\mathrm{D}^-}$ and $\mel{\mathrm{S}_1\mathrm{S}_1\mathrm{S}_0}{\hat{\mathcal{H}}}{\mathrm{S}_1\mathrm{D}^-\mathrm{D}^+}$, are of comparable magnitude in all aggregates and consistently large. The coupling $\mel{\mathrm{S}_1\mathrm{S}_1\mathrm{S}_0}{\hat{\mathcal{H}}}{\mathrm{D}^+\mathrm{D}^-\mathrm{S}_1}$ is also well reproduced in the Null-Type and Zero-Frenkel geometries, but underestimated by \symci{} in the H- and J-aggregates. A similar trend is observed for couplings between Frenkel-type and \ctR{}-type biexcitons, which are systematically smaller in \symci{} across all structures.
\added{However, the coupling from \lectR{} to \ctR{}-type biexcitons, $\mel{\mathrm{S}_0\mathrm{S}_1\mathrm{D}^+\mathrm{D}^-}{\hat{\mathcal{H}}}{\mathrm{D}^+\mathrm{D}^-\mathrm{D}^+\mathrm{D}^-}$, are in good agreement for both methods and all aggregates.}
\replaced{These discrepancies share a common feature: the underlying transitions involve configurations delocalized over more than two monomers and contain charge-transfer character.}{As a first conclusion, \symci{} seems to systematically underestimates the couplings between Frenkel-type and \ctR{}-type biexcitons in the ethene molecule.}
\added{Nonetheless, the impact of these underestimations is minimal for the doubly-excited adiabatic wavefunctions (see Supporting Information Section~\ref{SI-sec:ethene_adiabatic_analysis}).} \ctR{} states are not optimally represented in the HOMO–LUMO basis obtained from SA-CASSCF orbitals averaged over S$_0$, S$_1$, and S$_2$, \added{possibly} leading to the underestimation observed in \symci{}. In contrast, \noci{} \replaced{provides}{should provide} a more accurate account of these configurations by employing explicit cationic and anionic states, albeit at higher computational cost. A possible route to improve \symci{} results for such states would be to enlarge the active space and construct projected multiconfigurational anion and cation states.
In summary, both methods agree well on the dominant biexciton couplings, and capture consistent physical trends across the different aggregate geometries. The main deviations are confined to long-range, \replaced{\ctR{}-involving}{\leleR{}--\ctctR{}-type} couplings, where \symci{} systematically underestimates the interaction strength. Given the substantial energetic gap between \leleR{} and \ctctR{} states, however, the impact on the final adiabatic states is limited, as state mixing in this regime remains weak.

\added{Despite this underestimation, both methods agree that the \leleR{}--\lectR{} coupling (e.g., $\mel{\sone\sone\szero}{\hat{\mathcal H}}{\sone\dpm\dmm}$) and \lectR{}-\ctctR{} couplings are significantly stronger than the \leleR{}--\ctctR{} coupling (e.g.,  $\mel{\szero\sone\sone\szero}{\hat{\mathcal H}}{\dpm\dmm\dpm\dmm}$). This disparity can be elucidated through the underlying symbolic expressions derived within the \symci{} framework.

For the \leleR{} to \lectR{} transition, the coupling is expressed as:
\begin{align}
\begin{split}
    \mel{\sone\sone\szero}{\hat{\mathcal H}}{\sone\dpm\dmm} &= \oneE{l_2}{l_3} - \twoE{h_2}{l_2}{l_3}{h_2} + \frac{1}{2} \twoE{l_2}{h_1}{h_1}{l_3} + \frac{1}{2} \twoE{l_2}{l_2}{l_1}{l_3} \\
    &\quad - \twoE{l_2}{l_1}{l_3}{l_1} - \twoE{l_2}{h_2}{l_3}{h_2} - \twoE{l_2}{h_1}{l_3}{h_1} \\
    &\quad+ \sum_{i\in h \setminus \{h_1,h_2\}} \left( \twoE{h_2}{i}{i}{l_3} - 2 \twoE{h_2}{i}{l_3}{i} \right) \mathrm{.}
\end{split}
\end{align}
This coupling is primarily governed by the hopping integral $\oneE{l_2}{l_3}$, representing a direct one-electron transfer between the LUMOs of the donor and acceptor molecules. This term is augmented by various two-electron interactions; notably, the terms $\twoE{h_2}{l_2}{l_3}{h_2}$ and $\twoE{l_2}{h_2}{l_3}{h_2}$ are localized on the adjacent fragments $2$ and $3$, thus maintaining significant values when the monomers are in close proximity.

In contrast, the \leleR{} to \ctctR{} coupling is given by:
\begin{align}
    \mel{\szero\sone\sone\szero}{\hat{\mathcal H}}{\dpm\dmm\dpm\dmm} = - \twoE{h_1}{l_3}{h_2}{l_4} + \frac{1}{2} \twoE{h_1}{l_3}{l_4}{h_2} \mathrm{.}
\end{align}
In this case, both the Coulomb integral $\twoE{h_1}{l_3}{h_2}{l_4}$ and the exchange integral $\twoE{h_1}{l_3}{l_4}{h_2}$ involve four distinct molecular fragments. Due to the lack of spatial overlap between these four centers, these multi-center integrals are inherently small. This symbolic interpretation explains why \leleR{}--\ctctR{} couplings are significantly weaker than \leleR{}--\lectR{} couplings and suggests that such type of expressions could serve as the basis for simplified models in high-throughput screening applications.
}

\subsubsection{Anthracene}
For anthracene we examined the same four types of aggregates (depicted in \reffigure{fig:geometries}~b)), but restricted the analysis to pentamers due to the computational cost of the \noci{} method. The aggregate geometries were derived from a dimer scan and feature a fixed interplanar separation of \SI{3.75}{\angstrom}. The H-type aggregate exhibits no lateral displacement. The Null-Type aggregate, has an additional translation per monomer of \SI{2.30}{\angstrom} along the shorter molecular axis (y-axis). The Zero-Frenkel aggregate features an additional translation per monomer of \SI{3.00}{\angstrom} in y-direction. Lastly, the J-type aggregate features an additional translation per monomer of \SI{4.50}{\angstrom} along the short molecular axis.

The set of diabatic states used to construct the benchmark Hamiltonians is analogous to that employed for ethylene. Since all five monomers participate in the excitations, edge constraints are lifted. In addition, two spatially separated \ctctR{} configurations---D$^+$D$^-$S$_0$D$^+$D$^-$ and D$^-$D$^+$S$_0$D$^-$D$^+$---were included in the analysis.
Having already compared key couplings in the ethylene model, we now shift focus to the adiabatic wavefunctions obtained by diagonalizing the respective diabatic Hamiltonians. \reftable{tab:anthracene_adiabatic_wavefuncitons} reports the two lowest adiabatic double-excitation energies (relative to the ground state) and their dominant diabatic contributions, classified into \leleR{}, spatially separated \leleR{} (\sepleleR{}), \lectR{}, \ctctR{}, and separated \ctctR{} (\sepctctR{}), for each of the four geometries and for both methods.

\begin{table*}[!h]
    \centering
    \begin{tabular}{|llr|r|r|r|r|r|r|}\hline
 \removeTableShort{H Aggregate}{H-type} & $\Psi$ & $\Delta E$/\si{\eV} &\gsR{} & \leleR{} & \sepleleR{} & \lectR{} & \ctctR{} & \sepctctR{} \\ \hline
& $\Psi_{1}^{\mathrm{NOCI}}$ & \num[round-mode=places,round-precision=\precision]{9.2987781432} & \num[round-mode=places,round-precision=\precision]{0.0000802175} & \num[round-mode=places,round-precision=\precision]{0.2815997109} & \num[round-mode=places,round-precision=\precision]{0.3157066245} & \num[round-mode=places,round-precision=\precision]{0.2665038807} & \num[round-mode=places,round-precision=\precision]{0.0260922144} & \num[round-mode=places,round-precision=\precision]{0.0001700778}  \\
\ifbool{finalversion}{}{
\oldrow & $\Psi_{1}^{\mathrm{SymCI}}$ & \num[round-mode=places,round-precision=\precision]{8.0125248203} & \num[round-mode=places,round-precision=\precision]{0.0004947888} & \num[round-mode=places,round-precision=\precision]{0.3128638930} & \num[round-mode=places,round-precision=\precision]{0.3434795869} & \num[round-mode=places,round-precision=\precision]{0.3102707163} & \num[round-mode=places,round-precision=\precision]{0.0328910119} & \num[round-mode=places,round-precision=\precision]{0.0000000032}  \\
}
\newrow & $\Psi_{1}^{\mathrm{SymCI}}$ & \num[round-mode=places,round-precision=\precision]{8.1929581933} & \num[round-mode=places,round-precision=\precision]{0.0005022207} & \num[round-mode=places,round-precision=\precision]{0.3128675177} & \num[round-mode=places,round-precision=\precision]{0.3436194966} & \num[round-mode=places,round-precision=\precision]{0.3084769245} & \num[round-mode=places,round-precision=\precision]{0.0345338347} & \num[round-mode=places,round-precision=\precision]{0.0000000058}  \\
\hline
 & $\Psi_{2}^{\mathrm{NOCI}}$ & \num[round-mode=places,round-precision=\precision]{9.4255721160} & \num[round-mode=places,round-precision=\precision]{0.0000000000} & \num[round-mode=places,round-precision=\precision]{0.2982361653} & \num[round-mode=places,round-precision=\precision]{0.3166909023} & \num[round-mode=places,round-precision=\precision]{0.2657679410} & \num[round-mode=places,round-precision=\precision]{0.0201802476} & \num[round-mode=places,round-precision=\precision]{0.0000019880}  \\
\ifbool{finalversion}{}{
\oldrow & $\Psi_{2}^{\mathrm{SymCI}}$ & \num[round-mode=places,round-precision=\precision]{8.1579101736} & \num[round-mode=places,round-precision=\precision]{0.0000000000} & \num[round-mode=places,round-precision=\precision]{0.3201168527} & \num[round-mode=places,round-precision=\precision]{0.3395908822} & \num[round-mode=places,round-precision=\precision]{0.3120051462} & \num[round-mode=places,round-precision=\precision]{0.0282864231} & \num[round-mode=places,round-precision=\precision]{0.0000006958}  \\
}
\newrow & $\Psi_{2}^{\mathrm{SymCI}}$ & \num[round-mode=places,round-precision=\precision]{8.3374977728} & \num[round-mode=places,round-precision=\precision]{0.0000000000} & \num[round-mode=places,round-precision=\precision]{0.3198958986} & \num[round-mode=places,round-precision=\precision]{0.3368295707} & \num[round-mode=places,round-precision=\precision]{0.3122958154} & \num[round-mode=places,round-precision=\precision]{0.0309782294} & \num[round-mode=places,round-precision=\precision]{0.0000004859}  \\
\hline
\hline
 \removeTableShort{Zero Davydov}{Null-Type} & $\Psi$ & $\Delta E$/\si{\eV} &\gsR{} & \leleR{} & \sepleleR{} & \lectR{} & \ctctR{} & \sepctctR{} \\ \hline
 & $\Psi_{1}^{\mathrm{NOCI}}$ & \num[round-mode=places,round-precision=\precision]{10.0690997800} & \num[round-mode=places,round-precision=\precision]{0.0000000000} & \num[round-mode=places,round-precision=\precision]{0.8691598364} & \num[round-mode=places,round-precision=\precision]{0.0186127601} & \num[round-mode=places,round-precision=\precision]{0.0913903529} & \num[round-mode=places,round-precision=\precision]{0.0014230600} & \num[round-mode=places,round-precision=\precision]{0.0000000025}  \\
 \ifbool{finalversion}{}{
\oldrow & $\Psi_{1}^{\mathrm{SymCI}}$ & \num[round-mode=places,round-precision=\precision]{8.8956072912} & \num[round-mode=places,round-precision=\precision]{0.0000450448} & \num[round-mode=places,round-precision=\precision]{0.4903334742} & \num[round-mode=places,round-precision=\precision]{0.4789621037} & \num[round-mode=places,round-precision=\precision]{0.0306047130} & \num[round-mode=places,round-precision=\precision]{0.0000546635} & \num[round-mode=places,round-precision=\precision]{0.0000000009}  \\
}
\newrow & $\Psi_{1}^{\mathrm{SymCI}}$ & \num[round-mode=places,round-precision=\precision]{9.0723953763} & \num[round-mode=places,round-precision=\precision]{0.0000471175} & \num[round-mode=places,round-precision=\precision]{0.4904440795} & \num[round-mode=places,round-precision=\precision]{0.4845992098} & \num[round-mode=places,round-precision=\precision]{0.0248821897} & \num[round-mode=places,round-precision=\precision]{0.0000274027} & \num[round-mode=places,round-precision=\precision]{0.0000000008}  \\
\hline
 & $\Psi_{2}^{\mathrm{NOCI}}$ & \num[round-mode=places,round-precision=\precision]{10.0726378851} & \num[round-mode=places,round-precision=\precision]{0.0000052267} & \num[round-mode=places,round-precision=\precision]{0.2450836944} & \num[round-mode=places,round-precision=\precision]{0.6134260307} & \num[round-mode=places,round-precision=\precision]{0.1117878662} & \num[round-mode=places,round-precision=\precision]{0.0031627999} & \num[round-mode=places,round-precision=\precision]{0.0000078011}  \\
 \ifbool{finalversion}{}{
\oldrow & $\Psi_{2}^{\mathrm{SymCI}}$ & \num[round-mode=places,round-precision=\precision]{8.9100312857} & \num[round-mode=places,round-precision=\precision]{0.0000000000} & \num[round-mode=places,round-precision=\precision]{0.5939959622} & \num[round-mode=places,round-precision=\precision]{0.3528622570} & \num[round-mode=places,round-precision=\precision]{0.0525253126} & \num[round-mode=places,round-precision=\precision]{0.0006164676} & \num[round-mode=places,round-precision=\precision]{0.0000000005}  \\
}
\newrow & $\Psi_{2}^{\mathrm{SymCI}}$ & \num[round-mode=places,round-precision=\precision]{9.0880111717} & \num[round-mode=places,round-precision=\precision]{0.0000000000} & \num[round-mode=places,round-precision=\precision]{0.5888011308} & \num[round-mode=places,round-precision=\precision]{0.3633599853} & \num[round-mode=places,round-precision=\precision]{0.0471401291} & \num[round-mode=places,round-precision=\precision]{0.0006987543} & \num[round-mode=places,round-precision=\precision]{0.0000000004}  \\
\hline
\hline
 \removeTableShort{Zero Coupling}{Zero-Frenkel} & $\Psi$ & $\Delta E$/\si{\eV} &\gsR{} & \leleR{} & \sepleleR{} & \lectR{} & \ctctR{} & \sepctctR{} \\ \hline
 & $\Psi_{1}^{\mathrm{NOCI}}$ & \num[round-mode=places,round-precision=\precision]{10.0333510780} & \num[round-mode=places,round-precision=\precision]{0.0000000072} & \num[round-mode=places,round-precision=\precision]{0.4337723764} & \num[round-mode=places,round-precision=\precision]{0.4333608106} & \num[round-mode=places,round-precision=\precision]{0.1012402694} & \num[round-mode=places,round-precision=\precision]{0.0020912177} & \num[round-mode=places,round-precision=\precision]{0.0000073115}  \\
 \ifbool{finalversion}{}{
\oldrow & $\Psi_{1}^{\mathrm{SymCI}}$ & \num[round-mode=places,round-precision=\precision]{8.8761401908} & \num[round-mode=places,round-precision=\precision]{0.0000001923} & \num[round-mode=places,round-precision=\precision]{0.4555046180} & \num[round-mode=places,round-precision=\precision]{0.4254575032} & \num[round-mode=places,round-precision=\precision]{0.1167620352} & \num[round-mode=places,round-precision=\precision]{0.0022756488} & \num[round-mode=places,round-precision=\precision]{0.0000000025}  \\
}
\newrow & $\Psi_{1}^{\mathrm{SymCI}}$ & \num[round-mode=places,round-precision=\precision]{9.0608314925} & \num[round-mode=places,round-precision=\precision]{0.0000002066} & \num[round-mode=places,round-precision=\precision]{0.4528178261} & \num[round-mode=places,round-precision=\precision]{0.4269506221} & \num[round-mode=places,round-precision=\precision]{0.1178173140} & \num[round-mode=places,round-precision=\precision]{0.0024140299} & \num[round-mode=places,round-precision=\precision]{0.0000000014}  \\

\hline
 & $\Psi_{2}^{\mathrm{NOCI}}$ & \num[round-mode=places,round-precision=\precision]{10.0540537833} & \num[round-mode=places,round-precision=\precision]{0.0000000000} & \num[round-mode=places,round-precision=\precision]{0.4266569166} & \num[round-mode=places,round-precision=\precision]{0.4560645642} & \num[round-mode=places,round-precision=\precision]{0.0904838404} & \num[round-mode=places,round-precision=\precision]{0.0011684487} & \num[round-mode=places,round-precision=\precision]{0.0000000023}  \\
 \ifbool{finalversion}{}{
\oldrow & $\Psi_{2}^{\mathrm{SymCI}}$ & \num[round-mode=places,round-precision=\precision]{8.8984778155} & \num[round-mode=places,round-precision=\precision]{0.0000000000} & \num[round-mode=places,round-precision=\precision]{0.4504037495} & \num[round-mode=places,round-precision=\precision]{0.4435521895} & \num[round-mode=places,round-precision=\precision]{0.1045254664} & \num[round-mode=places,round-precision=\precision]{0.0015185899} & \num[round-mode=places,round-precision=\precision]{0.0000000047}  \\
}
\newrow & $\Psi_{2}^{\mathrm{SymCI}}$ & \num[round-mode=places,round-precision=\precision]{9.0822820028} & \num[round-mode=places,round-precision=\precision]{0.0000000000} & \num[round-mode=places,round-precision=\precision]{0.4434483061} & \num[round-mode=places,round-precision=\precision]{0.4490192538} & \num[round-mode=places,round-precision=\precision]{0.1058832204} & \num[round-mode=places,round-precision=\precision]{0.0016492142} & \num[round-mode=places,round-precision=\precision]{0.0000000055}  \\

\hline
\hline
 \removeTableShort{J Aggregate}{J-type} & $\Psi$ & $\Delta E$/\si{\eV} &\gsR{} & \leleR{} & \sepleleR{} & \lectR{} & \ctctR{} & \sepctctR{} \\ \hline
 & $\Psi_{1}^{\mathrm{NOCI}}$ & \num[round-mode=places,round-precision=\precision]{10.0989575512} & \num[round-mode=places,round-precision=\precision]{0.0000136628} & \num[round-mode=places,round-precision=\precision]{0.5064722895} & \num[round-mode=places,round-precision=\precision]{0.4677988980} & \num[round-mode=places,round-precision=\precision]{0.0195868284} & \num[round-mode=places,round-precision=\precision]{0.0000618308} & \num[round-mode=places,round-precision=\precision]{0.0000003770}  \\
 \ifbool{finalversion}{}{
\oldrow & $\Psi_{1}^{\mathrm{SymCI}}$ & \num[round-mode=places,round-precision=\precision]{8.9002416784} & \num[round-mode=places,round-precision=\precision]{0.0000644795} & \num[round-mode=places,round-precision=\precision]{0.5137496557} & \num[round-mode=places,round-precision=\precision]{0.4660767144} & \num[round-mode=places,round-precision=\precision]{0.0200662200} & \num[round-mode=places,round-precision=\precision]{0.0000429303} & \num[round-mode=places,round-precision=\precision]{0.0000000001}  \\
}
\newrow & $\Psi_{1}^{\mathrm{SymCI}}$ & \num[round-mode=places,round-precision=\precision]{9.0824806755} & \num[round-mode=places,round-precision=\precision]{0.0000660929} & \num[round-mode=places,round-precision=\precision]{0.5136513067} & \num[round-mode=places,round-precision=\precision]{0.4668354441} & \num[round-mode=places,round-precision=\precision]{0.0194063852} & \num[round-mode=places,round-precision=\precision]{0.0000407710} & \num[round-mode=places,round-precision=\precision]{0.0000000000}  \\
\hline
 & $\Psi_{2}^{\mathrm{NOCI}}$ & \num[round-mode=places,round-precision=\precision]{10.1190666397} & \num[round-mode=places,round-precision=\precision]{0.0000000000} & \num[round-mode=places,round-precision=\precision]{0.4855245312} & \num[round-mode=places,round-precision=\precision]{0.4923267218} & \num[round-mode=places,round-precision=\precision]{0.0169585897} & \num[round-mode=places,round-precision=\precision]{0.0000250604} & \num[round-mode=places,round-precision=\precision]{0.0000000000}  \\
 \ifbool{finalversion}{}{
\oldrow & $\Psi_{2}^{\mathrm{SymCI}}$ & \num[round-mode=places,round-precision=\precision]{8.9351240638} & \num[round-mode=places,round-precision=\precision]{0.0000000000} & \num[round-mode=places,round-precision=\precision]{0.4989079480} & \num[round-mode=places,round-precision=\precision]{0.4834109614} & \num[round-mode=places,round-precision=\precision]{0.0176588310} & \num[round-mode=places,round-precision=\precision]{0.0000222594} & \num[round-mode=places,round-precision=\precision]{0.0000000002}  \\
}
\newrow & $\Psi_{2}^{\mathrm{SymCI}}$ & \num[round-mode=places,round-precision=\precision]{9.1174251866} & \num[round-mode=places,round-precision=\precision]{0.0000000000} & \num[round-mode=places,round-precision=\precision]{0.4982738244} & \num[round-mode=places,round-precision=\precision]{0.4846372607} & \num[round-mode=places,round-precision=\precision]{0.0170680842} & \num[round-mode=places,round-precision=\precision]{0.0000208306} & \num[round-mode=places,round-precision=\precision]{0.0000000001}  \\
\hline
\end{tabular}
    \caption{Comparison of selected adiabatic doubly-excited wavefunctions obtained from diagonalization of the diabatic Hamiltonians using \symci{} and \noci{} for the anthracene aggregate. The table reports the energies relative to the ground state and the contributions of \leleR{}, separated \leleR{} (\sepleleR{}), \lectR{}, \ctctR{}, and separated \ctctR{} (\sepctctR{}) for the two lowest adiabatic states across all four aggregate types.}
    \label{tab:anthracene_adiabatic_wavefuncitons}
\end{table*}

As shown in \reftable{tab:anthracene_adiabatic_wavefuncitons}, the wavefunctions obtained from \noci{} and \symci{} are largely similar in terms of diabatic composition. Notably, for the Null-Type aggregate, \symci{} predicts somewhat larger contributions from spatially separated \leleR{} states than \noci{} for the first two excited biexcitonic states. \added{But since the \noci{} wavefunctions are degenerated, their order and mixing is arbitrary.} Regarding excitation energies, \noci{} yields consistently higher values, about \replaced{\SIrange{1.0}{1.3}{\eV}}{\SIrange{1.0}{1.1}{\eV}} above those from \symci{}.
\added{This systematic energy offsets between the methods are significantly reduced upon the application of a monomer-based diagonal shift to the \symci{} Hamiltonian, the details of which are provided in the Supporting Information (Section~\ref{SI-sec:diagonal_correction}). Running the same calculations with a bigger active space per fragment yielded very similar results (see Supporting Information Section~\ref{SI-sec:big_AS}).}
Despite this offset, the energy gap between the two lowest biexcitonic states, $\Psi_1$ and $\Psi_2$, is very similar for both methods, indicating good agreement on relative energetics. 
This mirrors the trends found for ethylene, where both methods captured biexciton couplings and wavefunction composition with good consistency overall, but where \symci{} tended to underestimate couplings involving charge transfer (\ctR{}) and delocalization across more than two monomers.
Taken together, these findings indicate that both methods yield qualitatively consistent results across chemically distinct systems. The agreement between \noci{} and \symci{} is slightly better for anthracene than for ethylene in terms of wavefunction composition, but systematic underestimation of \ctR{} couplings in \symci{} remains a shared limitation in both cases. The physical trends governing the character and energetics of the biexcitonic states—such as the role of packing geometry, overlap-driven stabilization, and \ctR{}-induced mixing—are closely parallel in the two systems.

\subsection{Physical Insights from the Model Aggregates}
\begin{figure}[ht]
    \centering
    \includegraphics[width=0.4\linewidth]{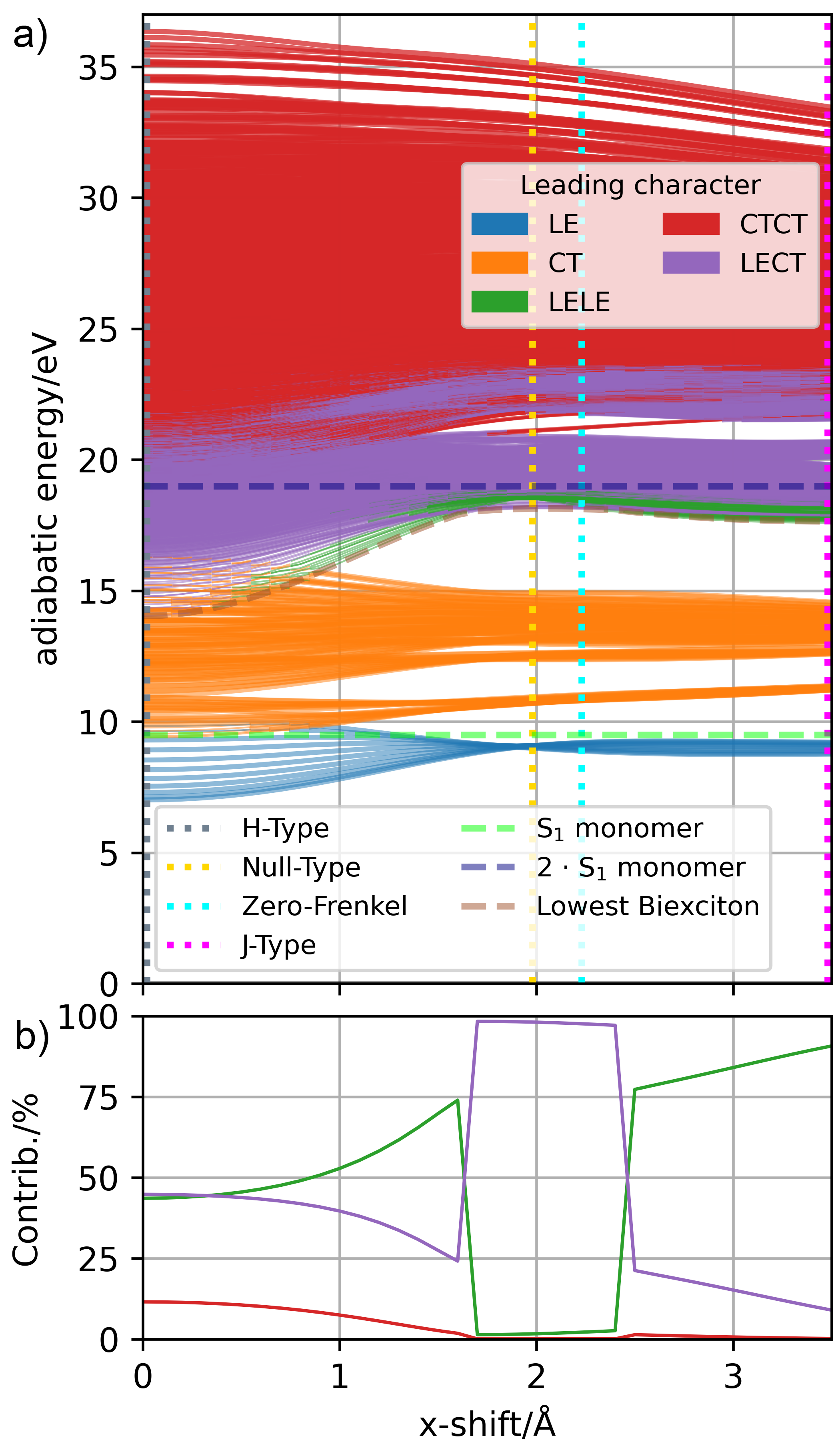}
    \caption{Adiabatic energies obtained from diagonalization of the diabatic \symci{} Hamiltonian along the scan coordinate. The scan systematically probes the transition from a perfectly stacked H-aggregate to a slipped J-aggregate by displacing the monomers in the x-direction (along the C--C bond) from \SIrange{0.0}{3.5}{\angstrom} in steps of \SI{0.1}{\angstrom}, while maintaining a constant interplanar distance of \SI{3.5}{\angstrom}. a) A total of 3146 excited states are represented, with each state color-coded according to its leading diabatic character. The ground state is set to \SI{0}{\eV}. b) The composition of to lowest biexcitonic state \added{(see discussion in the SI chapter~\ref{SI-sec:biexcitonic_section})} along the scan coordinate.}
    \label{fig:ethene_scan_adiabatic}
\end{figure}

Revisiting the obtained results presented \reftable{tab:ethene_couplings} with a scope on the physical interpretation, interesting aspects can be observed: First of all, it becomes evident that the type of aggregate exerts a substantial influence on the electronic couplings. For instance, the coupling associated with a non-parallel biexciton \replaced{diffusion}{transfer} process, $\mel{\mathrm{S}_1\mathrm{S}_1\mathrm{S}_0}{\hat{\mathcal{H}}}{\mathrm{S}_1\mathrm{S}_0\mathrm{S}_1}$, is particularly strong in the H-aggregate, reaching \SI{0.45}{\eV}, but nearly vanishes in the Zero-Frenkel aggregate, dropping below \SI{0.02}{\eV}. More generally, all selected couplings are largest in the H-aggregate, consistent with its maximized spatial orbital overlap. Furthermore, we find that couplings between Frenkel-type biexcitons and \lectR{} states, $\mel{\mathrm{S}_1\mathrm{S}_1\mathrm{S}_0}{\hat{\mathcal{H}}}{\mathrm{S}_1\mathrm{D}^+\mathrm{D}^-}$ and $\mel{\mathrm{S}_1\mathrm{S}_1\mathrm{S}_0}{\hat{\mathcal{H}}}{\mathrm{S}_1\mathrm{D}^-\mathrm{D}^+}$, are consistently large in all aggregate structures, underscoring the role of \lectR{} mixing in the biexciton regime.
To obtain a more global picture of how biexciton couplings and adiabatic energies evolve with molecular packing, we performed a continuous scan from the eclipsed H-aggregate to the fully slipped J-aggregate, displacing each monomer along the molecular x-axis (C–C bond direction) from \SI{0.0}{\angstrom} to \SI{3.5}{\angstrom} in steps of \SI{0.1}{\angstrom}, with the z-distance fixed at \SI{3.5}{\angstrom}. This scan was carried out exclusively with \symci{}, which allows efficient construction and diagonalization of the full biexcitonic Hamiltonian.
In contrast to the reduced 72-state model used for benchmarking, the full Hamiltonian now includes all \leR{} and \ctR{} single excitons as well as \leleR{}, \lectR{}, and \ctctR{} biexcitons and their range separated variants, totaling 3147 diabatic configurations. The resulting adiabatic energies along the scan coordinate are shown in \reffigure{fig:ethene_scan_adiabatic}. The color of each state reflects its dominant diabatic character, providing a clear visual impression of how the electronic structure evolves across the slip coordinate.

In the single-exciton domain, we observe the expected Davydov splitting of \leR{} states (blue) in the H-aggregate. As the monomer displacement increases toward the J-aggregate, this splitting diminishes and vanishes at the Null-Type geometry (indicated by the yellow dashed line), where dipolar couplings and \ctR{} couplings effectively cancel each other, leading to degenerate "monomer-like" states. Beyond this point, the splitting re-emerges in the J-aggregate regime, though it remains weaker than in the H-aggregate due to reduced spatial overlap and hence weaker excitonic coupling.
The single-exciton \ctR{} states (orange) lie energetically above the \leR{} manifold, but exhibit notable mixing in the H-aggregate region, where the lowest \ctR{} states approach the upper \leR{} states. As with the \leR{} states, \ctR{} states are most strongly split in the H-aggregate and gradually contract toward the J-aggregate, although no full degeneracy is observed. Within the \ctR{} manifold, configurations with adjacent \dpR{} and \dmR{} centers are energetically stabilized up to \SI{2}{\eV} relative to more delocalized \ctR{} configurations, forming a distinct low-energy subgroup.

Across most of the scan (except of the vicinity of the H-aggregate and Null aggregate geometries), the lowest biexcitonic states are dominated by \leleR{} character. In the immediate H-aggregate region, however, no low-lying biexcitonic state exhibits a leading \leleR{} contribution; instead, the lowest states are dominated by \lectR{} configurations. Upon increasing lateral displacement, the energy of the low-lying biexciton drastically rises and \leleR{} states gradually emerge from the \lectR{} manifold, becoming the energetically lowest biexcitons beyond an x-shift of approximately \SI{1}{\angstrom}. At the Null-Type aggregate, \lectR{} states again form the lowest biexcitonic states.
\added{Specifically, in the region between \SIrange{1.7}{2.4}{\angstrom}, a nearly pure manifold of \lectR{} states intersects the adiabatic mixture formed by the \leleR{} and \lectR{} bands. This energetic crossing dictates the character of the lowest-lying state; within this window, the \lectR{} configurations become energetically more favorable than the previously dominant \leleR{} state. Consequently, the rapid character shift visualized in \reffigure{fig:ethene_scan_adiabatic}~b) is a direct result of this diabatic state crossing, causing the tracking algorithm to switch to a different lowest-energy state. This behavior is driven by the near-degeneracy of the \leleR{} states in this spatial region, which leads to their full contraction and subsequent reordering within the energy hierarchy.}
The \lectR{} states largely follow the energetic behavior expected from a superposition of single-exciton \leR{} and \ctR{} states and therefore exhibit pronounced splitting in the H-aggregate region, where intermonomer coupling is strongest. At higher energies, the \ctctR{} states form a separate manifold. Although located well above the lowest biexcitons overall, the \ctctR{} states overlap substantially with the \lectR{} band, particularly near the H-aggregate, where the density of states increases and becomes quasi-continuous over a broad energy range.
As a consequence, the \leR{} states energetically overlap with \ctR{} states, the \ctR{} states overlap with \lectR{} biexcitons, and the \lectR{} states in turn overlap with \ctctR{} configurations, giving rise to a dense and interconnected manifold in the H-aggregate regime. This energetic overlap between the single-exciton and biexcitonic sectors progressively diminishes upon shifting toward the J-aggregate, where reduced intermonomer interactions lead to a contraction and separation of the respective bands.
In \reffigure{fig:ethene_scan_adiabatic}~b), the diabatic composition of the lowest biexcitonic state along the scan coordinate is shown. In the H-aggregate region, this state exhibits nearly equal leading contributions from \leleR{} and \lectR{} configurations. Concomitantly, it undergoes a pronounced stabilization in energy relative to neighboring packing arrangements. This combination of energetic lowering and balanced admixture of locally excited and charge-transfer character invites an analogy to excimer formation, where a strongly stabilized one-particle excited state arises from approximately equal contributions of \leR{} and \ctR{} configurations and is known to act as a trap for exciton migration.\cite{hoche_mechanism_2017} However, assessing the physical implications of such a potential “bi-excimer” state, including its formation mechanism, stability, and dynamical role,  requires further dedicated detailed theoretical analysis. \added{The concept of a bi-excimer introduced here provides a microscopic interpretation of stabilized biexciton states in terms of mixed \leleR{}/\lectR{} character, suggesting that experimentally observed bound biexcitons may, in some cases, originate from such hybridization.}
 
 For the anthracene aggregates, similar physical trends \added{are} observed: the H-aggregate again stabilizes the lowest biexcitons\added{, since the energy of $\Psi_{1}$ at the H-type geometry is smaller then the other energies of the lowest biexcitonic state $\Psi_{1}$ at the other three geometries,} and exhibits strong admixture of \lectR{} character, while the J-aggregate leads to more diabatically pure \leleR{} and \ctctR{} states. As in ethylene, the \ctR{} contributions are most pronounced in H-type geometries, and diminish upon slipping toward the J-regime.

\begin{figure}[!h]
    \centering
    \includegraphics[width=0.4\linewidth]{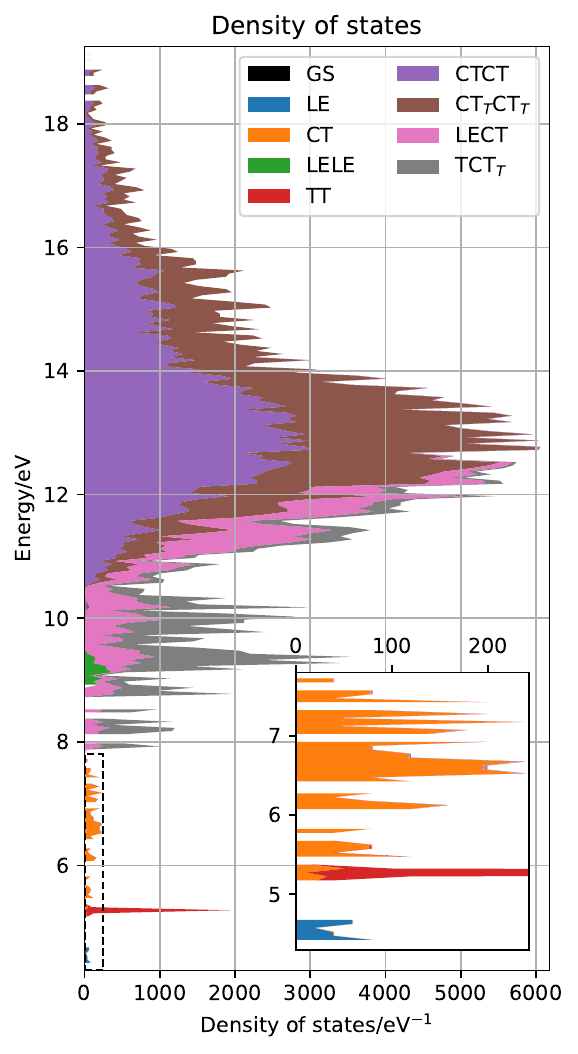}
    \caption{Density of adiabatic states obtained from diagonalization of the full Hamiltonian in the monomer frontier orbital space of the anthracene cut-off. Contributions from \leR{}, \ctR{}, \leleR{}, \lectR{}, \ctctR{}, \ttR{}, \tcttR{} and \cttcttR{} configurations are distinguished. The accumulated contributions give the complete density of states.}
    \label{fig:anthracene_crystall_DOS}
\end{figure}
To access a more realistic packing arrangement, we constructed a cutout structure from the anthracene crystal lattice containing 15 monomers, using the experimental geometry reported by Marciniak et al.\cite{marciniak_crystal_2002}. A side view of the aggregate is shown in \reffigure{fig:geometries}~c). \added{The anthracene monomer geometry was optimized at the MP2 level using the smaller \textit{cc-pVDZ}\cite{dunning1989gaussian} basis set. \symci{} calculations employed SA-CASSCF(2,2)\cite{roos_complete_1980,RN155,RN230} orbitals obtained from \textit{ORCA 6.0}\cite{RN204, RN205, RN84, RN139, RN21, RN171, RN200, RN96, RN178} and included only the frontier orbitals of each monomer in the active space.} The full Hamiltonian was constructed in the monomer-local frontier-orbital basis, restricted to spin-singlet configurations and the \remove{smaller} \textit{cc-pVDZ}\cite{dunning1989gaussian} basis set, using \symci{}. In addition to the ground state, the model includes all \leR{} and \ctR{} single excitons, biexcitonic states of \leleR{}, \lectR{}, and \ctctR{} type, as well as multiexciton states involving \added{local} triplets (\ttR{}, \tcttR{}, and \cttcttR{})\added{, which are paired to a spin-singlet}, yielding a diabatic Hamiltonian of 22 276 configurations. After diagonalization, adiabatic energies and leading diabatic characters were obtained. Given the dimensionality of the problem, the analysis focuses on the energy-resolved density of states (DOS) with color-coded diabatic character groups (see \reffigure{fig:anthracene_crystall_DOS}).
The resulting spectrum reproduces the energetic ordering of excitonic families observed in the idealized aggregates. The single-exciton manifold spans \SIrange{4.4}{7.7}{\eV}, with the lowest states (\SIrange{4.4}{4.7}{\eV}) being nearly pure \leR{} excitons. A band of nearest-neighbor \ctR{} states appears around \SI{5.3}{\eV} and coincides with the \ttR{} biexciton manifold, showing that singlet-coupled triplet pairs are energetically accessible already within the single-exciton energy window as observed in singlet fission, Furthermore, the narrow bandwidth of the latter indicates weak interaction between different triplet-pair configurations, in line with theoretical findings by Scholes, Singh and de Sousa. Higher-lying \ctR{} states associated with more distant charge separation extend up to \SI{7.7}{\eV}. The broad and dense \lectR{} and \tcttR{} state manifolds follows almost seamlessly, which extend up to \SI{10.5}{\eV}, bridging the one and two particle \ctR{} states and overlap with the comparatively narrow \leleR{} band that appears between \SIrange{9.0}{9.3}{\eV}. 
This identifies double excitations containing a single charge-transfer component, irrespective of their spin character, as the central connective layer between the single-exciton and biexcitonic regimes.

At higher energies, the DOS is increasingly dominated by \ctctR{} and \cttcttR{} configurations, forming a dense band that reaches \SI{18.9}{\eV}.
  Together, these features delineate a densely interconnected multiexcitonic landscape in which charge-transfer–containing states play a key structural role.

\begin{figure}[!htbp]
    \centering
    \includegraphics[width=0.4\linewidth]{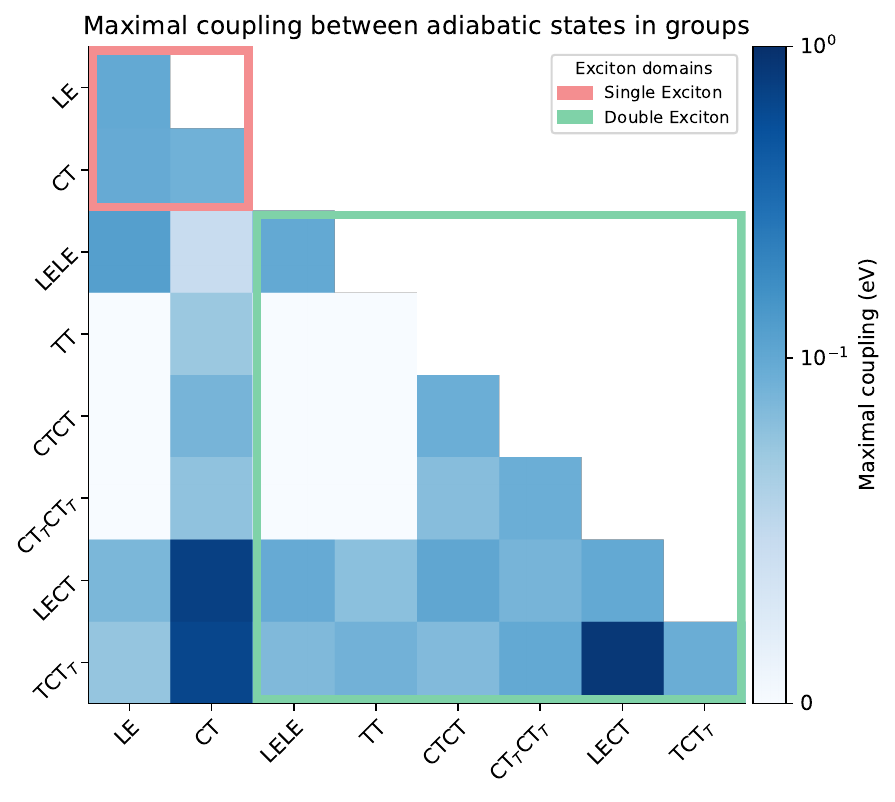}
    \caption{Coupling matrix of diabatic states for the Hamiltonian of the anthracene crystal cutout. The plot shows the maximal absolute couplings between groups of diabatic states in the lower triangle and intra-group couplings (excluding the diabatic energies) along the diagonal blocks. A color scale is used, with logarithmic scaling from \SIrange{1.0}{0.1}{\eV} and linear scaling from \SIrange{0.1}{0.0}{\eV}.}
    \label{fig:anthracene_crystall_couplings}
\end{figure}

To assess the potential mixing between different classes of diabatic states, such as \ctctR{} and \cttcttR{}, we analyze the couplings between all diabatic states grouped by their electronic character. For inter-group interactions, we focus on the maximal absolute coupling between any pair of states belonging to two different groups. To quantify intra-group coupling, we evaluate the couplings among states within the same group, excluding the diagonal elements corresponding to the diabatic energies. \reffigure{fig:anthracene_crystall_couplings} presents the resulting coupling matrix for the Hamiltonian of the anthracene crystal cut-off.

The strongest couplings occur between the \ctR{} and the \lectR{} states, between the \ctR{} and \tcttR{} states, as well as between \lectR{} and \tcttR{} states, all reaching magnitudes on the order of \SI{1}{\eV}. Hence, all states featuring one \ctR{} configuration are strongly coupled to one another. Couplings between other diabatic classes are generally weaker, on the order of \SI{100}{\milli\eV}, including the \leR{}--\ctR{}, \leR{}--\leleR{}, \leR{}--\lectR{}, \leleR{}--\lectR{}, \ttR{}--\tcttR{}, and \ctctR{}--\cttcttR{} interactions. Notable exceptions exist where coupling is essentially absent: \ttR{} states couple only weakly to other \ttR{} states and show negligible interaction with \ctctR{} and \cttcttR{} configurations. Similarly, \leleR{} and \leR{} states exhibit very small couplings to \ttR{}, \ctctR{}, and \cttcttR{}.
Importantly, the \lectR{} and \tcttR{} manifolds not only provide strong electronic coupling but also span the energy gap between the single-exciton \ctR{} band (extending up to \SI{7.7}{\eV}) and the \leleR{} biexciton states (located around \SIrange{9.0}{9.3}{\eV}). This dual role---energetic continuity combined with strong coupling---positions the \ctxR{} states as a central interface layer between the one-particle and two-particle manifolds.
This connectivity suggests mechanistic pathways beyond those commonly discussed. In SF, the transition from the single-exciton \leR{} state to the biexcitonic \ttR{} manifold is mediated by virtual \ctR{} configurations. The present analysis reveals an analogous but distinct pathway connecting the \leleR{} biexcitons to the single-exciton manifold: \leleR{} →
 \lectR{} → \ctR{} → \leR{}. This sequence is energetically downhill and proceeds via the strongest couplings identified in the system (\SI{\sim 1}{\eV}), suggesting a \ctR{}-mediated biexciton relaxation channel that could compete with conventional Frenkel-type exciton--exciton annihilation pathways involving higher-lying singlet states.

Furthermore, the established picture of triplet-pair separation posits that the conversion from adjacent \ttR{} to spatially separated \seponettR{} configurations proceeds via virtual \tcttR{} intermediates through a Dexter-type superexchange mechanism. An analogous process may operate for Frenkel biexcitons: the spatial separation of a correlated \leleR{} state into a \sepleleR{} configuration could be mediated by virtual \lectR{} intermediates, rendering a \ctR{}-mediated superexchange a plausible pathway for biexciton dissociation.

We note that the present model restricts the single-exciton manifold to the lowest local excitations ($S_1$-type LE states) and does not include higher-lying states ($S_n$ ($n>1$)) on individual monomers. While this basis captures the essential connectivity between single- and two-particle manifolds, a complete description of exciton-exciton annihilation would require explicit inclusion of higher excited states, which serve as the energetic sink for the annihilation process.

\section{Conclusion and Outlook}
We have introduced two complementary fragment-based configuration-interaction approaches, \symci{} and \noci{}, for the systematic construction of multiexcitonic Hamiltonians in molecular aggregates. While \symci{} enables efficient large-scale calculations through its use of orthogonal orbitals and single-configuration fragment states, \noci{} provides benchmark-quality couplings from fully relaxed multiconfigurational fragment wavefunctions. Benchmarking on ethylene and anthracene aggregates demonstrates that both methods capture the essential physics of biexciton formation and yield consistent trends across packing geometries, with \symci{} showing minor limitations for \replaced{couplings involving delocalized charge-transfer configurations}{\leleR{}--\ctctR{} couplings}.
The systematic slip scans on ethylene aggregates reveal a pronounced geometry dependence of biexcitonic character. In H-type aggregates, the lowest biexcitonic states are not dominated by \leleR{} configurations, but instead exhibit strong \lectR{} admixture. The resulting state, characterized by nearly equal \leleR{} and \lectR{} contributions and a pronounced energetic stabilization, invites an analogy to excimer formation in the one-particle manifold: just as single-exciton excimers arise from balanced \leR{} and \ctR{} mixing and act as traps for exciton migration, this ``bi-excimer'' state may play an analogous role in the two-particle manifold. Whether such states function as traps for biexciton transport, and how their formation influences multiexcitonic dynamics, remains an open question. In contrast, J-type aggregates yield diabatically purer biexciton states with reduced \ctR{} admixture, suggesting that packing geometry can serve as a design parameter to tune biexcitonic character and intermanifold connectivity.
Application to a realistic 15-monomer anthracene crystal cutout, comprising over 22,000 diabatic configurations, reveals a hierarchical connectivity structure within the multiexcitonic landscape. The \ctxR{} states---\lectR{} and \tcttR{} configurations containing exactly one charge-transfer component---emerge as electronic gateways that interconnect the one-particle and two-particle manifolds. These states provide both energetic continuity, bridging the gap between the single-exciton \ctR{} band and the \leleR{} biexcitons, and the strongest electronic couplings identified in the system. This positions the \ctxR{} manifold as a \replaced{corridor}{gateway} through which population can flow between excitonic regimes. The resulting picture extends beyond established mechanisms such as \ctR{}-mediated SF: an analogous pathway, \leleR{} → \lectR{} →
\ctR{} 
→ \leR{}, may provide a \replaced{\ctR{}}{\ctxR{}}-mediated channel for biexciton relaxation that competes with conventional exciton--exciton annihilation involving higher-lying singlet states. Furthermore, in analogy to triplet-pair separation mediated by virtual  \replaced{\tcttR{}}{\ctxR{} (\tcttR{})} intermediates, the spatial separation of Frenkel biexcitons (\leleR{} → \sepleleR{}) may proceed via virtual \replaced{\lectR{}}{\ctxR{} (\lectR{})} configurations.

Several open questions emerge from this analysis. First, the dynamical role of \ctxR{} states: Whether they act as real, transiently populated intermediates or as virtual states mediating superexchange, remains to be established and will likely depend on the specific system and energetic landscape. Second, conventional exciton--exciton annihilation proceeds through higher-lying singlet states that undergo rapid internal conversion, rendering the process irreversible and dissipative. The \replaced{\ctR{}-mediated pathway}{\ctxR{}-gateway} identified here raises the question of whether intermanifold transitions via \ctxR{} states could proceed more reversibly, potentially enabling back-and-forth interconversion between one-particle and two-particle manifolds without the energy loss associated with relaxation through $S_n$ states. Third, at donor--acceptor interfaces, where \ctR{} states are energetically stabilized and more directly accessible, the \ctxR{}-\replaced{mediated pathways}{gateway} may become increasingly relevant, suggesting implications for the photophysics of organic heterojunctions. Looking forward, the fragment-based CI framework established here provides a foundation for addressing these questions. Enlarging the monomer-local active space within \symci{} to include additional occupied and virtual orbitals would enable explicit representation of higher singlet excitations ($S_n$), required for a complete description of exciton--exciton annihilation. Crucially, because \symci{} constructs spin-adapted configurations from fragment-local building blocks, the diabatic character labels (\leR{}, \ctR{}, \ttR{}, etc.) remain directly interpretable even in extended active spaces---each aggregate configuration retains explicit information about its constituent fragment occupations. This interpretability, which is central to the present analysis, would be lost in conventional supermolecular CI approaches where excited-state character must be extracted \textit{a posteriori} through diabatization procedures. Beyond intermanifold transitions, the fragment-based CI framework also provides direct access to the couplings governing biexciton motion within the two-particle manifold. \replaced{In analogy to exciton transport in the one-particle picture, where \leR{}--\leR{} couplings determine exciton diffusion, biexciton transport is governed by the electronic couplings between spatially distinct biexcitonic configurations---for instance, couplings of the type $\langle \mathrm{S}_1\mathrm{S}_1\mathrm{S}_0 | \hat{\mathcal{H}} | \mathrm{S}_0\mathrm{S}_1\mathrm{S}_1 \rangle$
describe correlated two-exciton hopping. These couplings, which emerge naturally from the diabatic Hamiltonian, open the possibility to study how correlated excitations propagate through extended aggregates---}{By analyzing the underlying symbolic expressions, we demonstrate that the \leleR{}--\lectR{} coupling is primarily driven by large one-electron hopping integrals ($\oneE{l_2}{l_3}$), whereas the \leleR{}--\ctctR{} coupling is inherently suppressed as it depends exclusively on multi-center two-electron integrals with vanishing spatial overlap. This hierarchical disparity in coupling strengths, which emerges naturally from the \symci{} framework, provides a rigorous basis for identifying the dominant relaxation and transport pathways in multiexcitonic systems. This rigorous basis for identifying dominant relaxation and transport pathways opens the possibility to study how correlated excitations propagate through extended aggregates---}a question that has received far less attention than single-exciton diffusion but may be critical for understanding multiexcitonic processes under high excitation densities.
Finally, the pronounced sensitivity of biexcitonic character and intermanifold connectivity to packing geometry suggests that molecular design and supramolecular organization can be exploited to engineer the \ctxR{} gateway---enhancing or suppressing specific multiexcitonic pathways to optimize functional properties such as singlet fission yield, triplet-pair separation efficiency, or upconversion performance. With this Perspective, we hope to stimulate further exchange between the electronic structure and quantum dynamics communities, as a deeper understanding of multiexcitonic processes will require both accurate descriptions of these correlated manifolds and dynamical methods capable of capturing their formation, transport, and decay.

\begin{acknowledgement}

J. E. A. is funded by the Deutsche Forschungsgemeinschaft (DFG, German Research Foundation) - IRTG 2991 Photoluminescence in Supramolecular Matrices - Project number 517122340. C. d. G. acknowledges financial support provided by the ministry of science and innovation of the Spanish administration through the project PID2023-148238NB-I00 and by the Generalitat de Catalunya through the projects 2021SGR00110. M. I. S. R. are thankful for funding by the Deutsche Forschungsgemeinschaft (DFG, German Research
Foundation) - Project number 555181242. This work used resources of the Oak Ridge Leadership Computing Facility (OLCF) at the Oak Ridge National Laboratory, which is supported by the Office of Science of the U.S. Department of Energy (DOE) under Contract DE-AC05-00OR22725 through the INCITE Project CHM154.

\end{acknowledgement}

\begin{suppinfo}
\added{The Supporting Information provides extended dataset analysis for the examples discussed in the main manuscript. Specifically, we include additional symbolic expressions, insights into the lowest biexciton states, and adiabatic comparisons for ethene using \noci{} and \symci{}; we also evaluate diagonal corrections and the impact of expanded active spaces for \symci{}.}
The computational results are openly available. These resources can be found in the GitHub
repository located at \href{https://github.com/roehr-lab/Fragment-Based-CI_data/}{\url{github.com/roehr-lab/Fragment-Based-CI_data}} \replaced{. This repository includes}{ and \href{https://doi.org/10.5281/zenodo.19249985}{\url{doi.org/10.5281/zenodo.19249985}}. These repositories include} all necessary data files required for reproducing the findings reported in this article.
Users can freely access and download the materials under the terms of the repository’s license to facilitate further research and verification of the results presented herein.

\end{suppinfo}

\bibliography{references}

@Preamble{
"\providecommand{\noopsort}[1]{}"
# "\providecommand{\singleletter}[1]{#1}%"
}

@article{Bossanyi_JACS_AU,
  author  = {David G. Bossanyi and Yoichi Sasaki and Shuangqing Wang and Dimitri Chekulaev and Nobuo Kimizuka and Nobuhiro Yanai and Jenny Clark},
  title   = {Spin Statistics for Triplet--Triplet Annihilation Upconversion: Exchange Coupling, Intermolecular Orientation, and Reverse Intersystem Crossing},
  journal = {JACS Au},
  year    = {2021},
  volume  = {1},
  number  = {12},
  pages   = {2188--2201},
  doi     = {10.1021/jacsau.1c00322}
}

@article{Bossanyi_Nat_chem,
  author  = {Bossanyi, David G. and Matthiesen, Maik and Wang, Shuangqing and Smith, Joel A. and Kilbride, Rachel C. and Shipp, James D. and Chekulaev, Dimitri and Holland, Emma and Anthony, John E. and Zaumseil, Jana and Musser, Andrew J. and Clark, Jenny},
  title   = {Emissive spin-0 triplet-pairs are a direct product of triplet--triplet annihilation in pentacene single crystals and anthradithiophene films},
  journal = {Nat. Chem.},
  year    = {2021},
  volume  = {13},
  number  = {2},
  pages   = {163--171},
  doi     = {10.1038/s41557-020-00593-y}
}

@Article{marciniak_crystal_2002,
  author   = {Marciniak, B. and Pavlyuk, V.},
  journal  = {Mol. Cryst. Liq. Cryst.},
  title    = {Crystal {Structure} of a {Metastable} {Anthracene} {Modification}, {Grown} from the {Vapor} {Phase}},
  year     = {2002},
  issn     = {1542-1406},
  month    = jan,
  number   = {1},
  pages    = {237--250},
  volume   = {373},
  doi      = {10.1080/10587250210538},
  fjournal = {Molecular Crystals and Liquid Crystals},
  url      = {https://doi.org/10.1080/10587250210538},
  urldate  = {2024-08-01},
}

@article{Amos:1961,
	author = {Amos, A. T. and Hall, G. G.},
	journal = {Proc. Roy. Soc. London Ser. A},
	pages = {483},
	title = {Single determinant wave functions},
	volume = {263},
	year = {1961}}

@article{Broer:1981,
	author = {Broer, R. and Nieuwpoort, W. C.},
	journal = {Chem. Phys.},
	pages = {291-303},
	title = {Broken orbital symmetry and the description of hole states in tetrahedral [CrO$_4$]$^{2-}$ anion. I. Introductory considerations and calculations on oxygen 1s hole states},
	volume = {54},
	year = {1981}}

@article{Broer:1988,
	author = {Broer, R. and Nieuwpoort, W. C.},
	journal = {Theor. Chim. Acta},
	pages = {405-418},
	title = {Broken orbital symmetry and the description of valence hole states in the tetrahedral [{C}r{O}$_4$]$^{2-}$ anion},
	volume = {73},
    doi = {10.1007/BF00527744},
	year = {1988}}

@incollection{deGraaf:2023,
	author = {de Graaf, C. and Broer, R. and Straatsma, T. P.},
	booktitle = {Reference Module in Chemistry, Molecular Sciences and Chemical Engineering},
	editor = {Shaik, S. and Hiberty, P.},
	publisher = {Elsevier},
	title = {Non-Orthogonal Configuration Interaction for Fragments},
	year = {2023}}

@article{King:1967,
	author = {King, H. F. and Stanton, R. E. and Kim, H. and Wyatt, R. E. and Parr, R. G.},
	journal = {J. Chem. Phys.},
	pages = {1936-1941},
	title = {Corresponding orbitals and the nonorthogonality problem in molecular quantum mechanics},
	volume = {47},
	year = {1967}}

@article{LiManni:2023,
	author = {Li Manni, G. and Galv{\'a}n, I. F. and Alavi, A. and Aleotti, F. and Aquilante, F. and Autschbach, J. and Avagliano, D. and Baiardi, A. and Bao, J. J. and Battaglia, S. and Birnoschi, L. and Blanco-Gonz{\'a}lez, A. and Bokarev, S. I. and Broer, R. and Cacciari, R. and Calio, P. B. and Carlson, R. K. and Couto, R. C. and Cerd{\'a}n, L. and Chibotaru, L. F. and Chilton, N. F. and Church, J. R. and Conti, I. and Coriani, S. and Cu{\'e}llar-Zuquin, J. and Daoud, R. E. and Dattani, N. and Decleva, P. and de Graaf, C. and Delcey, M. and De Vico, L. and Dobrautz, W. and Dong, S. S. and Feng, R. and Ferr{\'e}, N. and Filatov, M. and Gagliardi, L. and Garavelli, M. and Gonz{\'a}lez, L. and Guan, Y. and Guo, M. and Hennefarth, M. R. and Hermes, M. R. and Hoyer, C. E. and Huix-Rotllant, M. and Jaiswal, V. K. and Kaiser, A. and Kaliakin, D. S. and Khamesian, M. and King, D. S. and Kochetov, V. and Kro\'snicki, M. and Kumaar, A. A. and Larsson, E. D. and Lehtola, S. and Lepetit, M. B. and Lischka, H. and L{\'o}pez R{\'\i}os, P. and Lundberg, M. and Ma, D. and Mai, S. and Marquetand, P. and Merritt, I. C. D. and Montorsi, F. and M{\"o}rchen, M. and Nenov, A. and Nguyen, V. H. A. and Nishimoto, Y. and Oakley, M. S. and Olivucci, M. and Oppel, M. and Padula, D. and Pandharkar, R. and Phung, Q. M. and Plasser, F. and Raggi, G. and Rebolini, E. and Reiher, M. and Rivalta, I. and Roca-Sanju{\'a}n, D. and Romig, T. and Safari, A. A. and S{\'a}nchez-Mansilla, A. and Sand, A. M. and Schapiro, I. and Scott, T. R. and Segarra-Mart{\'\i}, J. and Segatta, F. and Sergentu, D.-C. and Sharma, P. and Shepard, R. and Shu, Y. and Staab, J. K. and Straatsma, T. P. and S{\o}rensen, L. K. and Tenorio, B. N. C. and Truhlar, D. G. and Ungur, L. and Vacher, M. and Veryazov, V. and Vo{\ss}, T. A. and Weser, O. and Wu, D. and Yang, X. and Yarkony, D. and Zhou, C. and Zobel, J. P. and Lindh, R.},
	doi = {10.1021/acs.jctc.3c00182},
	journal = {J. Chem. Theory Comput.},
	pages = {6933-6991},
	title = {The OpenMolcas Web: A Community-Driven Approach to Advancing Computational Chemistry},
	volume = {19},
	year = {2023}}

@article{Lopez:2023a,
	author = {L{\'o}pez, X. and Straatsma, T. P. and S{\'a}nchez-Mansilla, A. and de Graaf, C.},
	doi = {10.1021/acs.jpcc.3c02083},
	journal = {J. Phys. Chem. C},
	pages = {16249-16258},
	title = {Non-orthogonal Configuration Interaction Study on the Effect of Thermal Distortions on the Singlet Fission Process in Photoexcited Pure and {B},{N}-Doped Pentacene Crystals},
	volume = {127},
	year = {2023}}

@article{Lowdin:1950,
	author = {L{\"o}wdin, P.-O.},
	journal = {J. Chem. Phys.},
	pages = {365-375},
	title = {On the non-orthogonality problem connected with the use of atomic wave functions in the theory of molecules and crystals},
	volume = {18},
	year = {1950}}

@phdthesis{Montfort:1980,
	author = {van Montfort, J. T.},
	school = {Universtity of Groningen},
	title = {Photo-electron spectroscopy. General theoretical aspects and the calculation of peak positions and intensities in some simple systems},
	year = {1980}}

@article{Sousa:2023,
	author = {Sousa, C. and S{\'a}nchez-Mansilla, A. and Broer, R. and Straatsma, T. P. and de Graaf, C.},
	doi = {10.1021/acs.jpca.3c04975},
	journal = {J. Phys. Chem. A},
	pages = {9944-9958},
	title = {A Nonorthogonal Configuration Interaction Approach to Singlet Fission in Perylenediimide Compounds},
	volume = {127},
	year = {2023}}

@article{Stan:2026,
	author = {Stan, I.-O. and Straatsma, T. P. and Broer, R. and de Graaf, C. and L{\'o}pez, X.},
	doi = {10.1021/acs.jctc.5c01695},
	journal = {J. Chem. Theory Comput.},
	pages = {1296-1311},
	title = {NOCI-F Electronic Couplings in Assemblies of Indolonaphthyridine Molecules: From Dimers to the Full Stack},
	volume = {22},
	year = {2026}}

@article{Straatsma:2022,
	author = {Straatsma, T. P. and Broer, R. and S{\'a}nchez-Mansilla, A. and Sousa, C. and de Graaf, C.},
	doi = {10.1021/acs.jctc.2c00266},
	journal = {J. Chem. Theory Comput.},
	pages = {3549-3565},
	title = {GronOR: Scalable and Accelerated Nonorthogonal Configuration Interaction for Molecular Fragment Wave Functions},
	volume = {18},
	year = {2022}}

@incollection{helgakerSpinSecondQuantization2000,
  title = {Spin in {{Second Quantization}}},
  booktitle = {Molecular {{Electronic-Structure Theory}}},
  author = {Helgaker, Trygve and Jørgensen, Poul and Olsen, Jeppe},
  date = {2000},
  publisher = {John Wiley \& Sons, Ltd},
  doi = {10.1002/9781119019572.ch2},
  url = {https://onlinelibrary.wiley.com/doi/abs/10.1002/9781119019572.ch2},
  urldate = {2024-08-13},
  isbn = {978-1-119-01957-2},
  langid = {english},
}

@incollection{weinbergSPINCETERA2015,
  title = {{{SPIN ET CETERA}}},
  booktitle = {Lectures on {{Quantum Mechanics}}},
  editor = {Weinberg, Steven},
  date = {2015},
  edition = {2},
  publisher = {Cambridge University Press},
  location = {Cambridge},
  doi = {10.1017/CBO9781316276105.006},
  url = {https://www.cambridge.org/core/books/lectures-on-quantum-mechanics/spin-et-cetera/752B4555E34A827682561BC374BA58B0},
  urldate = {2024-08-13},
  isbn = {978-1-107-11166-0},
}

@article{Spano1991,
  author    = {Spano, F. C. and Agranovich, V. M. and Mukamel, S.},
  title     = {Biexciton States and Two-Photon Absorption in Molecular Monolayers},
  journal   = {J. Chem. Phys.},
  year      = {1991},
  volume    = {95},
  pages     = {1400--1409}
}

@article{Casanova2018,
  author    = {Casanova, David},
  title     = {Theoretical Modeling of Singlet Fission},
  journal   = {Chem. Rev.},
  year      = {2018},
  volume    = {118},
  pages     = {7164--7207},
  doi       = {10.1021/acs.chemrev.7b00601}
}

@article{Berkelbach2013a,
  author    = {Berkelbach, Timothy C. and Hybertsen, Mark S. and Reichman, David R.},
  title     = {Microscopic Theory of Singlet Exciton Fission. I. General Formulation},
  journal   = {J. Chem. Phys.},
  year      = {2013},
  volume    = {138},
  pages     = {114102},
  doi       = {10.1063/1.4794425}
}

@article{Berkelbach2013b,
  author    = {Berkelbach, Timothy C. and Hybertsen, Mark S. and Reichman, David R.},
  title     = {Microscopic Theory of Singlet Exciton Fission. II. Application to Pentacene Dimers and the Role of Superexchange},
  journal   = {J. Chem. Phys.},
  year      = {2013},
  volume    = {138},
  pages     = {114103},
  doi       = {10.1063/1.4794427}
}

@article{Zeng2014,
  author    = {Zeng, Tao and Hoffmann, Roald and Ananth, Nandini},
  title     = {The Low-Lying Electronic States of Pentacene and Their Roles in Singlet Fission},
  journal   = {J. Am. Chem. Soc.},
  year      = {2014},
  volume    = {136},
  pages     = {5755--5764},
  doi       = {10.1021/ja500887a}
}

@article{Miyata2019,
  author    = {Miyata, Kiyoshi and Conrad-Burton, Felisa S. and Geyer, Florian L. and Zhu, X.-Y.},
  title     = {Triplet Pair States in Singlet Fission},
  journal   = {Chem. Rev.},
  year      = {2019},
  volume    = {119},
  pages     = {4261--4292},
  doi       = {10.1021/acs.chemrev.8b00572}
}

@article{Dreuw2005,
  author    = {Dreuw, Andreas and Head-Gordon, Martin},
  title     = {Single-Reference Ab Initio Methods for the Calculation of Excited States of Large Molecules},
  journal   = {Chem. Rev.},
  year      = {2005},
  volume    = {105},
  pages     = {4009--4037},
  doi       = {10.1021/cr0505627}
}

@article{GuttierrezMeza2021,
  author  = {Guti{\'e}rrez-Meza, Elizabeth and Malatesta, Ravyn and Li, Hongmo and Bargigia, Ilaria
             and Srimath Kandada, Ajay Ram and Valverde-Ch{\'a}vez, David A. and Kim, Seong-Min
             and Li, Hao and Stingelin, Natalie and Tretiak, Sergei and Bittner, Eric R. and Silva-Acu{\~n}a, Carlos},
  title   = {Frenkel biexcitons in hybrid HJ photophysical aggregates},
  journal = {Sci. Adv.},
  year    = {2021},
  volume  = {7},
  number  = {50},
  pages   = {eabi5197},
  doi     = {10.1126/sciadv.abi5197}
}

@article{BittnerSilva2022,
  author  = {Bittner, Eric R. and Silva, Carlos},
  title   = {Concerning the Stability of Biexcitons in Hybrid HJ Aggregates of $\pi$-Conjugated Polymers},
  journal = {J. Chem. Phys.},
  year    = {2022},
  volume  = {156},
  pages   = {181101},
  doi     = {10.1063/5.0090515}
}

@Article{Agranovich2003,
  author    = {Agranovich, Vladimir M and Litinskaia, M and Lidzey, David G},
  journal   = {Phys. Rev. B},
  title     = {Cavity polaritons in microcavities containing disordered organic semiconductors},
  year      = {2003},
  number    = {8},
  pages     = {085311},
  volume    = {67},
  doi       = {10.1103/PhysRevB.67.085311},
  fjournal  = {Physical Review B},
  publisher = {APS},
}

@article{Tempelaar2017,
	author = {Tempelaar, R. and Jansen, T. L. C. and Knoester, J.},
	doi = {10.1021/acs.jpclett.7b02745},
	journal = {J. Phys. Chem. Lett.},
	pages = {6113-6117},
	title = {Exciton-Exciton Annihilation Is Coherently Suppressed in H-Aggregates, but Not in J-Aggregates},
	volume = {8},
	year = {2017}}

@Article{Zimmerman2010,
  author   = {Zimmerman, Paul M. and Zhang, Zhiyong and Musgrave, Charles B.},
  journal  = {Nat. Chem.},
  title    = {Singlet fission in pentacene through multi-exciton quantum states},
  year     = {2010},
  number   = {8},
  pages    = {648--652},
  volume   = {2},
  doi      = {10.1038/nchem.694},
  fjournal = {Nature Chemistry},
}

@Article{Parker2014,
  author   = {Parker, Shane M. and Seideman, Tamar and Ratner, Mark A. and Shiozaki, Toru},
  journal  = {J. Phys. Chem. C},
  title    = {Model Hamiltonian Analysis of Singlet Fission from First Principles},
  year     = {2014},
  number   = {24},
  pages    = {12700--12705},
  volume   = {118},
  doi      = {10.1021/jp505082a},
  fjournal = {Journal of Physical Chemistry C},
}

@Article{Smith2010,
  author   = {Smith, Millicent B. and Michl, Josef},
  journal  = {Chem. Rev.},
  title    = {Singlet Fission},
  year     = {2010},
  number   = {11},
  pages    = {6891--6936},
  volume   = {110},
  doi      = {10.1021/cr1002613},
  fjournal = {Chemical Reviews},
}

@article{Zeng2022,
  author    = {Le Zeng and Ling Huang and Jinfeng Han and Gang Han},
  title     = {Enhancing Triplet–Triplet Annihilation Upconversion: From Molecular Design to Present Applications},
  journal   = {Acc. Chem. Res.},
  year      = {2022},
  volume    = {55},
  number    = {18},
  pages     = {2604--2615},
  doi       = {10.1021/acs.accounts.2c00307}
}

@Article{Scholes2015,
  author   = {Scholes, Gregory D.},
  journal  = {J. Phys. Chem. A},
  title    = {Correlated Pair States Formed by Singlet Fission and Exciton–Exciton Annihilation},
  year     = {2015},
  number   = {51},
  pages    = {12699--12705},
  volume   = {119},
  doi      = {10.1021/acs.jpca.5b09725},
  fjournal = {Journal of Physical Chemistry A},
}

@Article{hoche_mechanism_2017,
  author     = {Hoche, Joscha and Schmitt, Hans-Christian and Humeniuk, Alexander and Fischer, Ingo and Mitrić, Roland and S. Röhr, Merle I.},
  journal    = {Phys. Chem. Chem. Phys.},
  title      = {The mechanism of excimer formation: an experimental and theoretical study on the pyrene dimer},
  year       = {2017},
  note       = {Publisher: Royal Society of Chemistry},
  number     = {36},
  pages      = {25002--25015},
  volume     = {19},
  doi        = {10.1039/C7CP03990E},
  fjournal   = {Physical Chemistry Chemical Physics},
  language   = {en},
  shorttitle = {The mechanism of excimer formation},
  url        = {https://pubs.rsc.org/en/content/articlelanding/2017/cp/c7cp03990e},
  urldate    = {2023-08-03},
}

@Article{Lee2018,
  author   = {Lee, Tia S. and Lin, Yunhui L. and Kim, Hwon and Pensack, Ryan D. and Rand, Barry P. and Scholes, Gregory D.},
  journal  = {J. Phys. Chem. Lett.},
  title    = {Triplet Energy Transfer Governs the Dissociation of the Correlated Triplet Pair in Exothermic Singlet Fission},
  year     = {2018},
  number   = {14},
  pages    = {4087--4095},
  volume   = {9},
  doi      = {10.1021/acs.jpclett.8b01834},
  fjournal = {Journal of Physical Chemistry Letters},
}

@article{Monguzzi2008,
  author  = {Angelo Monguzzi and Roberto Tubino and Francesco Meinardi},
  title   = {Upconversion-induced delayed fluorescence in multicomponent organic systems: Role of Dexter energy transfer},
  journal = {Phys. Rev. B},
  year    = {2008},
  volume  = {77},
  pages   = {155122},
  doi     = {10.1103/PhysRevB.77.155122}
}

@article{Gray2014,
  author  = {Victor Gray and Damir Dzebo and Maria Abrahamsson and Bo Albinsson and Kasper Moth-Poulsen},
  title   = {Triplet--triplet annihilation photon-upconversion: towards solar energy applications},
  journal = {Phys. Chem. Chem. Phys.},
  year    = {2014},
  volume  = {16},
  pages   = {10345--10352},
  doi     = {10.1039/C4CP00744A}
}

@article{Gilligan2024,
  author  = {Alexander T. Gilligan and Raythe Owens and Ethan G. Miller and Nicholas F. Pompetti and Niels H. Damrauer},
  title   = {Enhancing NIR-to-visible upconversion in a rigidly coupled tetracene dimer: approaching statistical limits for triplet--triplet annihilation using intramolecular multiexciton states},
  journal = {Chem. Sci.},
  year    = {2024},
  volume  = {15},
  pages   = {1283--1296},
  doi     = {10.1039/D3SC04795D}
}

@article{Ambrosio2014,
  title = {Singlet fission in linear chains of molecules},
  volume = {141},
  ISSN = {1089-7690},
  url = {http://dx.doi.org/10.1063/1.4902135},
  DOI = {10.1063/1.4902135},
  number = {20},
  journal = {J. Chem. Phys.},
  publisher = {AIP Publishing},
  author = {Ambrosio,  Francesco and Troisi,  Alessandro},
  year = {2014},
  month = nov
}

@Article{maly_wavelike_2020,
  author   = {Malý, Pavel and Lüttig, Julian and Turkin, Arthur and Dostál, Jakub and Lambert, Christoph and Brixner, Tobias},
  journal  = {Chem. Sci.},
  title    = {From wavelike to sub-diffusive motion: exciton dynamics and interaction in squaraine copolymers of varying length},
  year     = {2020},
  number   = {2},
  pages    = {456--466},
  volume   = {11},
  doi      = {10.1039/C9SC04367E},
  fjournal = {Chemical Science},
}

@article{Dostal2018,
  author  = {Dost{\'a}l, Jakub and Fennel, Franziska and Koch, Federico and Herbst, Stefanie and W{\"u}rthner, Frank and Brixner, Tobias},
  title   = {Direct observation of exciton--exciton interactions},
  journal = {Nat. Commun.},
  year    = {2018},
  volume  = {9},
  pages   = {2466},
  doi     = {10.1038/s41467-018-04884-4}
}

@article{Schlesinger2020,
  author    = {Schlesinger, I. and Powers-Riggs, N. E. and Logsdon, J. L. and Qi, Y. and Miller, S. A. and Tempelaar, R. and Young, R. M. and Wasielewski, M. R.},
  title     = {Charge-transfer biexciton annihilation in a donor–acceptor co-crystal yields high-energy long-lived charge carriers},
  journal   = {Chem. Sci.},
  year      = {2020},
  volume    = {11},
  pages     = {9532--9541},
  doi       = {10.1039/d0sc03301d}
}

@Article{Mavroyannis1977,
  author   = {Mavroyannis, Constantine},
  journal  = {J. Low Temp. Phys.},
  title    = {On the theory of charge-transfer biexciton spectra of molecular crystals},
  year     = {1977},
  number   = {5--6},
  pages    = {669--683},
  volume   = {26},
  doi      = {10.1007/BF00654916},
  fjournal = {Journal of Low Temperature Physics},
}

@article{KuwataGonokami1994,
  author  = {Kuwata-Gonokami, M. and Peyghambarian, N. and Meissner, K. and Fluegel, B. and Sato, Y. and Ema, K. and Shimano, R. and Mazumdar, S. and Guo, F. and Tokihiro, T. and Ezaki, H. and Hanamura, E.},
  title   = {Exciton strings in an organic charge-transfer crystal},
  journal = {Nature},
  year    = {1994},
  volume  = {367},
  pages   = {47--48},
  doi     = {10.1038/367047a0}
}

@Article{Myong2021,
  author   = {Myong, Min Su and Song, Young-Kwang and Lee, Hyun Woo and Lee, Min Jae and Park, Hyeon-Jeong and Kim, Chang Su and Cho, Manki},
  journal  = {J. Mater. Chem. C},
  title    = {Ultrafast Photo-driven Charge Transfer Exciton Dynamics in Single Donor–Acceptor Co-crystals},
  year     = {2021},
  number   = {45},
  pages    = {16028--16037},
  volume   = {9},
  doi      = {10.1039/D1TC04313G},
  fjournal = {Journal of Materials Chemistry C},
}

@article{Dixit1991,
  author  = {Dixit, S. N. and Guo, Dandan and Mazumdar, Sumit},
  title   = {Essential-states mechanism of optical nonlinearity in $\pi$-conjugated polymers},
  journal = {Phys. Rev. B},
  year    = {1991},
  volume  = {43},
  pages   = {6781--6784},
  doi     = {10.1103/PhysRevB.43.6781}
}

@article{Chandross1997PRB,
  author  = {Chandross, Michael and Mazumdar, Sumit},
  title   = {Coulomb interactions and linear, nonlinear, and triplet absorption in poly(para-phenylenevinylene)},
  journal = {Phys. Rev. B},
  year    = {1997},
  volume  = {55},
  pages   = {1497--1504},
  doi     = {10.1103/PhysRevB.55.1497}
}

@article{Shukla2003,
  author  = {Shukla, Alok and Ghosh, Haranath and Mazumdar, Sumit},
  title   = {Theory of excited-state absorption in phenylene-based $\pi$-conjugated polymers},
  journal = {Phys. Rev. B},
  year    = {2003},
  volume  = {67},
  pages   = {245203},
  doi     = {10.1103/PhysRevB.67.245203}
}

@article{Psiachos2009,
  author  = {Psiachos, Dimitris and Mazumdar, Sumit},
  title   = {Correlated-electron description of the photophysics of thin films of $\pi$-conjugated polymers},
  journal = {Phys. Rev. B},
  year    = {2009},
  volume  = {79},
  pages   = {155106},
  doi     = {10.1103/PhysRevB.79.155106}
}

@article{Sanders2019,
  author  = {Sanders, Samuel N. and Pun, Alexander B. and Parenti, Kaitlyn R. and Kumarasamy, Elango and Yablon, Lauren M. and Sfeir, Matthew Y. and Campos, Luis M.},
  title   = {Understanding the Bound Triplet-Pair State in Singlet Fission},
  journal = {Chem},
  year    = {2019},
  volume  = {5},
  number  = {8},
  pages   = {1988--2005},
  doi     = {10.1016/j.chempr.2019.05.012}
}

@article{hudson_what_2022,
        title = {What Next for Singlet Fission in Photovoltaics? The Fate of Triplet and Triplet-Pair Excitons},
  volume = {126},
  ISSN = {1932-7455},
  url = {http://dx.doi.org/10.1021/acs.jpcc.2c00273},
  DOI = {10.1021/acs.jpcc.2c00273},
  number = {12},
  journal = {J. Phys. Chem. C},
  publisher = {American Chemical Society (ACS)},
  author = {Hudson,  Rohan J. and Stuart,  Alexandra N. and Huang,  David M. and Kee,  Tak W.},
  year = {2022},
  month = mar,
  pages = {5369–5377}
}

@article{folie_long-lived_2018,
	title = {Long-{Lived} {Correlated} {Triplet} {Pairs} in a $/pi$-{Stacked} {Crystalline} {Pentacene} {Derivative}},
	volume = {140},
	issn = {0002-7863, 1520-5126},
	url = {https://pubs.acs.org/doi/10.1021/jacs.7b12662},
	doi = {10.1021/jacs.7b12662},
	language = {en},
	number = {6},
	urldate = {2023-11-06},
	journal = {J. Am. Chem. Soc.},
	author = {Folie, Brendan D. and Haber, Jonah B. and Refaely-Abramson, Sivan and Neaton, Jeffrey B. and Ginsberg, Naomi S.},
	month = feb,
	year = {2018},
	pages = {2326--2335},
}

@article{Chan2011,
  author  = {Chan, Wai-Lun and Ligges, Manuel and Zhu, Xiaoyang},
  title   = {Observing the Multiexciton State in Singlet Fission and Ensuing Electron--Hole Pair Generation in Solution},
  journal = {Science},
  year    = {2011},
  volume  = {334},
  number  = {6062},
  pages   = {1541--1545},
  doi     = {10.1126/science.1213986}
}

@article{Pensack2016,
  author  = {Pensack, Ryan D. and Grieco, Christopher and Asbury, John B. and Scholes, Gregory D.},
  title   = {Observation of Two Triplet-Pair Intermediates in Singlet Exciton Fission},
  journal = {J. Phys. Chem. Lett.},
  year    = {2016},
  volume  = {7},
  number  = {12},
  pages   = {2370--2375},
  doi     = {10.1021/acs.jpclett.6b00947}
}

@article{taffet_overlap-driven_2020,
        title = {Overlap-{Driven} {Splitting} of {Triplet} {Pairs} in {Singlet} {Fission}},
        volume = {142},
        issn = {0002-7863},
        url = {https://doi.org/10.1021/jacs.0c09276},
        doi = {10.1021/jacs.0c09276},
        number = {47},
        urldate = {2023-08-02},
        journal = {J. Am. Chem. Soc.},
        author = {Taffet, Elliot J. and Beljonne, David and Scholes, Gregory D.},
        month = nov,
        year = {2020},
        pages = {20040--20047},
}

@article{abraham_revealing_2021,
        title = {Revealing the {Contest} between {Triplet}–{Triplet} {Exchange} and {Triplet}–{Triplet} {Energy} {Transfer} {Coupling} in {Correlated} {Triplet} {Pair} {States} in {Singlet} {Fission}},
        volume = {12},
        url = {https://doi.org/10.1021/acs.jpclett.1c03217},
        doi = {10.1021/acs.jpclett.1c03217},
        number = {43},
        urldate = {2023-08-01},
        journal = {J. Phys. Chem. Lett.},
        author = {Abraham, Vibin and Mayhall, Nicholas J.},
        month = nov,
        year = {2021},
        pages = {10505--10514},
}

@Article{Tayebjee2016,
  author    = {Murad J. Y. Tayebjee and Samuel N. Sanders and Elango Kumarasamy and Luis M. Campos and Matthew Y. Sfeir and Dane R. McCamey},
  journal   = {Nat. Phys.},
  title     = {Quintet multiexciton dynamics in singlet fission},
  year      = {2016},
  month     = oct,
  number    = {2},
  pages     = {182--188},
  volume    = {13},
  doi       = {10.1038/nphys3909},
  fjournal  = {Nature Physics},
  publisher = {Springer Science and Business Media {LLC}},
  url       = {https://doi.org/10.1038/nphys3909},
}

@article{Kundu2021,
  doi = {10.1021/acs.jpclett.0c03301},
  url = {https://doi.org/10.1021/acs.jpclett.0c03301},
  year = {2021},
  month = feb,
  publisher = {American Chemical Society ({ACS})},
  volume = {12},
  number = {5},
  pages = {1468--1474},
  author = {Arup Kundu and Jyotishman Dasgupta},
  title = {Photogeneration of Long-Lived Triplet States through Singlet Fission in Lycopene H-Aggregates},
  journal = {J. Phys. Chem. Lett.}
}

@article{Yong2017,
  author  = {Yong, Chaw-Keong and Musser, Andrew J. and Bayliss, Sam L. and Lukman, Steven and Tamura, Hiroyuki and Bubnova, Olga and Hallani, Rawad K. and Meneau, Aur{\'e}lie and Resel, Roland and Maruyama, Munetaka and Hotta, Shu and Herz, Laura M. and Beljonne, David and Anthony, John E. and Clark, Jenny and Sirringhaus, Henning},
  title   = {The entangled triplet pair state in acene and heteroacene materials},
  journal = {Nat. Commun.},
  year    = {2017},
  volume  = {8},
  pages   = {15953},
  doi     = {10.1038/ncomms15953}
}

@Article{basel_unified_2017,
  author   = {Basel, Bettina S. and Zirzlmeier, Johannes and Hetzer, Constantin and Phelan, Brian T. and Krzyaniak, Matthew D. and Reddy, S. Rajagopala and Coto, Pedro B. and Horwitz, Noah E. and Young, Ryan M. and White, Fraser J. and Hampel, Frank and Clark, Timothy and Thoss, Michael and Tykwinski, Rik R. and Wasielewski, Michael R. and Guldi, Dirk M.},
  journal  = {Nat. Commun.},
  title    = {Unified model for singlet fission within a non-conjugated covalent pentacene dimer},
  year     = {2017},
  issn     = {2041-1723},
  month    = may,
  number   = {1},
  pages    = {15171},
  volume   = {8},
  doi      = {10.1038/ncomms15171},
  fjournal = {Nature Communications},
  language = {en},
  url      = {https://www.nature.com/articles/ncomms15171},
  urldate  = {2023-11-06},
}

@Article{lubert-perquel_identifying_2018,
  author   = {Lubert-Perquel, Daphné and Salvadori, Enrico and Dyson, Matthew and Stavrinou, Paul N. and Montis, Riccardo and Nagashima, Hiroki and Kobori, Yasuhiro and Heutz, Sandrine and Kay, Christopher W. M.},
  journal  = {Nat. Commun.},
  title    = {Identifying triplet pathways in dilute pentacene films},
  year     = {2018},
  issn     = {2041-1723},
  month    = oct,
  number   = {1},
  pages    = {4222},
  volume   = {9},
  doi      = {10.1038/s41467-018-06330-x},
  fjournal = {Nature Communications},
  language = {en},
  url      = {https://www.nature.com/articles/s41467-018-06330-x},
  urldate  = {2023-11-06},
}

@article{weiss_strongly_2017,
	title = {Strongly exchange-coupled triplet pairs in an organic semiconductor},
	volume = {13},
	copyright = {2016 Springer Nature Limited},
	issn = {1745-2481},
	url = {https://www.nature.com/articles/nphys3908},
	doi = {10.1038/nphys3908},
	language = {en},
	number = {2},
	urldate = {2023-08-01},
	journal = {Nat. Phys.},
	author = {Weiss, Leah R. and Bayliss, Sam L. and Kraffert, Felix and Thorley, Karl J. and Anthony, John E. and Bittl, Robert and Friend, Richard H. and Rao, Akshay and Greenham, Neil C. and Behrends, Jan},
	month = feb,
	year = {2017},
	pages = {176--181},
}

@article{zirzlmeier_solution-based_2016,
    title = {Solution-based intramolecular singlet fission in cross-conjugated pentacene dimers},
    volume = {8},
    issn = {2040-3364, 2040-3372},
    url = {http://xlink.rsc.org/?DOI=C6NR02493A},
    doi = {10.1039/C6NR02493A},
    language = {en},
    number = {19},
    urldate = {2023-11-05},
    journal = {Nanoscale},
    author = {Zirzlmeier, Johannes and Casillas, Rubén and Reddy, S. Rajagopala and Coto, Pedro B. and Lehnherr, Dan and Chernick, Erin T. and Papadopoulos, Ilias and Thoss, Michael and Tykwinski, Rik R. and Guldi, Dirk M.},
    year = {2016},
    pages = {10113--10123},
}

@Article{tamura_first-principles_2015,
  author     = {Tamura, Hiroyuki and Huix-Rotllant, Miquel and Burghardt, Irene and Olivier, Yoann and Beljonne, David},
  journal    = {Phys. Rev. Lett.},
  title      = {First-{Principles} {Quantum} {Dynamics} of {Singlet} {Fission}: {Coherent} versus {Thermally} {Activated} {Mechanisms} {Governed} by {Molecular} $/pi$ {Stacking}},
  year       = {2015},
  issn       = {0031-9007, 1079-7114},
  month      = aug,
  number     = {10},
  pages      = {107401},
  volume     = {115},
  doi        = {10.1103/PhysRevLett.115.107401},
  fjournal   = {Physical Review Letters},
  language   = {en},
  shorttitle = {First-{Principles} {Quantum} {Dynamics} of {Singlet} {Fission}},
  url        = {https://link.aps.org/doi/10.1103/PhysRevLett.115.107401},
  urldate    = {2023-11-05},
}

@article{accomasso_diabatization_2019,
    title = {Diabatization by {Localization} in the {Framework} of {Configuration} {Interaction} {Based} on {Floating} {Occupation} {Molecular} {Orbitals} ({FOMO}-{CI})},
    volume = {3},
    issn = {2367-0932, 2367-0932},
    url = {https://chemistry-europe.onlinelibrary.wiley.com/doi/10.1002/cptc.201900056},
    doi = {10.1002/cptc.201900056},
    language = {en},
    number = {9},
    urldate = {2023-11-05},
    journal = {ChemPhotoChem},
    author = {Accomasso, Davide and Persico, Maurizio and Granucci, Giovanni},
    month = sep,
    year = {2019},
    pages = {933--944},
}

@Article{tamura_triplet_2020,
  author   = {Tamura, Hiroyuki},
  journal  = {J. Phys. Chem. A},
  title    = {Triplet {Exciton} {Transfers} and {Triplet}–{Triplet} {Annihilation} in {Anthracene} {Derivatives} via {Direct} versus {Superexchange} {Pathways} {Governed} by {Molecular} {Packing}},
  year     = {2020},
  issn     = {1089-5639, 1520-5215},
  month    = oct,
  number   = {39},
  pages    = {7943--7949},
  volume   = {124},
  doi      = {10.1021/acs.jpca.0c06835},
  fjournal = {Journal of Physical Chemistry A},
  language = {en},
  url      = {https://pubs.acs.org/doi/10.1021/acs.jpca.0c06835},
  urldate  = {2023-11-05},
}

@Article{yang_first-principle_2015,
  author   = {Yang, Chou-Hsun and Hsu, Chao-Ping},
  journal  = {J. Phys. Chem. Lett.},
  title    = {First-{Principle} {Characterization} for {Singlet} {Fission} {Couplings}},
  year     = {2015},
  issn     = {1948-7185, 1948-7185},
  month    = may,
  number   = {10},
  pages    = {1925--1929},
  volume   = {6},
  doi      = {10.1021/acs.jpclett.5b00437},
  fjournal = {Journal of Physical Chemistry Letters},
  language = {en},
  url      = {https://pubs.acs.org/doi/10.1021/acs.jpclett.5b00437},
  urldate  = {2023-11-05},
}

@Article{manjanath_enhancing_2022,
  author   = {Manjanath, Aaditya and Yang, Chou-Hsun and Kue, Karl and Wang, Chun-I and Claudio, Gil C. and Hsu, Chao-Ping},
  journal  = {J. Chem. Theory Comput.},
  title    = {Enhancing {Singlet} {Fission} {Coupling} with {Nonbonding} {Orbitals}},
  year     = {2022},
  issn     = {1549-9618, 1549-9626},
  month    = feb,
  number   = {2},
  pages    = {1017--1029},
  volume   = {18},
  doi      = {10.1021/acs.jctc.1c00868},
  fjournal = {Journal of Chemical Theory and Computation},
  language = {en},
  url      = {https://pubs.acs.org/doi/10.1021/acs.jctc.1c00868},
  urldate  = {2023-11-05},
}

@Article{chan_quantum_2013,
  author     = {Chan, Wai-Lun and Berkelbach, Timothy C. and Provorse, Makenzie R. and Monahan, Nicholas R. and Tritsch, John R. and Hybertsen, Mark S. and Reichman, David R. and Gao, Jiali and Zhu, X.-Y.},
  journal    = {Acc. Chem. Res.},
  title      = {The {Quantum} {Coherent} {Mechanism} for {Singlet} {Fission}: {Experiment} and {Theory}},
  year       = {2013},
  issn       = {0001-4842, 1520-4898},
  month      = jun,
  number     = {6},
  pages      = {1321--1329},
  volume     = {46},
  doi        = {10.1021/ar300286s},
  fjournal   = {Accounts of Chemical Research},
  language   = {en},
  shorttitle = {The {Quantum} {Coherent} {Mechanism} for {Singlet} {Fission}},
  url        = {https://pubs.acs.org/doi/10.1021/ar300286s},
  urldate    = {2023-11-05},
}

@Article{wibowo_rigorous_2017,
  author   = {Wibowo, Meilani and Broer, Ria and Havenith, Remco W.A.},
  journal  = {Comput. Theor. Chem.},
  title    = {A rigorous nonorthogonal configuration interaction approach for the calculation of electronic couplings between diabatic states applied to singlet fission},
  year     = {2017},
  issn     = {2210271X},
  month    = sep,
  pages    = {190--194},
  volume   = {1116},
  doi      = {10.1016/j.comptc.2017.03.013},
  fjournal = {Computational and Theoretical Chemistry},
  language = {en},
  url      = {https://linkinghub.elsevier.com/retrieve/pii/S2210271X1730110X},
  urldate  = {2023-11-05},
}

@Article{berkelbach_microscopic_2014,
  author   = {Berkelbach, Timothy C. and Hybertsen, Mark S. and Reichman, David R.},
  journal  = {J. Chem. Phys.},
  title    = {Microscopic theory of singlet exciton fission. {III}. {Crystalline} pentacene},
  year     = {2014},
  issn     = {0021-9606},
  month    = aug,
  number   = {7},
  pages    = {074705},
  volume   = {141},
  doi      = {10.1063/1.4892793},
  fjournal = {J. Chem. Phys.},
  url      = {https://doi.org/10.1063/1.4892793},
  urldate  = {2023-07-26},
}

@Article{nakano_quantum_2019,
  author     = {Nakano, Masayoshi and Nagami, Takanori and Tonami, Takayoshi and Okada, Kenji and Ito, Soichi and Kishi, Ryohei and Kitagawa, Yasutaka and Kubo, Takashi},
  journal    = {J. Comput. Chem.},
  title      = {Quantum {Master} {Equation} {Approach} to {Singlet} {Fission} {Dynamics} in {Pentacene} {Linear} {Aggregate} {Models}: {Size} {Dependences} of {Excitonic} {Coupling} {Effects}},
  year       = {2019},
  issn       = {1096-987X},
  note       = {\_eprint: https://onlinelibrary.wiley.com/doi/pdf/10.1002/jcc.25539},
  number     = {1},
  pages      = {89--104},
  volume     = {40},
  doi        = {10.1002/jcc.25539},
  fjournal   = {J. Comput. Chem.},
  language   = {en},
  shorttitle = {Quantum {Master} {Equation} {Approach} to {Singlet} {Fission} {Dynamics} in {Pentacene} {Linear} {Aggregate} {Models}},
  url        = {https://onlinelibrary.wiley.com/doi/abs/10.1002/jcc.25539},
  urldate    = {2023-07-27},
}

@Article{ryerson_structure_2019,
  author     = {Ryerson, Joseph L. and Zaykov, Alexandr and Aguilar Suarez, Luis E. and Havenith, Remco W. A. and Stepp, Brian R. and Dron, Paul I. and Kaleta, Jiří and Akdag, Akin and Teat, Simon J. and Magnera, Thomas F. and Miller, John R. and Havlas, Zdeněk and Broer, Ria and Faraji, Shirin and Michl, Josef and Johnson, Justin C.},
  journal    = {J. Chem. Phys.},
  title      = {Structure and photophysics of indigoids for singlet fission: {Cibalackrot}},
  year       = {2019},
  issn       = {0021-9606, 1089-7690},
  month      = nov,
  number     = {18},
  pages      = {184903},
  volume     = {151},
  doi        = {10.1063/1.5121863},
  fjournal = {J. Chem. Phys.},
  language   = {en},
  shorttitle = {Structure and photophysics of indigoids for singlet fission},
  url        = {https://pubs.aip.org/jcp/article/151/18/184903/198029/Structure-and-photophysics-of-indigoids-for},
  urldate    = {2023-11-05},
}

@article{korovina_singlet_2016,
        title = {Singlet {Fission} in a {Covalently} {Linked} {Cofacial} {Alkynyltetracene} {Dimer}},
        volume = {138},
        issn = {0002-7863},
        url = {https://doi.org/10.1021/jacs.5b10550},
        doi = {10.1021/jacs.5b10550},
        number = {2},
        urldate = {2023-08-01},
        journal = {J. Am. Chem. Soc.},
        author = {Korovina, Nadezhda V. and Das, Saptaparna and Nett, Zachary and Feng, Xintian and Joy, Jimmy and Haiges, Ralf and Krylov, Anna I. and Bradforth, Stephen E. and Thompson, Mark E.},
        month = jan,
        year = {2016},
        pages = {617--627},
}

@Article{Korovina2020,
  author   = {Korovina, Nadezhda V. and Chang, Christopher H. and Johnson, Justin C.},
  journal  = {Nat. Chem.},
  title    = {Spatial separation of triplet excitons drives endothermic singlet fission},
  year     = {2020},
  number   = {4},
  pages    = {391--398},
  volume   = {12},
  doi      = {10.1038/s41557-020-0424-3},
  fjournal = {Nature Chemistry},
  url      = {https://doi.org/10.1038/s41557-020-0424-3},
}

@article{Baronas,
author = {Baronas, P. and Kreiza, G. and Naimovičius, L. and Radiunas, E. and Kazlauskas, K. and Orentas, E. and Juršėnas, S.},
title = {Sweet Spot of Intermolecular Coupling in Crystalline Rubrene: Intermolecular Separation to Minimize Singlet Fission and Retain Triplet–Triplet Annihilation},
journal = {J. Phys. Chem. C},
volume = {126},
number = {36},
pages = {15327-15335},
year = {2022},
doi = {10.1021/acs.jpcc.2c04572},
}

@article{Qian,
author = {Qian, Yuqin and Huang-Fu, Zhi-Chao and Zhang, Tong and Li, Xia and Harutyunyan, Avetik R. and Chen, Gugang and Chen, Hanning and Rao, Yi},
title = {Temperature-Dependent Recombination of Triplet Biexcitons in Singlet Fission of Hexacene},
journal = {J. Phys. Chem. C},
volume = {126},
number = {19},
pages = {8377-8383},
year = {2022},
doi = {10.1021/acs.jpcc.1c10691},
}

@article{Tan2021,
  doi = {10.1021/acs.jpca.0c07832},
  url = {https://doi.org/10.1021/acs.jpca.0c07832},
  year = {2021},
  month = mar,
  publisher = {American Chemical Society ({ACS})},
  volume = {125},
  number = {9},
  pages = {1972--1980},
  author = {Yunshu Tan and Guohua Tao},
  title = {Exploring the State Space Structure of Multiple Spins via Modular Tensor Diagram Approach: Going beyond the Exciton Pair State},
  journal = {J. Phys. Chem. A}
}

@article{Tao2020,
  doi = {10.1021/acs.jpca.0c00263},
  url = {https://doi.org/10.1021/acs.jpca.0c00263},
  year = {2020},
  month = jun,
  publisher = {American Chemical Society ({ACS})},
  volume = {124},
  number = {26},
  pages = {5435--5443},
  author = {Guohua Tao and Yunshu Tan},
  title = {Modular Tensor Diagram Approach for the Construction of Spin Eigenfunctions: The Case Study of Exciton Pair States},
  journal = {J. Phys. Chem. A}
}

@Article{singh2024,
  author    = {Singh, Anurag and Röhr, Merle I. S.},
  journal   = {J. Chem. Theory Comput.},
  title     = {Configuration {Interaction} in {Frontier} {Molecular} {Orbital} {Basis} for {Screening} the {Spin}-{Correlated}, {Spatially} {Separated} {Triplet} {Pair} {State}$^{\textrm{1}}$ ({T}···{T}) {Formation}},
  year      = {2024},
  issn      = {1549-9618, 1549-9626},
  month     = oct,
  number    = {19},
  pages     = {8624--8633},
  volume    = {20},
  copyright = {https://doi.org/10.15223/policy-029},
  doi       = {10.1021/acs.jctc.4c00473},
  fjournal  = {Journal of Chemical Theory and Computation},
  language  = {en},
  url       = {https://pubs.acs.org/doi/10.1021/acs.jctc.4c00473},
  urldate   = {2025-09-17},
}

@Article{sousa2025,
  author     = {Sousa, C. and López, X. and Dong, X. and Broer, R. and Straatsma, T. P. and De Graaf, C.},
  journal    = {J. Phys. Chem. C},
  title      = {Nonorthogonal {Configuration} {Interaction} for {Singlet} {Fission}: {Beyond} the {Dimer}},
  year       = {2025},
  issn       = {1932-7447, 1932-7455},
  month      = feb,
  number     = {8},
  pages      = {4290--4302},
  volume     = {129},
  copyright  = {https://creativecommons.org/licenses/by/4.0/},
  doi        = {10.1021/acs.jpcc.4c08656},
  fjournal   = {Journal of Physical Chemistry C},
  language   = {en},
  shorttitle = {Nonorthogonal {Configuration} {Interaction} for {Singlet} {Fission}},
  url        = {https://pubs.acs.org/doi/10.1021/acs.jpcc.4c08656},
  urldate    = {2025-09-17},
}

@article{Wang2023,
  author  = {Wang, Yu-Chen and Feng, Shishi and Kong, Yi and Huang, Xunkun and Liang, WanZhen and Zhao, Yi},
  title   = {Electronic Couplings for Singlet Fission Processes Based on the Fragment Particle-Hole Densities},
  journal = {J. Chem. Theory Comput.},
  year    = {2023},
  volume  = {19},
  number  = {13},
  pages   = {3900--3914},
  doi     = {10.1021/acs.jctc.3c00243}
}

@Article{chilkuriComparisonManyParticleRepresentations2021,
  author     = {Chilkuri, Vijay Gopal and Neese, Frank},
  journal    = {J. Chem. Theory Comput.},
  title      = {Comparison of {{Many-Particle Representations}} for {{Selected Configuration Interaction}}: {{II}}. {{Numerical Benchmark Calculations}}},
  year       = {2021},
  month      = apr,
  copyright  = {{\copyright} 2021 The Authors. Published by American Chemical Society},
  doi        = {10.1021/acs.jctc.1c00081},
  fjournal   = {Journal of Chemical Theory and Computation},
  langid     = {english},
  publisher  = {American Chemical Society},
  shorttitle = {Comparison of {{Many-Particle Representations}} for {{Selected Configuration Interaction}}},
  urldate    = {2025-10-06},
}

@Article{falesFastTransformationsConfiguration2020a,
  author    = {Fales, B. Scott and Mart{\'i}nez, Todd J.},
  journal   = {J. Chem. Phys.},
  title     = {Fast Transformations between Configuration State Function and {{Slater}} Determinant Bases for Direct Configuration Interaction},
  year      = {2020},
  issn      = {0021-9606},
  month     = apr,
  number    = {16},
  volume    = {152},
  doi       = {10.1063/5.0005155},
  fjournal = {J. Chem. Phys.},
  langid    = {english},
  publisher = {AIP Publishing},
  urldate   = {2025-10-06},
}

@article{ugandi2023configuration,
  title={A configuration-based heatbath-CI for spin-adapted multireference electronic structure calculations with large active spaces},
  author={Ugandi, Mihkel and Roemelt, Michael},
  journal={J. Comput. Chem.},
  volume={44},
  number={31},
  pages={2374--2390},
  year={2023},
  publisher={Wiley Online Library}
}

@Article{grabenstetterGenerationGenealogicalSpin1976,
  author   = {Grabenstetter, J. E. and Tseng, T. J. and Grein, F.},
  journal  = {Int. J. Quantum Chem.},
  title    = {Generation of Genealogical Spin Eigenfunctions},
  year     = {1976},
  issn     = {1097-461X},
  number   = {1},
  pages    = {143--149},
  volume   = {10},
  doi      = {10.1002/qua.560100112},
  fjournal = {International Journal of Quantum Chemistry},
  langid   = {english},
  urldate  = {2025-10-06},
}

@Article{kotaniTablesIntegralsUseful1953,
  author    = {Kotani, Masao and Ishiguro, Eiichi and Hijikata, Katsunori and Nakamura, Tafeashi and Amemiya, Ayao},
  journal   = {J. Phys. Soc. Japan},
  title     = {Tables of {{Integrals Useful}} for the {{Calculations}} of {{Molecular Energies}}. {{III}}},
  year      = {1953},
  issn      = {0031-9015},
  month     = jul,
  number    = {4},
  pages     = {463--476},
  volume    = {8},
  doi       = {10.1143/JPSJ.8.463},
  fjournal  = {Journal of the Physical Society of Japan},
  publisher = {The Physical Society of Japan},
  urldate   = {2025-10-06},
}

@Article{paunczBranchingDiagramSerbertype1977,
  author   = {Pauncz, Ruben},
  journal  = {Int. J. Quantum Chem.},
  title    = {{Branching diagram and serber-type spin functions. Algorithms for their construction and special properties}},
  year     = {1977},
  issn     = {1097-461X},
  number   = {2},
  pages    = {369--382},
  volume   = {12},
  doi      = {10.1002/qua.560120213},
  fjournal = {International Journal of Quantum Chemistry},
  langid   = {french},
  urldate  = {2025-10-06},
}

@article{yamanouchiAtomicEnergyLevels1938,
  title = {On {{Atomic Energy Levels}} of Pnp {{Configurations}}},
  author = {Yamanouchi, Takahiko},
  year = {1938},
  journal = {Proc. Phys. Math. Soc. Jpn.},
  volume = {20},
  pages = {547--562},
  doi = {10.11429/ppmsj1919.20.0_547}
}

@article{yamanouchiCalculationAtomicEnergy1936,
  title = {On the {{Calculation}} of {{Atomic Energy Lerels}}. {{IV}}},
  author = {Yamanouchi, Takahiko},
  year = {1936},
  journal = {Proc. Phys. Math. Soc. Jpn.},
  volume = {18},
  pages = {623--640},
  doi = {10.11429/ppmsj1919.18.0_623}
}

@article{yamanouchiConstructionUnitaryIrreducible1937,
  title = {On the {{Construction}} of {{Unitary Irreducible Representations}} of the {{Symmetric Group}}},
  author = {Yamanouchi, Takahiko},
  year = {1937},
  journal = {Proc. Phys. Math. Soc. Jpn.},
  volume = {19},
  pages = {436--450},
  doi = {10.11429/ppmsj1919.19.0_436}
}

@article{RN204,
author = {Neese,F.},
title = {Software update: the ORCA program system, version 5.0},
journal = {WIRES Comput. Molec. Sci.},
volume = {12},
number = {1},
pages = {e1606},
DOI = {10.1002/wcms.1606},
year = {2022},
type = {journal Article}
}

@article{RN205,
author = {Neese,F.},
title = {The SHARK Integral Generation and Digestion System},
journal = {J. Comput. Chem.},
pages = {1-16},
DOI = {10.1002/jcc.26942},
year = {2022},
type = {journal Article}
}

@article{RN84,
author = {Neese,F.},
title = {The ORCA program system},
journal = {WIRES Comput. Molec. Sci.},
volume = {2},
number = {1},
pages = {73-78},
DOI = {10.1002/wcms.81},
year = {2012},
type = {journal Article}
}

@article{RN139,
author = {Neese,F.},
title = {Software update: the ORCA program system, version 4.0},
journal = {WIRES Comput. Molec. Sci.},
volume = {8},
number = {1},
pages = {1-6},
DOI = {10.1002/wcms.1327},
year = {2018},
type = {journal Article}
}

@article{RN21,
author = {Neese,F.},
title = {Approximate second-order SCF convergence for spin unrestricted wavefunctions},
journal = {Chem. Phys. Lett.},
volume = {325},
number = {1-3},
pages = {93-98},
DOI = {10.1016/s0009-2614(00)00662-x},
year = {2000},
type = {journal Article}
}

@article{RN155,
author = {Kollmar,C. and Sivalingam,K. and Helmich-Paris,B. and Angeli,C. and Neese,F.},
title = {A perturbation-based super-CI approach for the orbital optimization of a CASSCF wave function},
journal = {J. Comput. Chem.},
volume = {40},
pages = {1463-1470},
DOI = {10.1002/jcc.25801},
year = {2019},
type = {journal Article}
}

@article{RN171,
author = {Kollmar,C. and Sivalingam,K. and Neese,F.},
title = {An efficient implementation of the NEVPT2 and CASPT2 methods avoiding higher-order density matrices},
journal = {J. Chem. Phys.},
volume = {155},
pages = {234104},
DOI = {10.1063/5.0072129},
year = {2021},
type = {journal Article}
}

@article{RN200,
author = {Guo,Y. and Sivalingam,K. and Neese,F.},
title = {Approximations of density matrices in N-electron valence state second-order perturbation theory (NEVPT2). I. Revisiting the NEVPT2 construction},
journal = {J. Chem. Phys.},
volume = {154},
pages = {214111},
year = {2021},
type = {journal Article}
}

@article{RN230,
author = {Ugandi,M. and Roemelt,M.},
title = {A recursive formulation of one-electron coupling coefficients for spin-adapted configuration interaction calculations featuring many unpaired electrons},
journal = {Int. J. Quantum Chem.},
volume = {123},
pages = {e27045},
DOI = {10.1002/qua.27045},
year = {2023},
type = {journal Article}
}

@article{RN96,
author = {Schapiro,I. and Sivalingam,K. and Neese,F.},
title = {Assessment of n-Electron Valence State Perturbation Theory for Vertical Excitation Energies},
journal = {J. Theor. Comput. Chem.},
volume = {9},
pages = {3567-3580},
DOI = {10.1021/ct400136y},
year = {2013},
type = {journal Article}
}

@article{RN178,
author = {Neese,F. and Wennmohs,F. and Becker,U. and Riplinger,C.},
title = {The ORCA quantum chemistry program package},
journal = {J. Chem. Phys.},
volume = {152},
pages = {Art. No. L224108},
DOI = {10.1063/5.0004608},
year = {2020},
type = {journal Article}
}

@article{dunning1989gaussian,
  title={Gaussian basis sets for use in correlated molecular calculations. I. The atoms boron through neon and hydrogen},
  author={Dunning Jr, Thom H},
  journal = {J. Chem. Phys.},
  volume={90},
  number={2},
  pages={1007--1023},
  year={1989},
  publisher={American Institute of Physics}
}

@Article{roos_complete_1980,
  author   = {Roos, Björn O. and Taylor, Peter R. and Sigbahn, Per E. M.},
  journal  = {Chem. Phys.},
  title    = {A complete active space {SCF} method ({CASSCF}) using a density matrix formulated super-{CI} approach},
  year     = {1980},
  issn     = {0301-0104},
  month    = may,
  number   = {2},
  pages    = {157--173},
  volume   = {48},
  doi      = {10.1016/0301-0104(80)80045-0},
  fjournal = {Chemical Physics},
  url      = {https://www.sciencedirect.com/science/article/pii/0301010480800450},
  urldate  = {2025-05-09},
}

@article{Vektaris1994,
  title={A new approach to the molecular biexciton theory},
  author={Vektaris, G},
  journal = {J. Chem. Phys.},
  volume={101},
  number={4},
  pages={3031--3040},
  year={1994},
  publisher={American Institute of Physics}
}

@Article{Gallagher1996,
  author    = {Gallagher, Frank B and Spano, Frank C},
  journal   = {Phys. Rev. B},
  title     = {Theory of biexcitons in one-dimensional polymers},
  year      = {1996},
  number    = {7},
  pages     = {3790},
  volume    = {53},
  fjournal  = {Physical Review B},
  publisher = {APS},
}

@article{Clark2021,
  title={Spin statistics for triplet--triplet annihilation upconversion: Exchange coupling, intermolecular orientation, and reverse intersystem crossing},
  author={Bossanyi, David G and Sasaki, Yoichi and Wang, Shuangqing and Chekulaev, Dimitri and Kimizuka, Nobuo and Yanai, Nobuhiro and Clark, Jenny},
  journal={JACS Au},
  volume={1},
  number={12},
  pages={2188--2201},
  year={2021},
  publisher={ACS Publications}
}

@article{zaykov_singlet_2019,
	title = {Singlet {Fission} {Rate}: {Optimized} {Packing} of a {Molecular} {Pair}. {Ethylene} as a {Model}},
	volume = {141},
	issn = {0002-7863},
	shorttitle = {Singlet {Fission} {Rate}},
	url = {https://doi.org/10.1021/jacs.9b08173},
	doi = {10.1021/jacs.9b08173},
	number = {44},
	urldate = {2024-08-13},
	journal = {J. Am. Chem. Soc.},
	publisher = {American Chemical Society},
	author = {Zaykov, Alexandr and Felkel, Petr and Buchanan, Eric A. and Jovanovic, Milena and Havenith, Remco W. A. and Kathir, R. K. and Broer, Ria and Havlas, Zdeněk and Michl, Josef},
	month = nov,
	year = {2019},
	pages = {17729--17743},
}

\end{document}


\maketitle

\tableofcontents
\newpage

\section{Analytical expressions calculated with \symci{}}

Some selected symbolic expressions calculated with \symci{}:
\begin{equation}
    \mel{\sone\sone\szero}{\hat{\mathcal H}}{\szero\sone\sone} = 2\twoE{l_1}{h_3}{h_1}{l_3} - \twoE{l_1}{h_3}{l_3}{h_1}
\end{equation}

\begin{equation}
    \mel{\sone\sone\szero}{\hat{\mathcal H}}{\sone\szero\sone} = 2\twoE{l_2}{h_3}{h_2}{l_3} - \twoE{l_2}{h_3}{l_3}{h_2}
\end{equation}

\begin{align}
\begin{split}
    \mel{\sone\sone\szero}{\hat{\mathcal H}}{\sone\dpm\dmm} = \oneE{l_2}{l_3} - \twoE{h_2}{l_2}{l_3}{h_2} + \frac{1}{2} \twoE{l_2}{h_1}{h_1}{l_3} + \frac{1}{2} \twoE{l_2}{l_2}{l_1}{l_3}\\ - \twoE{l_2}{l_1}{l_3}{l_1} - \twoE{l_2}{h_2}{l_3}{h_2} - \twoE{l_2}{h_1}{l_3}{h_1} + \sum_{i\in h \setminus \{h_1,h_2\}} \twoE{h_2}{i}{i}{l_3} - 2 \twoE{h_2}{i}{l_3}{i}
    \end{split}
\end{align}

\begin{align}
\begin{split}
    \mel{\sone\sone\szero}{\hat{\mathcal H}}{\sone\dmm\dpm} = \oneE{l_3}{l_2} - \twoE{h_3}{l_3}{h_2}{h_3} + \frac{1}{2} \twoE{l_3}{h_1}{h_1}{h_2} + \frac{1}{2} \twoE{l_3}{l_3}{l_1}{h_2}\\ - \twoE{l_3}{l_1}{h_2}{l_1} - \twoE{l_3}{h_3}{l_2}{h_3} - \twoE{l_3}{h_1}{h_2}{h_1} + 
    \sum_{i\in h \setminus \{h_1,h_3\}} \twoE{l_3}{i}{i}{h_2} - 2 \twoE{l_3}{i}{l_3}{i}
\end{split}
\end{align}

\begin{align}
     \mel{\sone\sone\szero}{\hat{\mathcal H}}{\dpm\dmm\sone} = \twoE{l_1}{h_3}{h_2}{l_3} - \frac{1}{2} \twoE{l_1}{h_3}{l_3}{h_2}
\end{align}

\begin{align}
    \mel{\szero\sone\sone\szero}{\hat{\mathcal H}}{\dpm\dmm\dpm\dmm} = - \twoE{h_1}{l_3}{h_2}{l_4} + \frac{1}{2} \twoE{h_1}{l_3}{l_4}{h_2}
\end{align}

\begin{align}
    \mel{\sone\szero\sone\szero}{\hat{\mathcal H}}{\dpm\dmm\dpm\dmm} = \twoE{l_1}{l_3}{l_2}{l_4} - \frac{1}{2} \twoE{l_1}{l_3}{l_4}{l_2}
\end{align}

\section{Lowest biexcitonic state in the stacked aggregates}
\label{sec:biexcitonic_section}

In response to the query regarding the identity and potential degeneracy of the "lowest" biexcitonic state, we provide a detailed analysis of the excited-state manifold for the ethene 15-mer.

\begin{figure}[ht]
     \centering
     \begin{subfigure}[b]{0.49\textwidth}
             \centering
    \includegraphics[width=\linewidth]{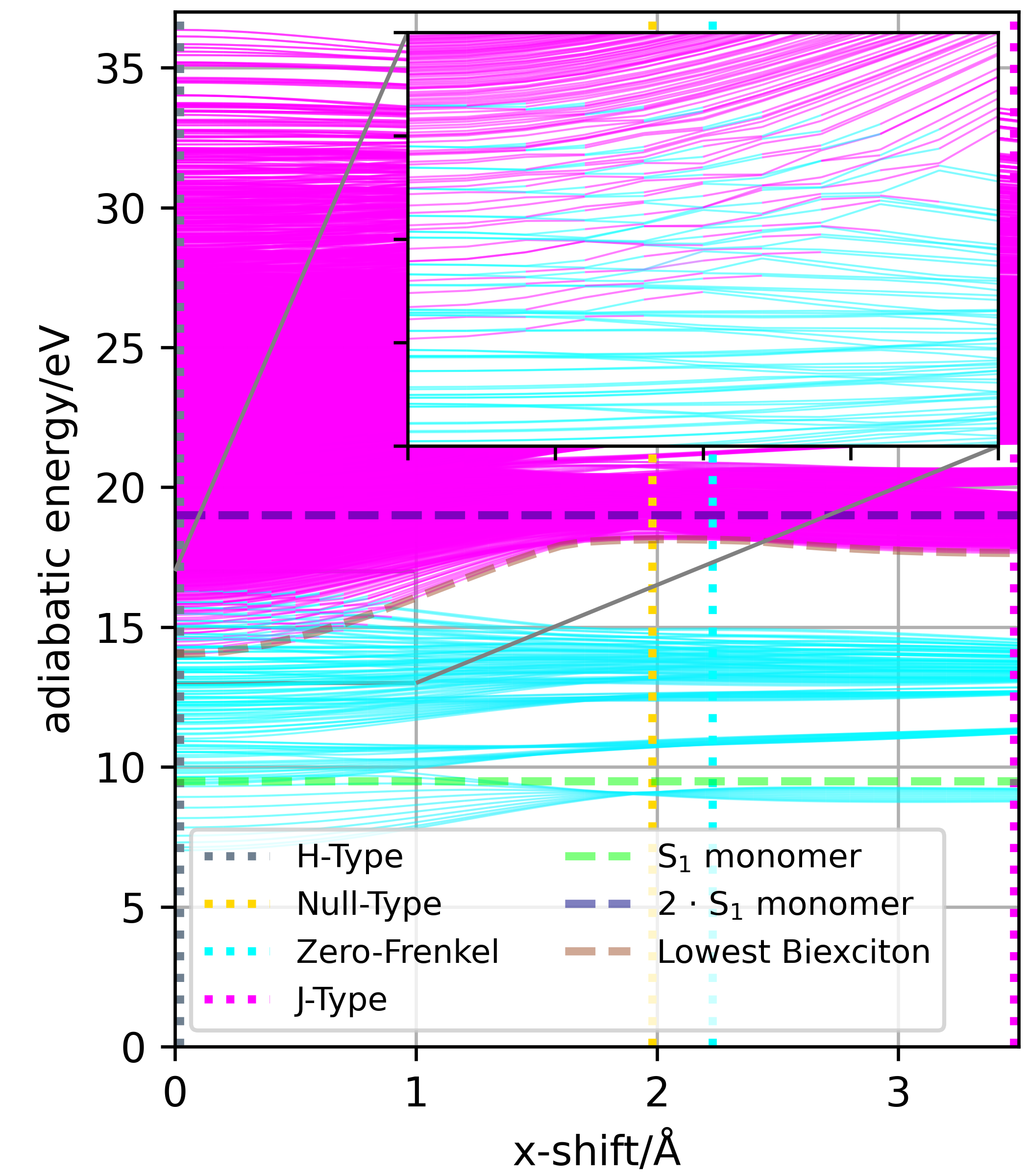}
    \caption{}
    \label{fig:ethene_scan_biexciton_character}
     \end{subfigure}
     \hfill
     \begin{subfigure}[b]{0.49\textwidth}
        \centering
    \includegraphics[width=\linewidth]{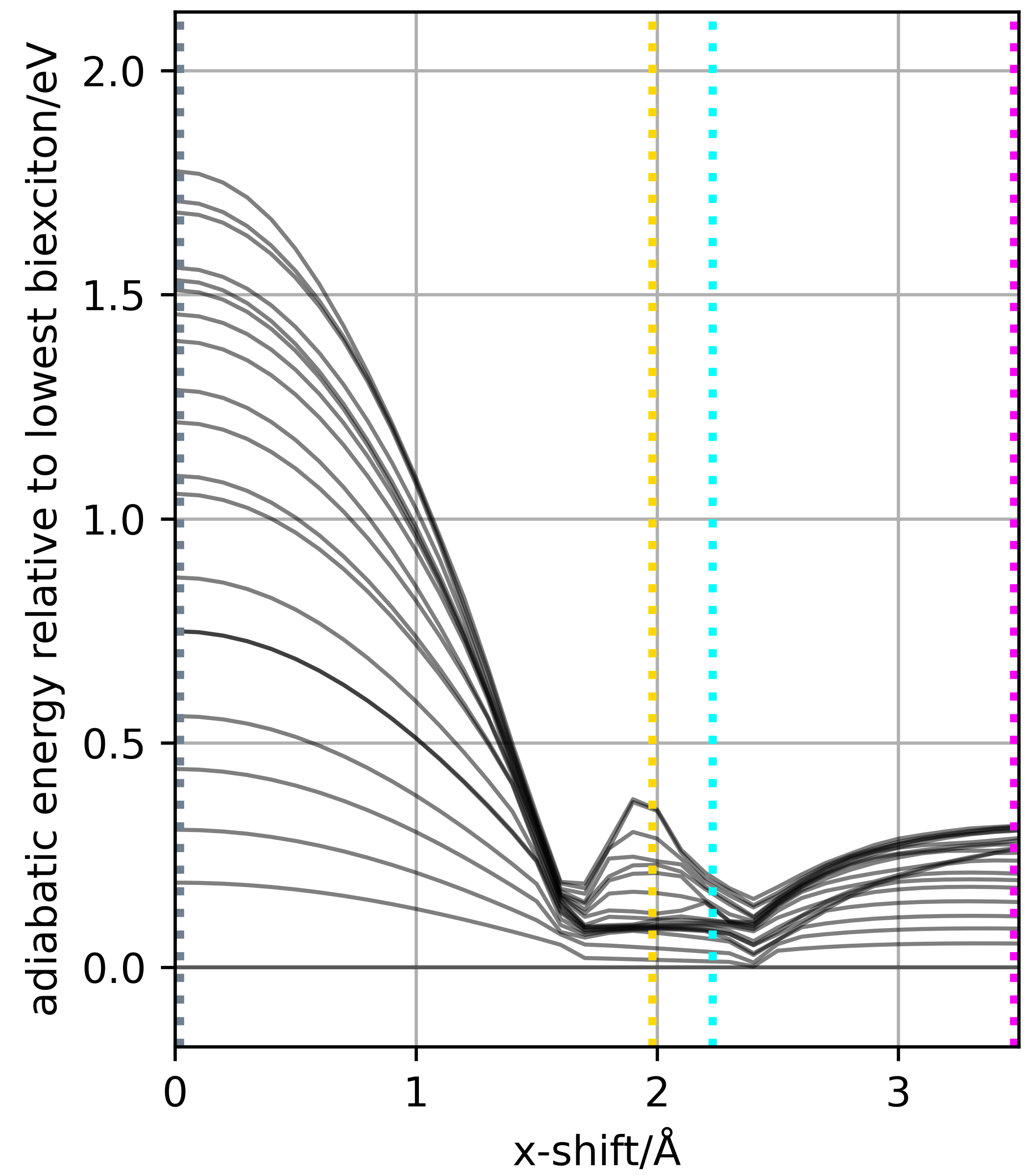}
    \caption{}
    \label{fig:ethene_scan_biexciton_density}
     \end{subfigure}
        \caption{Adiabatic energies obtained from diagonalization of the diabatic \symci{} Hamiltonian along the scan coordinate of ethene 15-mer. The scan systematically probes the transition from a perfectly stacked H-aggregate to a slipped J-aggregate by displacing the monomers in the x-direction (along the C--C bond) from \SIrange{0.0}{3.5}{\angstrom} in steps of \SI{0.1}{\angstrom}, while maintaining a constant interplanar distance of \SI{3.5}{\angstrom}. a) A total of 3146 excited states are represented, with each state color-coded according to the exictonic level (blue: single excited; pink: double excited). The ground state is set to \SI{0}{\eV}. b) The 20 lowest biexcitonic states are represented. The lowest  biexcitonic state is set to \SI{0}{\eV}. }
        \label{fig:ethene_scan_biexciton}
\end{figure}

As shown in \ref{fig:ethene_scan_biexciton_character}, while the energy levels of the single-exciton and biexciton manifolds may appear adjacent in certain regions of the displacement coordinate, the states remain fundamentally distinct in character. The color-coding reveals a clear-cut separation between the two domains without significant mixing of character, allowing for an unambiguous classification of the biexcitonic manifold across the entire scan.

Regarding the question of degeneracy, \ref{fig:ethene_scan_biexciton_density} illustrates the energy gaps between the 20 lowest biexcitonic states. While the density of states is high, a unique mathematical "lowest" state is well-defined for the vast majority of the scan. Significant quasi-degeneracy is only observed in a localized region near a displacement of \SI{2.4}{\angstrom}, where three states become nearly energetically equivalent.

\begin{figure}[ht]
    \centering
    \includegraphics[width=0.5\linewidth]{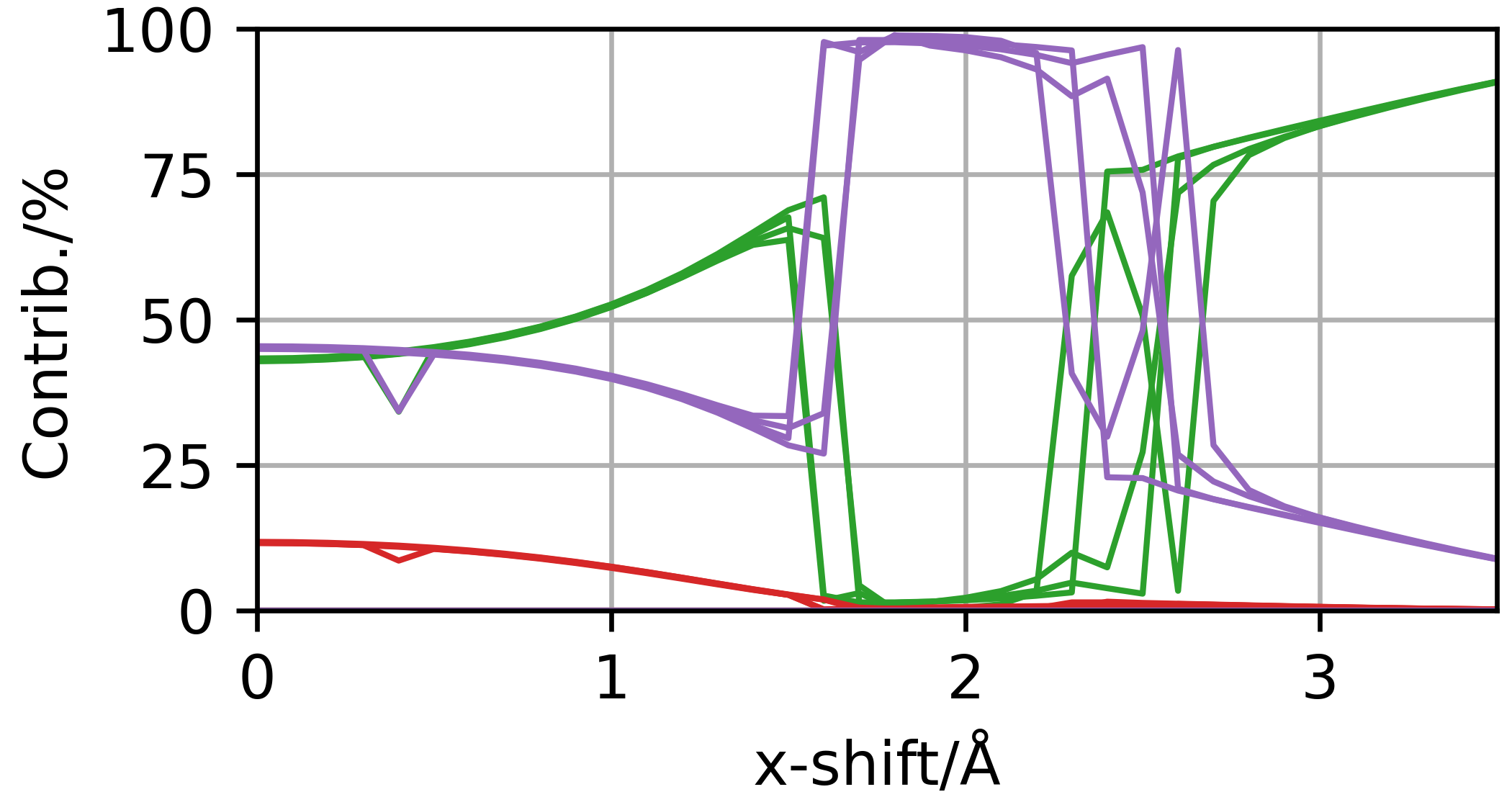}
    \caption{The composition of to 5 lowest biexcitonic state along the scan coordinate.}
    \label{fig:ethene_scan_paper_5_states}
\end{figure}

The composition of these states, visualized in \ref{fig:ethene_scan_paper_5_states}, reveals that the five lowest biexcitonic states share similar diabatic contributions. However, they remain energetically distinct (exhibiting small but finite splitting) except at specific crossing points. The apparent complexity in the spectrum arises primarily from the intersection of diabatic wavefunctions, which triggers a change in the identity of the adiabatic lowest biexcitonic state rather than a broad collapse into a degenerate manifold. Thus, while the states are close in energy, they remain distinct electronic entities.

\section{Lowest Adiabatic Doubly-Excited Wavefunctions for Ethene}
\label{sec:ethene_adiabatic_analysis}

In \reftable{tab:ethene_adiabatic_wavefuncitons}, we compare the energies and compositions of the three lowest biexcitonic states for the four ethene aggregate types as calculated by \noci{} and \symci{}. While a systematic energy offset exists between the two methods, the relative energy spacings and the underlying diabatic contributions remain highly consistent. This agreement confirms that \symci{} accurately captures the electronic nature of the doubly-excited manifold.

\begin{table*}[ht]
    \centering
    \begin{tabular}{|llr|r|r|r|r|r|r|}\hline
 H Aggregate & $\Psi$ & $\Delta E$/\si{\eV} &GS & LELE & LE..LE & LECT & CTCT & CT..CT \\ \hline
 & $\Psi_{1}^{\mathrm{NOCI}}$ & \num[round-mode=places,round-precision=\precision]{16.6557989330} & \num[round-mode=places,round-precision=\precision]{0.0002003793} & \num[round-mode=places,round-precision=\precision]{0.1289336096} & \num[round-mode=places,round-precision=\precision]{0.1821925875} & \num[round-mode=places,round-precision=\precision]{0.2483743009} & \num[round-mode=places,round-precision=\precision]{0.1325440189} & \num[round-mode=places,round-precision=\precision]{0.0000000000}  \\
& $\Psi_{1}^{\mathrm{SymCI}}$ & \num[round-mode=places,round-precision=\precision]{16.0420882607} & \num[round-mode=places,round-precision=\precision]{0.0013572007} & \num[round-mode=places,round-precision=\precision]{0.2973031642} & \num[round-mode=places,round-precision=\precision]{0.3442747044} & \num[round-mode=places,round-precision=\precision]{0.3172452814} & \num[round-mode=places,round-precision=\precision]{0.0398196492} & \num[round-mode=places,round-precision=\precision]{0.0000000000}  \\
\hline
 & $\Psi_{2}^{\mathrm{NOCI}}$ & \num[round-mode=places,round-precision=\precision]{16.6921313058} & \num[round-mode=places,round-precision=\precision]{0.0000000002} & \num[round-mode=places,round-precision=\precision]{0.1315133646} & \num[round-mode=places,round-precision=\precision]{0.1804781797} & \num[round-mode=places,round-precision=\precision]{0.2492475474} & \num[round-mode=places,round-precision=\precision]{0.1342181546} & \num[round-mode=places,round-precision=\precision]{0.0000000000}  \\
& $\Psi_{2}^{\mathrm{SymCI}}$ & \num[round-mode=places,round-precision=\precision]{16.0629957581} & \num[round-mode=places,round-precision=\precision]{0.0000000000} & \num[round-mode=places,round-precision=\precision]{0.3087373769} & \num[round-mode=places,round-precision=\precision]{0.3373447309} & \num[round-mode=places,round-precision=\precision]{0.3149634161} & \num[round-mode=places,round-precision=\precision]{0.0389544761} & \num[round-mode=places,round-precision=\precision]{0.0000000000}  \\
\hline
 & $\Psi_{3}^{\mathrm{NOCI}}$ & \num[round-mode=places,round-precision=\precision]{16.7603425494} & \num[round-mode=places,round-precision=\precision]{0.0000267215} & \num[round-mode=places,round-precision=\precision]{0.1305028424} & \num[round-mode=places,round-precision=\precision]{0.1787455979} & \num[round-mode=places,round-precision=\precision]{0.2519501551} & \num[round-mode=places,round-precision=\precision]{0.1394770133} & \num[round-mode=places,round-precision=\precision]{0.0000000000}  \\
& $\Psi_{3}^{\mathrm{SymCI}}$ & \num[round-mode=places,round-precision=\precision]{16.1389963696} & \num[round-mode=places,round-precision=\precision]{0.0002665976} & \num[round-mode=places,round-precision=\precision]{0.3063086195} & \num[round-mode=places,round-precision=\precision]{0.3367294919} & \num[round-mode=places,round-precision=\precision]{0.3170563560} & \num[round-mode=places,round-precision=\precision]{0.0396389351} & \num[round-mode=places,round-precision=\precision]{0.0000000000}  \\
\hline
\hline
 Zero Davydov & $\Psi$ & $\Delta E$/\si{\eV} &GS & LELE & LE..LE & LECT & CTCT & CT..CT \\ \hline
 & $\Psi_{1}^{\mathrm{NOCI}}$ & \num[round-mode=places,round-precision=\precision]{18.7706737601} & \num[round-mode=places,round-precision=\precision]{0.0006911004} & \num[round-mode=places,round-precision=\precision]{0.2660267553} & \num[round-mode=places,round-precision=\precision]{0.0840842271} & \num[round-mode=places,round-precision=\precision]{0.1777324614} & \num[round-mode=places,round-precision=\precision]{0.3665901539} & \num[round-mode=places,round-precision=\precision]{0.0000000000}  \\
& $\Psi_{1}^{\mathrm{SymCI}}$ & \num[round-mode=places,round-precision=\precision]{17.9015939501} & \num[round-mode=places,round-precision=\precision]{0.0000000054} & \num[round-mode=places,round-precision=\precision]{0.7942864008} & \num[round-mode=places,round-precision=\precision]{0.1031766318} & \num[round-mode=places,round-precision=\precision]{0.1005683208} & \num[round-mode=places,round-precision=\precision]{0.0019686411} & \num[round-mode=places,round-precision=\precision]{0.0000000000}  \\
\hline
 & $\Psi_{2}^{\mathrm{NOCI}}$ & \num[round-mode=places,round-precision=\precision]{18.8289810773} & \num[round-mode=places,round-precision=\precision]{0.0000000011} & \num[round-mode=places,round-precision=\precision]{0.2937621646} & \num[round-mode=places,round-precision=\precision]{0.1114314308} & \num[round-mode=places,round-precision=\precision]{0.1939288002} & \num[round-mode=places,round-precision=\precision]{0.2953872158} & \num[round-mode=places,round-precision=\precision]{0.0000000000}  \\
& $\Psi_{2}^{\mathrm{SymCI}}$ & \num[round-mode=places,round-precision=\precision]{17.9016824260} & \num[round-mode=places,round-precision=\precision]{0.0000000000} & \num[round-mode=places,round-precision=\precision]{0.8374905107} & \num[round-mode=places,round-precision=\precision]{0.0648856192} & \num[round-mode=places,round-precision=\precision]{0.0958280365} & \num[round-mode=places,round-precision=\precision]{0.0017958336} & \num[round-mode=places,round-precision=\precision]{0.0000000000}  \\
\hline
 & $\Psi_{3}^{\mathrm{NOCI}}$ & \num[round-mode=places,round-precision=\precision]{18.8940782673} & \num[round-mode=places,round-precision=\precision]{0.0000461538} & \num[round-mode=places,round-precision=\precision]{0.3217075327} & \num[round-mode=places,round-precision=\precision]{0.1564805674} & \num[round-mode=places,round-precision=\precision]{0.2105119909} & \num[round-mode=places,round-precision=\precision]{0.2073999748} & \num[round-mode=places,round-precision=\precision]{0.0000000000}  \\
& $\Psi_{3}^{\mathrm{SymCI}}$ & \num[round-mode=places,round-precision=\precision]{17.9021729173} & \num[round-mode=places,round-precision=\precision]{0.0000000000} & \num[round-mode=places,round-precision=\precision]{0.7590892690} & \num[round-mode=places,round-precision=\precision]{0.1339077026} & \num[round-mode=places,round-precision=\precision]{0.1048621009} & \num[round-mode=places,round-precision=\precision]{0.0021409275} & \num[round-mode=places,round-precision=\precision]{0.0000000000}  \\
\hline
\hline
 Zero Coupling & $\Psi$ & $\Delta E$/\si{\eV} &GS & LELE & LE..LE & LECT & CTCT & CT..CT \\ \hline
 & $\Psi_{1}^{\mathrm{NOCI}}$ & \num[round-mode=places,round-precision=\precision]{18.7772930039} & \num[round-mode=places,round-precision=\precision]{0.0004208400} & \num[round-mode=places,round-precision=\precision]{0.3280527218} & \num[round-mode=places,round-precision=\precision]{0.1715405621} & \num[round-mode=places,round-precision=\precision]{0.2035732242} & \num[round-mode=places,round-precision=\precision]{0.1741311087} & \num[round-mode=places,round-precision=\precision]{0.0000000000}  \\
& $\Psi_{1}^{\mathrm{SymCI}}$ & \num[round-mode=places,round-precision=\precision]{17.8442504611} & \num[round-mode=places,round-precision=\precision]{0.0000001110} & \num[round-mode=places,round-precision=\precision]{0.4530590085} & \num[round-mode=places,round-precision=\precision]{0.4087248299} & \num[round-mode=places,round-precision=\precision]{0.1348488629} & \num[round-mode=places,round-precision=\precision]{0.0033671878} & \num[round-mode=places,round-precision=\precision]{0.0000000000}  \\
\hline
 & $\Psi_{2}^{\mathrm{NOCI}}$ & \num[round-mode=places,round-precision=\precision]{18.8129505231} & \num[round-mode=places,round-precision=\precision]{0.0000000556} & \num[round-mode=places,round-precision=\precision]{0.3445136480} & \num[round-mode=places,round-precision=\precision]{0.1940066145} & \num[round-mode=places,round-precision=\precision]{0.2066025843} & \num[round-mode=places,round-precision=\precision]{0.1350696851} & \num[round-mode=places,round-precision=\precision]{0.0000000000}  \\
& $\Psi_{2}^{\mathrm{SymCI}}$ & \num[round-mode=places,round-precision=\precision]{17.8484008397} & \num[round-mode=places,round-precision=\precision]{0.0000000000} & \num[round-mode=places,round-precision=\precision]{0.4555073619} & \num[round-mode=places,round-precision=\precision]{0.4083815034} & \num[round-mode=places,round-precision=\precision]{0.1328659697} & \num[round-mode=places,round-precision=\precision]{0.0032451649} & \num[round-mode=places,round-precision=\precision]{0.0000000000}  \\
\hline
 & $\Psi_{3}^{\mathrm{NOCI}}$ & \num[round-mode=places,round-precision=\precision]{18.8586228199} & \num[round-mode=places,round-precision=\precision]{0.0000339431} & \num[round-mode=places,round-precision=\precision]{0.3623858269} & \num[round-mode=places,round-precision=\precision]{0.2221964254} & \num[round-mode=places,round-precision=\precision]{0.2076802492} & \num[round-mode=places,round-precision=\precision]{0.0927137179} & \num[round-mode=places,round-precision=\precision]{0.0000000000}  \\
& $\Psi_{3}^{\mathrm{SymCI}}$ & \num[round-mode=places,round-precision=\precision]{17.8555159465} & \num[round-mode=places,round-precision=\precision]{0.0000000131} & \num[round-mode=places,round-precision=\precision]{0.4598679718} & \num[round-mode=places,round-precision=\precision]{0.4074859197} & \num[round-mode=places,round-precision=\precision]{0.1295971366} & \num[round-mode=places,round-precision=\precision]{0.0030489589} & \num[round-mode=places,round-precision=\precision]{0.0000000000}  \\
\hline
\hline
 J Aggregate & $\Psi$ & $\Delta E$/\si{\eV} &GS & LELE & LE..LE & LECT & CTCT & CT..CT \\ \hline
 & $\Psi_{1}^{\mathrm{NOCI}}$ & \num[round-mode=places,round-precision=\precision]{18.8310043265} & \num[round-mode=places,round-precision=\precision]{0.0001048534} & \num[round-mode=places,round-precision=\precision]{0.4047117815} & \num[round-mode=places,round-precision=\precision]{0.3765191320} & \num[round-mode=places,round-precision=\precision]{0.1309936052} & \num[round-mode=places,round-precision=\precision]{0.0119205395} & \num[round-mode=places,round-precision=\precision]{0.0000000000}  \\
& $\Psi_{1}^{\mathrm{SymCI}}$ & \num[round-mode=places,round-precision=\precision]{17.8108922106} & \num[round-mode=places,round-precision=\precision]{0.0001041225} & \num[round-mode=places,round-precision=\precision]{0.4818772400} & \num[round-mode=places,round-precision=\precision]{0.4460250189} & \num[round-mode=places,round-precision=\precision]{0.0713206672} & \num[round-mode=places,round-precision=\precision]{0.0006729514} & \num[round-mode=places,round-precision=\precision]{0.0000000000}  \\
\hline
 & $\Psi_{2}^{\mathrm{NOCI}}$ & \num[round-mode=places,round-precision=\precision]{18.8427396373} & \num[round-mode=places,round-precision=\precision]{0.0000000043} & \num[round-mode=places,round-precision=\precision]{0.4064772593} & \num[round-mode=places,round-precision=\precision]{0.3789797383} & \num[round-mode=places,round-precision=\precision]{0.1296027886} & \num[round-mode=places,round-precision=\precision]{0.0104359807} & \num[round-mode=places,round-precision=\precision]{0.0000000000}  \\
& $\Psi_{2}^{\mathrm{SymCI}}$ & \num[round-mode=places,round-precision=\precision]{17.8205004178} & \num[round-mode=places,round-precision=\precision]{0.0000000000} & \num[round-mode=places,round-precision=\precision]{0.4839049192} & \num[round-mode=places,round-precision=\precision]{0.4448727269} & \num[round-mode=places,round-precision=\precision]{0.0705782359} & \num[round-mode=places,round-precision=\precision]{0.0006441179} & \num[round-mode=places,round-precision=\precision]{0.0000000000}  \\
\hline
 & $\Psi_{3}^{\mathrm{NOCI}}$ & \num[round-mode=places,round-precision=\precision]{18.8625031955} & \num[round-mode=places,round-precision=\precision]{0.0000130482} & \num[round-mode=places,round-precision=\precision]{0.4054607711} & \num[round-mode=places,round-precision=\precision]{0.3857178581} & \num[round-mode=places,round-precision=\precision]{0.1276494522} & \num[round-mode=places,round-precision=\precision]{0.0085611182} & \num[round-mode=places,round-precision=\precision]{0.0000000000}  \\
& $\Psi_{3}^{\mathrm{SymCI}}$ & \num[round-mode=places,round-precision=\precision]{17.8391209557} & \num[round-mode=places,round-precision=\precision]{0.0000137375} & \num[round-mode=places,round-precision=\precision]{0.4812257475} & \num[round-mode=places,round-precision=\precision]{0.4485898295} & \num[round-mode=places,round-precision=\precision]{0.0695643252} & \num[round-mode=places,round-precision=\precision]{0.0006063603} & \num[round-mode=places,round-precision=\precision]{0.0000000000}  \\
\hline
\end{tabular}

    \caption{Comparison of selected adiabatic doubly-excited wavefunctions for the ethene aggregates. The table reports energies relative to the ground state and the percentage contributions from local-excitation (\leleR{} and \sepleleR{}), charge-transfer (\lectR{}), and double charge-transfer (\ctctR{} and \sepctctR{}) configurations for the three lowest adiabatic states.}
    \label{tab:ethene_adiabatic_wavefuncitons}
\end{table*}

\section{Diagonal Correction of Adiabatic Doubly-Excited Wavefunctions}
\label{sec:diagonal_correction}

A systematic energy shift is observed between the doubly-excited wavefunctions of the four anthracene aggregates. To mitigate this, a diagonal correction was applied to the \symci{} Hamiltonian. This correction accounts for the energy discrepancies between the monomer species as calculated by \noci{} and \symci{} for the \dpR{}, \dmR{}, \gsR{}, and \soneR{} states. These offsets are mapped onto the diabatic basis and added to the diagonal elements of the \symci{} Hamiltonian. 

This feature is implemented such that the \symci{} program automatically applies a provided list of corrections, streamlining the evaluation process. Beyond addressing method-specific shifts, this approach can also incorporate dynamic correlation at the monomer level, shifting the resulting spectra into more physically meaningful ranges. However, as this correction does not arise from a strictly derived perturbation theory, it is treated as a heuristic refinement and is therefore omitted from the main text.

Upon applying this correction to the anthracene aggregates, the agreement with \noci{} results improves significantly. The remaining energy difference between the two methods is reduced to a maximum of \SI{0.15}{\eV}, as detailed in \reftable{tab:anthracene_adiabatic_table_CASSCF}.
Note that for the Zero Davydov aggregate, the agreement in diabatic contributions between the two methods initially appears poor. However, as the \noci{} wavefunctions are degenerate, their specific ordering and mixing are mathematically arbitrary. Consequently, the correspondence between the methods is significantly more consistent than a preliminary inspection suggests.

\begin{table*}[ht]
    \centering
    \begin{tabular}{|llr|r|r|r|r|r|r|}\hline
 H Aggregate & $\Psi$ & $\Delta E$/\si{\eV} &GS & LELE & LE..LE & LECT & CTCT & CT..CT \\ \hline
 & $\Psi_{1}^{\mathrm{NOCI}}$ & \num[round-mode=places,round-precision=\precision]{9.2987781432} & \num[round-mode=places,round-precision=\precision]{0.0000802175} & \num[round-mode=places,round-precision=\precision]{0.2815997109} & \num[round-mode=places,round-precision=\precision]{0.3157066245} & \num[round-mode=places,round-precision=\precision]{0.2665038807} & \num[round-mode=places,round-precision=\precision]{0.0260922144} & \num[round-mode=places,round-precision=\precision]{0.0001700778}  \\
& $\Psi_{1}^{\mathrm{SymCI}}$ & \num[round-mode=places,round-precision=\precision]{9.1456333889} & \num[round-mode=places,round-precision=\precision]{0.0003505878} & \num[round-mode=places,round-precision=\precision]{0.2776985649} & \num[round-mode=places,round-precision=\precision]{0.2996535920} & \num[round-mode=places,round-precision=\precision]{0.3518834352} & \num[round-mode=places,round-precision=\precision]{0.0704138019} & \num[round-mode=places,round-precision=\precision]{0.0000000181}  \\
\hline
 & $\Psi_{2}^{\mathrm{NOCI}}$ & \num[round-mode=places,round-precision=\precision]{9.4255721160} & \num[round-mode=places,round-precision=\precision]{0.0000000000} & \num[round-mode=places,round-precision=\precision]{0.2982361653} & \num[round-mode=places,round-precision=\precision]{0.3166909023} & \num[round-mode=places,round-precision=\precision]{0.2657679410} & \num[round-mode=places,round-precision=\precision]{0.0201802476} & \num[round-mode=places,round-precision=\precision]{0.0000019880}  \\
& $\Psi_{2}^{\mathrm{SymCI}}$ & \num[round-mode=places,round-precision=\precision]{9.2892902388} & \num[round-mode=places,round-precision=\precision]{0.0000000000} & \num[round-mode=places,round-precision=\precision]{0.2818127215} & \num[round-mode=places,round-precision=\precision]{0.2795035849} & \num[round-mode=places,round-precision=\precision]{0.3624658116} & \num[round-mode=places,round-precision=\precision]{0.0762170084} & \num[round-mode=places,round-precision=\precision]{0.0000008735}  \\
\hline
 & $\Psi_{3}^{\mathrm{NOCI}}$ & \num[round-mode=places,round-precision=\precision]{9.6156737374} & \num[round-mode=places,round-precision=\precision]{0.0000210429} & \num[round-mode=places,round-precision=\precision]{0.2178535862} & \num[round-mode=places,round-precision=\precision]{0.4611518042} & \num[round-mode=places,round-precision=\precision]{0.2402056159} & \num[round-mode=places,round-precision=\precision]{0.0015384883} & \num[round-mode=places,round-precision=\precision]{0.0001126641}  \\
& $\Psi_{3}^{\mathrm{SymCI}}$ & \num[round-mode=places,round-precision=\precision]{9.5625048249} & \num[round-mode=places,round-precision=\precision]{0.0000986169} & \num[round-mode=places,round-precision=\precision]{0.2385257565} & \num[round-mode=places,round-precision=\precision]{0.4084531346} & \num[round-mode=places,round-precision=\precision]{0.3415868772} & \num[round-mode=places,round-precision=\precision]{0.0113356127} & \num[round-mode=places,round-precision=\precision]{0.0000000021}  \\
\hline
\hline
 Zero Davydov & $\Psi$ & $\Delta E$/\si{\eV} &GS & LELE & LE..LE & LECT & CTCT & CT..CT \\ \hline
 & $\Psi_{1}^{\mathrm{NOCI}}$ & \num[round-mode=places,round-precision=\precision]{10.0690997800} & \num[round-mode=places,round-precision=\precision]{0.0000000000} & \num[round-mode=places,round-precision=\precision]{0.8691598364} & \num[round-mode=places,round-precision=\precision]{0.0186127601} & \num[round-mode=places,round-precision=\precision]{0.0913903529} & \num[round-mode=places,round-precision=\precision]{0.0014230600} & \num[round-mode=places,round-precision=\precision]{0.0000000025}  \\
& $\Psi_{1}^{\mathrm{SymCI}}$ & \num[round-mode=places,round-precision=\precision]{10.1097217694} & \num[round-mode=places,round-precision=\precision]{0.0000368112} & \num[round-mode=places,round-precision=\precision]{0.5348178580} & \num[round-mode=places,round-precision=\precision]{0.4157920581} & \num[round-mode=places,round-precision=\precision]{0.0492180764} & \num[round-mode=places,round-precision=\precision]{0.0001351932} & \num[round-mode=places,round-precision=\precision]{0.0000000030}  \\
\hline
 & $\Psi_{2}^{\mathrm{NOCI}}$ & \num[round-mode=places,round-precision=\precision]{10.0726378851} & \num[round-mode=places,round-precision=\precision]{0.0000052267} & \num[round-mode=places,round-precision=\precision]{0.2450836944} & \num[round-mode=places,round-precision=\precision]{0.6134260307} & \num[round-mode=places,round-precision=\precision]{0.1117878662} & \num[round-mode=places,round-precision=\precision]{0.0031627999} & \num[round-mode=places,round-precision=\precision]{0.0000078011}  \\
& $\Psi_{2}^{\mathrm{SymCI}}$ & \num[round-mode=places,round-precision=\precision]{10.1156800747} & \num[round-mode=places,round-precision=\precision]{0.0000000000} & \num[round-mode=places,round-precision=\precision]{0.7650227318} & \num[round-mode=places,round-precision=\precision]{0.1030235563} & \num[round-mode=places,round-precision=\precision]{0.1226944897} & \num[round-mode=places,round-precision=\precision]{0.0092592150} & \num[round-mode=places,round-precision=\precision]{0.0000000072}  \\
\hline
 & $\Psi_{3}^{\mathrm{NOCI}}$ & \num[round-mode=places,round-precision=\precision]{10.0747391331} & \num[round-mode=places,round-precision=\precision]{0.0000109715} & \num[round-mode=places,round-precision=\precision]{0.6431643623} & \num[round-mode=places,round-precision=\precision]{0.3028190555} & \num[round-mode=places,round-precision=\precision]{0.0461205702} & \num[round-mode=places,round-precision=\precision]{0.0001717047} & \num[round-mode=places,round-precision=\precision]{0.0000031929}  \\
& $\Psi_{3}^{\mathrm{SymCI}}$ & \num[round-mode=places,round-precision=\precision]{10.1416935782} & \num[round-mode=places,round-precision=\precision]{0.0000065130} & \num[round-mode=places,round-precision=\precision]{0.0521220959} & \num[round-mode=places,round-precision=\precision]{0.8244519053} & \num[round-mode=places,round-precision=\precision]{0.1174501414} & \num[round-mode=places,round-precision=\precision]{0.0059693445} & \num[round-mode=places,round-precision=\precision]{0.0000000000}  \\
\hline
\hline
 Zero Coupling & $\Psi$ & $\Delta E$/\si{\eV} &GS & LELE & LE..LE & LECT & CTCT & CT..CT \\ \hline
 & $\Psi_{1}^{\mathrm{NOCI}}$ & \num[round-mode=places,round-precision=\precision]{10.0333510780} & \num[round-mode=places,round-precision=\precision]{0.0000000072} & \num[round-mode=places,round-precision=\precision]{0.4337723764} & \num[round-mode=places,round-precision=\precision]{0.4333608106} & \num[round-mode=places,round-precision=\precision]{0.1012402694} & \num[round-mode=places,round-precision=\precision]{0.0020912177} & \num[round-mode=places,round-precision=\precision]{0.0000073115}  \\
& $\Psi_{1}^{\mathrm{SymCI}}$ & \num[round-mode=places,round-precision=\precision]{10.0747405132} & \num[round-mode=places,round-precision=\precision]{0.0000001573} & \num[round-mode=places,round-precision=\precision]{0.4234842851} & \num[round-mode=places,round-precision=\precision]{0.4036531161} & \num[round-mode=places,round-precision=\precision]{0.1658387369} & \num[round-mode=places,round-precision=\precision]{0.0070237013} & \num[round-mode=places,round-precision=\precision]{0.0000000033}  \\
\hline
 & $\Psi_{2}^{\mathrm{NOCI}}$ & \num[round-mode=places,round-precision=\precision]{10.0540537833} & \num[round-mode=places,round-precision=\precision]{0.0000000000} & \num[round-mode=places,round-precision=\precision]{0.4266569166} & \num[round-mode=places,round-precision=\precision]{0.4560645642} & \num[round-mode=places,round-precision=\precision]{0.0904838404} & \num[round-mode=places,round-precision=\precision]{0.0011684487} & \num[round-mode=places,round-precision=\precision]{0.0000000023}  \\
& $\Psi_{2}^{\mathrm{SymCI}}$ & \num[round-mode=places,round-precision=\precision]{10.0993969027} & \num[round-mode=places,round-precision=\precision]{0.0000000000} & \num[round-mode=places,round-precision=\precision]{0.4185384790} & \num[round-mode=places,round-precision=\precision]{0.4243519852} & \num[round-mode=places,round-precision=\precision]{0.1519143345} & \num[round-mode=places,round-precision=\precision]{0.0051951879} & \num[round-mode=places,round-precision=\precision]{0.0000000134}  \\
\hline
 & $\Psi_{3}^{\mathrm{NOCI}}$ & \num[round-mode=places,round-precision=\precision]{10.0792315067} & \num[round-mode=places,round-precision=\precision]{0.0000000034} & \num[round-mode=places,round-precision=\precision]{0.1747352299} & \num[round-mode=places,round-precision=\precision]{0.7363374551} & \num[round-mode=places,round-precision=\precision]{0.0694674647} & \num[round-mode=places,round-precision=\precision]{0.0000145988} & \num[round-mode=places,round-precision=\precision]{0.0000007744}  \\
& $\Psi_{3}^{\mathrm{SymCI}}$ & \num[round-mode=places,round-precision=\precision]{10.1324685899} & \num[round-mode=places,round-precision=\precision]{0.0000000173} & \num[round-mode=places,round-precision=\precision]{0.1653323052} & \num[round-mode=places,round-precision=\precision]{0.7162606164} & \num[round-mode=places,round-precision=\precision]{0.1182423087} & \num[round-mode=places,round-precision=\precision]{0.0001647505} & \num[round-mode=places,round-precision=\precision]{0.0000000018}  \\
\hline
\hline
 J Aggregate & $\Psi$ & $\Delta E$/\si{\eV} &GS & LELE & LE..LE & LECT & CTCT & CT..CT \\ \hline
 & $\Psi_{1}^{\mathrm{NOCI}}$ & \num[round-mode=places,round-precision=\precision]{10.0989575512} & \num[round-mode=places,round-precision=\precision]{0.0000136628} & \num[round-mode=places,round-precision=\precision]{0.5064722895} & \num[round-mode=places,round-precision=\precision]{0.4677988980} & \num[round-mode=places,round-precision=\precision]{0.0195868284} & \num[round-mode=places,round-precision=\precision]{0.0000618308} & \num[round-mode=places,round-precision=\precision]{0.0000003770}  \\
& $\Psi_{1}^{\mathrm{SymCI}}$ & \num[round-mode=places,round-precision=\precision]{10.1223515305} & \num[round-mode=places,round-precision=\precision]{0.0000526187} & \num[round-mode=places,round-precision=\precision]{0.5093209767} & \num[round-mode=places,round-precision=\precision]{0.4641995073} & \num[round-mode=places,round-precision=\precision]{0.0263401734} & \num[round-mode=places,round-precision=\precision]{0.0000867238} & \num[round-mode=places,round-precision=\precision]{0.0000000001}  \\
\hline
 & $\Psi_{2}^{\mathrm{NOCI}}$ & \num[round-mode=places,round-precision=\precision]{10.1190666397} & \num[round-mode=places,round-precision=\precision]{0.0000000000} & \num[round-mode=places,round-precision=\precision]{0.4855245312} & \num[round-mode=places,round-precision=\precision]{0.4923267218} & \num[round-mode=places,round-precision=\precision]{0.0169585897} & \num[round-mode=places,round-precision=\precision]{0.0000250604} & \num[round-mode=places,round-precision=\precision]{0.0000000000}  \\
& $\Psi_{2}^{\mathrm{SymCI}}$ & \num[round-mode=places,round-precision=\precision]{10.1579158873} & \num[round-mode=places,round-precision=\precision]{0.0000000000} & \num[round-mode=places,round-precision=\precision]{0.4929827644} & \num[round-mode=places,round-precision=\precision]{0.4835775758} & \num[round-mode=places,round-precision=\precision]{0.0233940541} & \num[round-mode=places,round-precision=\precision]{0.0000456054} & \num[round-mode=places,round-precision=\precision]{0.0000000002}  \\
\hline
 & $\Psi_{3}^{\mathrm{NOCI}}$ & \num[round-mode=places,round-precision=\precision]{10.1424367479} & \num[round-mode=places,round-precision=\precision]{0.0000017360} & \num[round-mode=places,round-precision=\precision]{0.3845813934} & \num[round-mode=places,round-precision=\precision]{0.5982190157} & \num[round-mode=places,round-precision=\precision]{0.0132428274} & \num[round-mode=places,round-precision=\precision]{0.0000026656} & \num[round-mode=places,round-precision=\precision]{0.0000000146}  \\
& $\Psi_{3}^{\mathrm{SymCI}}$ & \num[round-mode=places,round-precision=\precision]{10.2001873324} & \num[round-mode=places,round-precision=\precision]{0.0000064993} & \num[round-mode=places,round-precision=\precision]{0.4084092356} & \num[round-mode=places,round-precision=\precision]{0.5725381498} & \num[round-mode=places,round-precision=\precision]{0.0190373092} & \num[round-mode=places,round-precision=\precision]{0.0000088060} & \num[round-mode=places,round-precision=\precision]{0.0000000000}  \\
\hline
\end{tabular}

    \caption{Comparison of selected adiabatic doubly-excited wavefunctions obtained from diagonalization of the diabatic Hamiltonians using \symci{} with diagonal corrections and \noci{} for the anthracene aggregate. The table reports the energies relative to the ground state and the contributions of \leleR{}, separated \leleR{} (\sepleleR{}), \lectR{}, \ctctR{}, and separated \ctctR{} (\sepctctR{}) for the two lowest adiabatic states across all four aggregate types.}
    \label{tab:anthracene_adiabatic_table_CASSCF}
\end{table*}

\section{Expansion of the Active Space in \symci{}}
\label{sec:big_AS}

This section explores the implications and computational feasibility of extending \symci{} calculations beyond the minimal HOMO-LUMO active space per fragment utilized in the primary study. While the current implementation natively supports arbitrary active spaces per monomer, the practical application of this feature is inherently governed by the exponential scaling of Configuration Interaction (CI) methods.

To validate the robustness of the HOMO-LUMO approximation, we performed comparative benchmarks on the four anthracene pentamer aggregates discussed in the main text. Specifically, results obtained using the standard HOMO/LUMO active space (Table~\ref{tab:anthracene_adiabatic_table_newNormal}) were compared against an expanded (HOMO-1, HOMO, LUMO, LUMO+1) active space (Table~\ref{tab:anthracene_adiabatic_table_4_AS}). Our findings indicate that both configurations yield nearly identical adiabatic energies and wave function compositions, justifying the use of the minimal active space for these systems.

Furthermore, we conducted a computational stress test on a 15-mer anthracene crystal cutout. This test compared the $(4\text{e}, 4\text{orb})$ active space against a further expanded $(6\text{e}, 6\text{orb})$ active space per fragment (spanning HOMO-2 to LUMO+2). While the former calculation converged successfully, the latter exceeded available memory limits when attempting to construct a sparse Hamiltonian matrix exceeding dimensions of $2 \times 10^6$ by $2 \times 10^6$. Although the integral transformation and state generation remained tractable, the sheer magnitude of the resulting Hilbert space underscores a bottleneck in matrix processing and storage. These results suggest that for larger clusters or wider active spaces, the implementation of intelligent state-selection algorithms and meaningful monomer-based multiconfigurational projections will be essential to maintain interpretability and computational efficiency.

\begin{table*}[ht]
    \centering
    \begin{tabular}{|llr|r|r|r|r|r|r|}\hline
 H Aggregate & $\Psi$ & $\Delta E$/\si{\eV} &GS & LELE & LE..LE & LECT & CTCT & CT..CT \\ \hline
 & $\Psi_{1}^{\mathrm{NOCI}}$ & \num[round-mode=places,round-precision=\precision]{9.2987781432} & \num[round-mode=places,round-precision=\precision]{0.0000802175} & \num[round-mode=places,round-precision=\precision]{0.2815997109} & \num[round-mode=places,round-precision=\precision]{0.3157066245} & \num[round-mode=places,round-precision=\precision]{0.2665038807} & \num[round-mode=places,round-precision=\precision]{0.0260922144} & \num[round-mode=places,round-precision=\precision]{0.0001700778}  \\
& $\Psi_{1}^{\mathrm{SymCI}}$ & \num[round-mode=places,round-precision=\precision]{8.1929581933} & \num[round-mode=places,round-precision=\precision]{0.0005022207} & \num[round-mode=places,round-precision=\precision]{0.3128675177} & \num[round-mode=places,round-precision=\precision]{0.3436194966} & \num[round-mode=places,round-precision=\precision]{0.3084769245} & \num[round-mode=places,round-precision=\precision]{0.0345338347} & \num[round-mode=places,round-precision=\precision]{0.0000000058}  \\
\hline
 & $\Psi_{2}^{\mathrm{NOCI}}$ & \num[round-mode=places,round-precision=\precision]{9.4255721160} & \num[round-mode=places,round-precision=\precision]{0.0000000000} & \num[round-mode=places,round-precision=\precision]{0.2982361653} & \num[round-mode=places,round-precision=\precision]{0.3166909023} & \num[round-mode=places,round-precision=\precision]{0.2657679410} & \num[round-mode=places,round-precision=\precision]{0.0201802476} & \num[round-mode=places,round-precision=\precision]{0.0000019880}  \\
& $\Psi_{2}^{\mathrm{SymCI}}$ & \num[round-mode=places,round-precision=\precision]{8.3374977728} & \num[round-mode=places,round-precision=\precision]{0.0000000000} & \num[round-mode=places,round-precision=\precision]{0.3198958986} & \num[round-mode=places,round-precision=\precision]{0.3368295707} & \num[round-mode=places,round-precision=\precision]{0.3122958154} & \num[round-mode=places,round-precision=\precision]{0.0309782294} & \num[round-mode=places,round-precision=\precision]{0.0000004859}  \\
\hline
 & $\Psi_{3}^{\mathrm{NOCI}}$ & \num[round-mode=places,round-precision=\precision]{9.6156737374} & \num[round-mode=places,round-precision=\precision]{0.0000210429} & \num[round-mode=places,round-precision=\precision]{0.2178535862} & \num[round-mode=places,round-precision=\precision]{0.4611518042} & \num[round-mode=places,round-precision=\precision]{0.2402056159} & \num[round-mode=places,round-precision=\precision]{0.0015384883} & \num[round-mode=places,round-precision=\precision]{0.0001126641}  \\
& $\Psi_{3}^{\mathrm{SymCI}}$ & \num[round-mode=places,round-precision=\precision]{8.5885133255} & \num[round-mode=places,round-precision=\precision]{0.0001171639} & \num[round-mode=places,round-precision=\precision]{0.2653664544} & \num[round-mode=places,round-precision=\precision]{0.4406220895} & \num[round-mode=places,round-precision=\precision]{0.2898063616} & \num[round-mode=places,round-precision=\precision]{0.0040879301} & \num[round-mode=places,round-precision=\precision]{0.0000000005}  \\
\hline
\hline
 Zero Davydov & $\Psi$ & $\Delta E$/\si{\eV} &GS & LELE & LE..LE & LECT & CTCT & CT..CT \\ \hline
 & $\Psi_{1}^{\mathrm{NOCI}}$ & \num[round-mode=places,round-precision=\precision]{10.0690997800} & \num[round-mode=places,round-precision=\precision]{0.0000000000} & \num[round-mode=places,round-precision=\precision]{0.8691598364} & \num[round-mode=places,round-precision=\precision]{0.0186127601} & \num[round-mode=places,round-precision=\precision]{0.0913903529} & \num[round-mode=places,round-precision=\precision]{0.0014230600} & \num[round-mode=places,round-precision=\precision]{0.0000000025}  \\
& $\Psi_{1}^{\mathrm{SymCI}}$ & \num[round-mode=places,round-precision=\precision]{9.0723953763} & \num[round-mode=places,round-precision=\precision]{0.0000471175} & \num[round-mode=places,round-precision=\precision]{0.4904440795} & \num[round-mode=places,round-precision=\precision]{0.4845992098} & \num[round-mode=places,round-precision=\precision]{0.0248821897} & \num[round-mode=places,round-precision=\precision]{0.0000274027} & \num[round-mode=places,round-precision=\precision]{0.0000000008}  \\
\hline
 & $\Psi_{2}^{\mathrm{NOCI}}$ & \num[round-mode=places,round-precision=\precision]{10.0726378851} & \num[round-mode=places,round-precision=\precision]{0.0000052267} & \num[round-mode=places,round-precision=\precision]{0.2450836944} & \num[round-mode=places,round-precision=\precision]{0.6134260307} & \num[round-mode=places,round-precision=\precision]{0.1117878662} & \num[round-mode=places,round-precision=\precision]{0.0031627999} & \num[round-mode=places,round-precision=\precision]{0.0000078011}  \\
& $\Psi_{2}^{\mathrm{SymCI}}$ & \num[round-mode=places,round-precision=\precision]{9.0880111717} & \num[round-mode=places,round-precision=\precision]{0.0000000000} & \num[round-mode=places,round-precision=\precision]{0.5888011308} & \num[round-mode=places,round-precision=\precision]{0.3633599853} & \num[round-mode=places,round-precision=\precision]{0.0471401291} & \num[round-mode=places,round-precision=\precision]{0.0006987543} & \num[round-mode=places,round-precision=\precision]{0.0000000004}  \\
\hline
 & $\Psi_{3}^{\mathrm{NOCI}}$ & \num[round-mode=places,round-precision=\precision]{10.0747391331} & \num[round-mode=places,round-precision=\precision]{0.0000109715} & \num[round-mode=places,round-precision=\precision]{0.6431643623} & \num[round-mode=places,round-precision=\precision]{0.3028190555} & \num[round-mode=places,round-precision=\precision]{0.0461205702} & \num[round-mode=places,round-precision=\precision]{0.0001717047} & \num[round-mode=places,round-precision=\precision]{0.0000031929}  \\
& $\Psi_{3}^{\mathrm{SymCI}}$ & \num[round-mode=places,round-precision=\precision]{9.1141277896} & \num[round-mode=places,round-precision=\precision]{0.0000066307} & \num[round-mode=places,round-precision=\precision]{0.1511720161} & \num[round-mode=places,round-precision=\precision]{0.7908148452} & \num[round-mode=places,round-precision=\precision]{0.0576505244} & \num[round-mode=places,round-precision=\precision]{0.0003559835} & \num[round-mode=places,round-precision=\precision]{0.0000000001}  \\
\hline
\hline
 Zero Coupling & $\Psi$ & $\Delta E$/\si{\eV} &GS & LELE & LE..LE & LECT & CTCT & CT..CT \\ \hline
 & $\Psi_{1}^{\mathrm{NOCI}}$ & \num[round-mode=places,round-precision=\precision]{10.0333510780} & \num[round-mode=places,round-precision=\precision]{0.0000000072} & \num[round-mode=places,round-precision=\precision]{0.4337723764} & \num[round-mode=places,round-precision=\precision]{0.4333608106} & \num[round-mode=places,round-precision=\precision]{0.1012402694} & \num[round-mode=places,round-precision=\precision]{0.0020912177} & \num[round-mode=places,round-precision=\precision]{0.0000073115}  \\
& $\Psi_{1}^{\mathrm{SymCI}}$ & \num[round-mode=places,round-precision=\precision]{9.0608314925} & \num[round-mode=places,round-precision=\precision]{0.0000002066} & \num[round-mode=places,round-precision=\precision]{0.4528178261} & \num[round-mode=places,round-precision=\precision]{0.4269506221} & \num[round-mode=places,round-precision=\precision]{0.1178173140} & \num[round-mode=places,round-precision=\precision]{0.0024140299} & \num[round-mode=places,round-precision=\precision]{0.0000000014}  \\
\hline
 & $\Psi_{2}^{\mathrm{NOCI}}$ & \num[round-mode=places,round-precision=\precision]{10.0540537833} & \num[round-mode=places,round-precision=\precision]{0.0000000000} & \num[round-mode=places,round-precision=\precision]{0.4266569166} & \num[round-mode=places,round-precision=\precision]{0.4560645642} & \num[round-mode=places,round-precision=\precision]{0.0904838404} & \num[round-mode=places,round-precision=\precision]{0.0011684487} & \num[round-mode=places,round-precision=\precision]{0.0000000023}  \\
& $\Psi_{2}^{\mathrm{SymCI}}$ & \num[round-mode=places,round-precision=\precision]{9.0822820028} & \num[round-mode=places,round-precision=\precision]{0.0000000000} & \num[round-mode=places,round-precision=\precision]{0.4434483061} & \num[round-mode=places,round-precision=\precision]{0.4490192538} & \num[round-mode=places,round-precision=\precision]{0.1058832204} & \num[round-mode=places,round-precision=\precision]{0.0016492142} & \num[round-mode=places,round-precision=\precision]{0.0000000055}  \\
\hline
 & $\Psi_{3}^{\mathrm{NOCI}}$ & \num[round-mode=places,round-precision=\precision]{10.0792315067} & \num[round-mode=places,round-precision=\precision]{0.0000000034} & \num[round-mode=places,round-precision=\precision]{0.1747352299} & \num[round-mode=places,round-precision=\precision]{0.7363374551} & \num[round-mode=places,round-precision=\precision]{0.0694674647} & \num[round-mode=places,round-precision=\precision]{0.0000145988} & \num[round-mode=places,round-precision=\precision]{0.0000007744}  \\
& $\Psi_{3}^{\mathrm{SymCI}}$ & \num[round-mode=places,round-precision=\precision]{9.1082371970} & \num[round-mode=places,round-precision=\precision]{0.0000000225} & \num[round-mode=places,round-precision=\precision]{0.1626881227} & \num[round-mode=places,round-precision=\precision]{0.7553528680} & \num[round-mode=places,round-precision=\precision]{0.0819078334} & \num[round-mode=places,round-precision=\precision]{0.0000511526} & \num[round-mode=places,round-precision=\precision]{0.0000000007}  \\
\hline
\hline
 J Aggregate & $\Psi$ & $\Delta E$/\si{\eV} &GS & LELE & LE..LE & LECT & CTCT & CT..CT \\ \hline
 & $\Psi_{1}^{\mathrm{NOCI}}$ & \num[round-mode=places,round-precision=\precision]{10.0989575512} & \num[round-mode=places,round-precision=\precision]{0.0000136628} & \num[round-mode=places,round-precision=\precision]{0.5064722895} & \num[round-mode=places,round-precision=\precision]{0.4677988980} & \num[round-mode=places,round-precision=\precision]{0.0195868284} & \num[round-mode=places,round-precision=\precision]{0.0000618308} & \num[round-mode=places,round-precision=\precision]{0.0000003770}  \\
& $\Psi_{1}^{\mathrm{SymCI}}$ & \num[round-mode=places,round-precision=\precision]{9.0824806755} & \num[round-mode=places,round-precision=\precision]{0.0000660929} & \num[round-mode=places,round-precision=\precision]{0.5136513067} & \num[round-mode=places,round-precision=\precision]{0.4668354441} & \num[round-mode=places,round-precision=\precision]{0.0194063852} & \num[round-mode=places,round-precision=\precision]{0.0000407710} & \num[round-mode=places,round-precision=\precision]{0.0000000000}  \\
\hline
 & $\Psi_{2}^{\mathrm{NOCI}}$ & \num[round-mode=places,round-precision=\precision]{10.1190666397} & \num[round-mode=places,round-precision=\precision]{0.0000000000} & \num[round-mode=places,round-precision=\precision]{0.4855245312} & \num[round-mode=places,round-precision=\precision]{0.4923267218} & \num[round-mode=places,round-precision=\precision]{0.0169585897} & \num[round-mode=places,round-precision=\precision]{0.0000250604} & \num[round-mode=places,round-precision=\precision]{0.0000000000}  \\
& $\Psi_{2}^{\mathrm{SymCI}}$ & \num[round-mode=places,round-precision=\precision]{9.1174251866} & \num[round-mode=places,round-precision=\precision]{0.0000000000} & \num[round-mode=places,round-precision=\precision]{0.4982738244} & \num[round-mode=places,round-precision=\precision]{0.4846372607} & \num[round-mode=places,round-precision=\precision]{0.0170680842} & \num[round-mode=places,round-precision=\precision]{0.0000208306} & \num[round-mode=places,round-precision=\precision]{0.0000000001}  \\
\hline
 & $\Psi_{3}^{\mathrm{NOCI}}$ & \num[round-mode=places,round-precision=\precision]{10.1424367479} & \num[round-mode=places,round-precision=\precision]{0.0000017360} & \num[round-mode=places,round-precision=\precision]{0.3845813934} & \num[round-mode=places,round-precision=\precision]{0.5982190157} & \num[round-mode=places,round-precision=\precision]{0.0132428274} & \num[round-mode=places,round-precision=\precision]{0.0000026656} & \num[round-mode=places,round-precision=\precision]{0.0000000146}  \\
& $\Psi_{3}^{\mathrm{SymCI}}$ & \num[round-mode=places,round-precision=\precision]{9.1589018908} & \num[round-mode=places,round-precision=\precision]{0.0000081081} & \num[round-mode=places,round-precision=\precision]{0.4154215154} & \num[round-mode=places,round-precision=\precision]{0.5708247420} & \num[round-mode=places,round-precision=\precision]{0.0137415376} & \num[round-mode=places,round-precision=\precision]{0.0000040969} & \num[round-mode=places,round-precision=\precision]{0.0000000000}  \\
\hline
\end{tabular}

    \caption{Comparison of selected adiabatic doubly-excited wavefunctions obtained via diabatic Hamiltonian diagonalization using \symci{} (HOMO/LUMO active space) and \noci{} for the anthracene 5-mer. Reported values include excitation energies and the percentage contributions of \leleR{}, \sepleleR{}, \lectR{}, \ctctR{}, and \sepctctR{} for the two lowest adiabatic states across all aggregate types.}
    \label{tab:anthracene_adiabatic_table_newNormal}
\end{table*}

\begin{table*}[ht]
    \centering
    \begin{tabular}{|llr|r|r|r|r|r|r|}\hline
 H Aggregate & $\Psi$ & $\Delta E$/\si{\eV} &GS & LELE & LE..LE & LECT & CTCT & CT..CT \\ \hline
 & $\Psi_{1}^{\mathrm{NOCI}}$ & \num[round-mode=places,round-precision=\precision]{9.2987781432} & \num[round-mode=places,round-precision=\precision]{0.0000802175} & \num[round-mode=places,round-precision=\precision]{0.2815997109} & \num[round-mode=places,round-precision=\precision]{0.3157066245} & \num[round-mode=places,round-precision=\precision]{0.2665038807} & \num[round-mode=places,round-precision=\precision]{0.0260922144} & \num[round-mode=places,round-precision=\precision]{0.0001700778}  \\
& $\Psi_{1}^{\mathrm{SymCI}}$ & \num[round-mode=places,round-precision=\precision]{8.2163453307} & \num[round-mode=places,round-precision=\precision]{0.0001698318} & \num[round-mode=places,round-precision=\precision]{0.3425639699} & \num[round-mode=places,round-precision=\precision]{0.3626327793} & \num[round-mode=places,round-precision=\precision]{0.2763207357} & \num[round-mode=places,round-precision=\precision]{0.0183126769} & \num[round-mode=places,round-precision=\precision]{0.0000000064}  \\
\hline
 & $\Psi_{2}^{\mathrm{NOCI}}$ & \num[round-mode=places,round-precision=\precision]{9.4255721160} & \num[round-mode=places,round-precision=\precision]{0.0000000000} & \num[round-mode=places,round-precision=\precision]{0.2982361653} & \num[round-mode=places,round-precision=\precision]{0.3166909023} & \num[round-mode=places,round-precision=\precision]{0.2657679410} & \num[round-mode=places,round-precision=\precision]{0.0201802476} & \num[round-mode=places,round-precision=\precision]{0.0000019880}  \\
& $\Psi_{2}^{\mathrm{SymCI}}$ & \num[round-mode=places,round-precision=\precision]{8.3227040219} & \num[round-mode=places,round-precision=\precision]{0.0000000000} & \num[round-mode=places,round-precision=\precision]{0.3470586904} & \num[round-mode=places,round-precision=\precision]{0.3695156786} & \num[round-mode=places,round-precision=\precision]{0.2694875728} & \num[round-mode=places,round-precision=\precision]{0.0139379038} & \num[round-mode=places,round-precision=\precision]{0.0000001544}  \\
\hline
 & $\Psi_{3}^{\mathrm{NOCI}}$ & \num[round-mode=places,round-precision=\precision]{9.6156737374} & \num[round-mode=places,round-precision=\precision]{0.0000210429} & \num[round-mode=places,round-precision=\precision]{0.2178535862} & \num[round-mode=places,round-precision=\precision]{0.4611518042} & \num[round-mode=places,round-precision=\precision]{0.2402056159} & \num[round-mode=places,round-precision=\precision]{0.0015384883} & \num[round-mode=places,round-precision=\precision]{0.0001126641}  \\
& $\Psi_{3}^{\mathrm{SymCI}}$ & \num[round-mode=places,round-precision=\precision]{8.5022791785} & \num[round-mode=places,round-precision=\precision]{0.0000314341} & \num[round-mode=places,round-precision=\precision]{0.2522761685} & \num[round-mode=places,round-precision=\precision]{0.5113239843} & \num[round-mode=places,round-precision=\precision]{0.2352151704} & \num[round-mode=places,round-precision=\precision]{0.0011532418} & \num[round-mode=places,round-precision=\precision]{0.0000000009}  \\
\hline
\hline
 Zero Davydov & $\Psi$ & $\Delta E$/\si{\eV} &GS & LELE & LE..LE & LECT & CTCT & CT..CT \\ \hline
 & $\Psi_{1}^{\mathrm{NOCI}}$ & \num[round-mode=places,round-precision=\precision]{10.0690997800} & \num[round-mode=places,round-precision=\precision]{0.0000000000} & \num[round-mode=places,round-precision=\precision]{0.8691598364} & \num[round-mode=places,round-precision=\precision]{0.0186127601} & \num[round-mode=places,round-precision=\precision]{0.0913903529} & \num[round-mode=places,round-precision=\precision]{0.0014230600} & \num[round-mode=places,round-precision=\precision]{0.0000000025}  \\
& $\Psi_{1}^{\mathrm{SymCI}}$ & \num[round-mode=places,round-precision=\precision]{8.8171337323} & \num[round-mode=places,round-precision=\precision]{0.0000116166} & \num[round-mode=places,round-precision=\precision]{0.5288685658} & \num[round-mode=places,round-precision=\precision]{0.4543641421} & \num[round-mode=places,round-precision=\precision]{0.0167410809} & \num[round-mode=places,round-precision=\precision]{0.0000145942} & \num[round-mode=places,round-precision=\precision]{0.0000000004}  \\
\hline
 & $\Psi_{2}^{\mathrm{NOCI}}$ & \num[round-mode=places,round-precision=\precision]{10.0726378851} & \num[round-mode=places,round-precision=\precision]{0.0000052267} & \num[round-mode=places,round-precision=\precision]{0.2450836944} & \num[round-mode=places,round-precision=\precision]{0.6134260307} & \num[round-mode=places,round-precision=\precision]{0.1117878662} & \num[round-mode=places,round-precision=\precision]{0.0031627999} & \num[round-mode=places,round-precision=\precision]{0.0000078011}  \\
& $\Psi_{2}^{\mathrm{SymCI}}$ & \num[round-mode=places,round-precision=\precision]{8.8198850952} & \num[round-mode=places,round-precision=\precision]{0.0000000000} & \num[round-mode=places,round-precision=\precision]{0.8467168159} & \num[round-mode=places,round-precision=\precision]{0.1136074664} & \num[round-mode=places,round-precision=\precision]{0.0393756815} & \num[round-mode=places,round-precision=\precision]{0.0003000353} & \num[round-mode=places,round-precision=\precision]{0.0000000008}  \\
\hline
 & $\Psi_{3}^{\mathrm{NOCI}}$ & \num[round-mode=places,round-precision=\precision]{10.0747391331} & \num[round-mode=places,round-precision=\precision]{0.0000109715} & \num[round-mode=places,round-precision=\precision]{0.6431643623} & \num[round-mode=places,round-precision=\precision]{0.3028190555} & \num[round-mode=places,round-precision=\precision]{0.0461205702} & \num[round-mode=places,round-precision=\precision]{0.0001717047} & \num[round-mode=places,round-precision=\precision]{0.0000031929}  \\
& $\Psi_{3}^{\mathrm{SymCI}}$ & \num[round-mode=places,round-precision=\precision]{8.8305258117} & \num[round-mode=places,round-precision=\precision]{0.0000026223} & \num[round-mode=places,round-precision=\precision]{0.0720981126} & \num[round-mode=places,round-precision=\precision]{0.8829282121} & \num[round-mode=places,round-precision=\precision]{0.0447401726} & \num[round-mode=places,round-precision=\precision]{0.0002308803} & \num[round-mode=places,round-precision=\precision]{0.0000000002}  \\
\hline
\hline
 Zero Coupling & $\Psi$ & $\Delta E$/\si{\eV} &GS & LELE & LE..LE & LECT & CTCT & CT..CT \\ \hline
 & $\Psi_{1}^{\mathrm{NOCI}}$ & \num[round-mode=places,round-precision=\precision]{10.0333510780} & \num[round-mode=places,round-precision=\precision]{0.0000000072} & \num[round-mode=places,round-precision=\precision]{0.4337723764} & \num[round-mode=places,round-precision=\precision]{0.4333608106} & \num[round-mode=places,round-precision=\precision]{0.1012402694} & \num[round-mode=places,round-precision=\precision]{0.0020912177} & \num[round-mode=places,round-precision=\precision]{0.0000073115}  \\
& $\Psi_{1}^{\mathrm{SymCI}}$ & \num[round-mode=places,round-precision=\precision]{8.7573886003} & \num[round-mode=places,round-precision=\precision]{0.0000000807} & \num[round-mode=places,round-precision=\precision]{0.4734153122} & \num[round-mode=places,round-precision=\precision]{0.4547715913} & \num[round-mode=places,round-precision=\precision]{0.0711612766} & \num[round-mode=places,round-precision=\precision]{0.0006517388} & \num[round-mode=places,round-precision=\precision]{0.0000000005}  \\
\hline
 & $\Psi_{2}^{\mathrm{NOCI}}$ & \num[round-mode=places,round-precision=\precision]{10.0540537833} & \num[round-mode=places,round-precision=\precision]{0.0000000000} & \num[round-mode=places,round-precision=\precision]{0.4266569166} & \num[round-mode=places,round-precision=\precision]{0.4560645642} & \num[round-mode=places,round-precision=\precision]{0.0904838404} & \num[round-mode=places,round-precision=\precision]{0.0011684487} & \num[round-mode=places,round-precision=\precision]{0.0000000023}  \\
& $\Psi_{2}^{\mathrm{SymCI}}$ & \num[round-mode=places,round-precision=\precision]{8.7736882358} & \num[round-mode=places,round-precision=\precision]{0.0000000000} & \num[round-mode=places,round-precision=\precision]{0.4673875888} & \num[round-mode=places,round-precision=\precision]{0.4694428258} & \num[round-mode=places,round-precision=\precision]{0.0627509941} & \num[round-mode=places,round-precision=\precision]{0.0004185901} & \num[round-mode=places,round-precision=\precision]{0.0000000013}  \\
\hline
 & $\Psi_{3}^{\mathrm{NOCI}}$ & \num[round-mode=places,round-precision=\precision]{10.0792315067} & \num[round-mode=places,round-precision=\precision]{0.0000000034} & \num[round-mode=places,round-precision=\precision]{0.1747352299} & \num[round-mode=places,round-precision=\precision]{0.7363374551} & \num[round-mode=places,round-precision=\precision]{0.0694674647} & \num[round-mode=places,round-precision=\precision]{0.0000145988} & \num[round-mode=places,round-precision=\precision]{0.0000007744}  \\
& $\Psi_{3}^{\mathrm{SymCI}}$ & \num[round-mode=places,round-precision=\precision]{8.7932645151} & \num[round-mode=places,round-precision=\precision]{0.0000000106} & \num[round-mode=places,round-precision=\precision]{0.1750790058} & \num[round-mode=places,round-precision=\precision]{0.7766833144} & \num[round-mode=places,round-precision=\precision]{0.0482271581} & \num[round-mode=places,round-precision=\precision]{0.0000105109} & \num[round-mode=places,round-precision=\precision]{0.0000000002}  \\
\hline
\hline
 J Aggregate & $\Psi$ & $\Delta E$/\si{\eV} &GS & LELE & LE..LE & LECT & CTCT & CT..CT \\ \hline
 & $\Psi_{1}^{\mathrm{NOCI}}$ & \num[round-mode=places,round-precision=\precision]{10.0989575512} & \num[round-mode=places,round-precision=\precision]{0.0000136628} & \num[round-mode=places,round-precision=\precision]{0.5064722895} & \num[round-mode=places,round-precision=\precision]{0.4677988980} & \num[round-mode=places,round-precision=\precision]{0.0195868284} & \num[round-mode=places,round-precision=\precision]{0.0000618308} & \num[round-mode=places,round-precision=\precision]{0.0000003770}  \\
& $\Psi_{1}^{\mathrm{SymCI}}$ & \num[round-mode=places,round-precision=\precision]{8.7581780357} & \num[round-mode=places,round-precision=\precision]{0.0000192167} & \num[round-mode=places,round-precision=\precision]{0.5131534874} & \num[round-mode=places,round-precision=\precision]{0.4735472197} & \num[round-mode=places,round-precision=\precision]{0.0132636150} & \num[round-mode=places,round-precision=\precision]{0.0000164613} & \num[round-mode=places,round-precision=\precision]{0.0000000000}  \\
\hline
 & $\Psi_{2}^{\mathrm{NOCI}}$ & \num[round-mode=places,round-precision=\precision]{10.1190666397} & \num[round-mode=places,round-precision=\precision]{0.0000000000} & \num[round-mode=places,round-precision=\precision]{0.4855245312} & \num[round-mode=places,round-precision=\precision]{0.4923267218} & \num[round-mode=places,round-precision=\precision]{0.0169585897} & \num[round-mode=places,round-precision=\precision]{0.0000250604} & \num[round-mode=places,round-precision=\precision]{0.0000000000}  \\
& $\Psi_{2}^{\mathrm{SymCI}}$ & \num[round-mode=places,round-precision=\precision]{8.7777377864} & \num[round-mode=places,round-precision=\precision]{0.0000000000} & \num[round-mode=places,round-precision=\precision]{0.4944627608} & \num[round-mode=places,round-precision=\precision]{0.4941009048} & \num[round-mode=places,round-precision=\precision]{0.0114281188} & \num[round-mode=places,round-precision=\precision]{0.0000082155} & \num[round-mode=places,round-precision=\precision]{0.0000000000}  \\
\hline
 & $\Psi_{3}^{\mathrm{NOCI}}$ & \num[round-mode=places,round-precision=\precision]{10.1424367479} & \num[round-mode=places,round-precision=\precision]{0.0000017360} & \num[round-mode=places,round-precision=\precision]{0.3845813934} & \num[round-mode=places,round-precision=\precision]{0.5982190157} & \num[round-mode=places,round-precision=\precision]{0.0132428274} & \num[round-mode=places,round-precision=\precision]{0.0000026656} & \num[round-mode=places,round-precision=\precision]{0.0000000146}  \\
& $\Psi_{3}^{\mathrm{SymCI}}$ & \num[round-mode=places,round-precision=\precision]{8.8005034906} & \num[round-mode=places,round-precision=\precision]{0.0000024820} & \num[round-mode=places,round-precision=\precision]{0.3979372364} & \num[round-mode=places,round-precision=\precision]{0.5931757522} & \num[round-mode=places,round-precision=\precision]{0.0088832756} & \num[round-mode=places,round-precision=\precision]{0.0000012538} & \num[round-mode=places,round-precision=\precision]{0.0000000000}  \\
\hline
\end{tabular}

    \caption{Comparison of adiabatic doubly-excited wavefunctions using an expanded (HOMO-1, HOMO, LUMO, LUMO+1) active space per monomer. The data illustrates the convergence of state compositions and energies relative to the minimal active space presented in Table~\ref{tab:anthracene_adiabatic_table_newNormal}.}
    \label{tab:anthracene_adiabatic_table_4_AS}
\end{table*}